\documentclass[aps,prb,amsmath,amssymb,citeautoscript,reprint]{revtex4-2}
\usepackage{graphicx, float, xcolor, pgf}
\usepackage{physics, siunitx}
\usepackage{textgreek}
\usepackage{enumitem}
\usepackage[frozencache,cachedir=.]{minted}
\usepackage[caption=false, justification=centerlast]{subfig}

\usepackage{hyperref}
\hypersetup{
    pdfauthor={Amartya Bose},
    pdftitle={QuantumDynamics.jl: A modular approach to simulations of dynamics of open-quantum systems},
    colorlinks=true,
}

\begin{document}
\title{QuantumDynamics.jl: A modular approach to simulations of dynamics of open quantum systems}
\author{Amartya Bose}
\affiliation{Department of Chemical Sciences, Tata Institute of Fundamental Research, Mumbai 400005, India}
\email{amartya.bose@tifr.res.in}
\begin{abstract}
    Simulation of non-adiabatic dynamics of a quantum system coupled to
    dissipative environments poses significant challenges. New sophisticated
    methods are regularly being developed with an eye towards moving to larger
    systems and more complicated description of solvents. Many of these methods,
    however, are quite difficult to implement and debug. Furthermore, trying to
    make the individual algorithms work together through a modular application
    programming interface (API) can be quite difficult. We present a new,
    open-source software framework, QuantumDynamics.jl, designed to address
    these challenges. It provides implementations of a variety of perturbative
    and non-perturbative methods for simulating the dynamics of these sytems.
    Most prominently, QuantumDynamics.jl supports hierarchical equations of
    motion and the family of methods based on path integrals. Effort has been
    made to ensure maximum compatibility of interface between the various
    methods. Additionally, QuantumDynamics.jl, being built on a high-level
    programming language, brings a host of modern features to explorations of
    systems such as usage of Jupyter notebooks and high level plotting, possibility
    of leveraging high-performance machine learning libraries for further
    development. Thus, while the built-in methods can be used as end-points in
    themselves, the package provides an integrated platform for experimentation,
    exploration, and method development.
\end{abstract}
\maketitle

\section{Introduction}
Understanding the evolution of a system over time is at the heart of chemistry
and physics. While many systems can indeed be treated classically, there are
several important problems where the quantum mechanical mechanism of tunneling
becomes inescapable. Some of the most ubiquitous of these are charge transfer
problems, excitation energy transfer processes and spin dynamics. Additionally,
in all these cases the dynamics may be severely modulated by the existence of
solvent degrees of freedom which exist at a given temperature. The necessity of
simulating the systems quantum mechanically while accounting for the environment
and solvents accurately prove to be significantly challenging.

Various approaches exist to tackle this problem. On the end of approximate
approaches there is the perturbative Bloch-Redfield Master
Equations~\cite{blochGeneralizedTheoryRelaxation1957,
    redfieldTheoryRelaxationProcesses1957} (BRME) and methods based on empirical
Lindbladians. However, these are uncontrolled approximations with no good error
bounds. Therefore it becomes important to be able to obtain the exact dynamics
for these systems. Various path integral-based techniques like the
quasi-adiabatic propagator path
integral~\cite{makriTensorPropagatorIterativeI1995,
    makriTensorPropagatorIterativeII1995, makriNumericalPathIntegral1995} (QuAPI)
family of methods and hierarchy equations of
motion~\cite{tanimuraTimeEvolutionQuantum1989,
    ishizakiQuantumDynamicsSystem2005, tanimuraStochasticLiouvilleLangevin2006}
(HEOM) family of methods exist which at greater costs can simulate the full
non-Markovian dynamics of a quantum system coupled with dissipative media using
the Feynman-Vernon influence functional~\cite{feynmanTheoryGeneralQuantum1963}.
Over the years, methods of unparallelled sophistication have been built on both
these frameworks which reduce the computational costs of
simulations~\cite{makriBlipDecompositionPath2014,
    makriIterativeBlipsummedPath2017, bosePairwiseConnectedTensor2022,
    boseMultisiteDecompositionTensor2022, makriModularPathIntegral2018,
    boseTensorNetworkRepresentation2021, jorgensenExploitingCausalTensor2019,
    makriSmallMatrixDisentanglement2020, makriSmallMatrixPath2020,
    makriSmallMatrixModular2021, makriSmallMatrixPath2021a,
    strathearnEfficientNonMarkovianQuantum2018, huPadeSpectrumDecompositions2011,
    shiEfficientHierarchicalLiouville2009, shiEfficientPropagationHierarchical2018,
    yanEfficientPropagationHierarchical2021,
    ikedaGeneralizationHierarchicalEquations2020}.

Despite the existence of a multitude of rigorous methods, software support for
quantum dynamics is relatively sparse. The situation becomes especially stark
when put in comparison to the plethora of alternatives, both open-source and
proprietary, that exist for electronic structure
theory~\cite{apraNWChemPresentFuture2020, frischGaussian16Rev2016,
    smithPSI4OpensourceSoftware2020, sunPySCFPythonbasedSimulations2018,
    balasubramaniTURBOMOLEModularProgram2020, kuhneCP2KElectronicStructure2020,
    giannozziQUANTUMESPRESSOModular2009,
    giannozziAdvancedCapabilitiesMaterials2017}. The lack of easily available
implementation of the latest methods prevent their widespread adoption. In
addition to preventing people from both being able to apply these novel ideas to
a variety of problems, this has the inadvertent disadvantage of preventing
critical comparison and evaluation of the different methods. In terms of
providing access to multiple state-of-the-art algorithms for dynamics in a
single package, i-PI~\cite{kapilIPIUniversalForce2019} is exemplary, providing
flexible implementations of various methods based on imaginary time path
integral and approximate quantum dynamics using ring-polymer. However, it does
not support approaches for simulating non-adiabatic processes.

Amongst exact methods for simulating processes that can be decomposed in terms
of a quantum system interacting with a thermal bath, HEOM has a fair number of
implementations~\cite{kreisbeckExcitonDynamicsLab,
    strumpferOpenQuantumDynamics2012, tsuchimotoSpinsDynamicsDissipative2015,
    temenHierarchicalEquationsMotion2020}.
QuTiP~\cite{johanssonQuTiPPythonFramework2013}, which supports a plethora of
approximate methods for simulating open quantum systems, has an implementation
of HEOM. A C++/Python software called
Libra~\cite{akimovLibraOpensourceMethodology2016} and a new Julia package called
NQCDynamics.jl~\cite{gardnerNQCDynamicsJlJulia2022} have been developed
primarily for using classical trajectory-based methods for simulating
non-adiabatic quantum dynamics. However, when it comes to numerically ``exact''
simulation of these systems and supporting the variety of state-of-the-art real
time path integral-based methods in a modular fashion, there is a severe dearth
of software. This has been a significant impediment in the approachability,
adoption and further development of these powerful methods.

Most computational codes have been historically written in C or C++ or Fortran.
While performant, these languages are low-level and their use significantly adds
to the code complexity and raises the bar for others contributing to the
frameworks. Of late, Python is being used for writing scientific code, with the
most performance intensive parts written in C or C++.  Prime examples of
programs and packages using this ``two-language'' infrastructure are
PySCF~\cite{sunPySCFPythonbasedSimulations2018},
Psi4~\cite{smithPSI4OpensourceSoftware2020},
i-PI~\cite{kapilIPIUniversalForce2019}, etc. A relatively new language called
Julia~\cite{bezansonJuliaFreshApproach2017}, with promise in terms of balancing
performance with ease of use, has been gaining popularity in the scientific
community. It features a just-in-time compilation scheme that solves the
two-language problem, where the API exposes features to a high-level language
but the performance-critical parts are coded in a different low-level language.
It consequently becomes easy to have scientific packages written completely in
Julia without sacrificing performance. There has been an explosion of packages
for computational chemistry in Julia in the recent
past~\cite{aroeiraFermiJlModern2022, gardnerNQCDynamicsJlJulia2022,
    herbstDFTKJulianApproach2021, kramerQuantumOpticsJlJulia2018,
    pooleNewKidBlock2020, pooleTaskBasedApproachParallel2022}.

We introduce a new open-source software package for the Julia language called
\href{https://github.com/amartyabose/QuantumDynamics.jl}{QuantumDynamics.jl} for
providing easy access to the state-of-the-art tools for rigorous simulation of
non-adiabatic systems to the community. An implementation in a high-performance,
high-level language is convenient for widespread adoption and easy development
in the future. Though it supports some approximate methods, the primary focus of
QuantumDynamics.jl is methods for numerically exact simulations of non-adiabatic
problems. The design aims at providing atomic concepts that help maximize reuse
of code between a diverse set of path integral-based methods. The paper is
organized as follows.
\href{https://amartyabose.github.io/QuantumDynamics.jl/dev/}{Online
    documentation} has already been provided. It will continue to be maintained,
updated, and improved upon as the package changes. In
Sec.~\ref{sec:methods_structure}, we discuss the methods supported in
QuantumDynamics.jl and the structure of the package. We demonstrate the usage of
the package through representative examples of the methods. While code snippets
have been provided in this paper, the full examples are there in the
\href{https://github.com/amartyabose/QuantumDynamics.jl/tree/main/examples}{examples
    folder of the repository}. Some concluding remarks are provided in
Sec.~\ref{sec:conclusions}.

\section{Methods Supported and Structure of the Code}\label{sec:methods_structure}
\subsection{Methods Supported}
The main focus of QuantumDynamics.jl is the simulation of dynamics of open
quantum systems with non-adiabatic processes. These are characterized a
relatively small dimensional quantum system, described by a Hamiltonian,
$\hat{H}_0$, interacting with $N_\text{env}$ large thermal environments.
\begin{align}
    \hat{H}                  & = \hat{H}_0 + \sum_b^{N_\text{env}}\hat{H}^{(b)}_\text{env}\label{eq:sys_bath}                                                                                  \\
    \hat{H}^{(b)}_\text{env} & = \sum_j \frac{p_{jb}^2}{2m_{jb}} + \frac{1}{2}m_{jb}\omega_{jb}^2\left( x_{jb} - \frac{c_{jb} \hat{s}_b}{m_{jb}\omega_{jb}^2} \right)^2, \label{eq:quapi_bath}
\end{align}
where $\omega_{jb}$ and $c_{jb}$ are the frequency and coupling of the $j$th
mode of the $b$th environment. The interaction between the system and the $b$th
environment is described by $\hat{H}^{(b)}_\text{env}$ and happens through the
system operator $\hat{s}_b$. In general the environments are atomistically
defined. However, under Gaussian response limit, it is possible to map the
effects of the atomistic environment onto a bath of harmonic
oscillators~\cite{makriLinearResponseApproximation1999,
    allenDirectComputationInfluence2016, waltersDirectDeterminationDiscrete2017,
    boseZerocostCorrectionsInfluence2022} through the energy gap
auto-correlation function and its spectral density,
\begin{align}
    J_b(\omega) & = \frac{\pi}{2}\sum_j \frac{c_{jb}^2}{m_{jb}\omega_{jb}}\delta(\omega-\omega_{jb}).
\end{align}
The famous spin-boson model is a specialization of Eq.~\ref{eq:sys_bath} for the
case of $\hat{H}_0 = \epsilon\sigma_z - \hbar\Omega\sigma_x$, where $\epsilon$
is the asymmetry between the two states and $\Omega$ is the coupling strength.
Spin-boson models typically have a single harmonic bath as an environment.
Two of the most common model spectral densities are
\begin{align}
    J_\text{ExpCutoff}(\omega, n)             & = \frac{2 \pi}{\Delta s^2}\hbar\xi\frac{\omega^n}{\omega_c^{n-1}}\exp\left(-\frac{\omega}{\omega_c}\right),\label{eq:expcutoff} \\
    \text{and } J_\text{DrudeLorentz}(\omega) & = \frac{2 \lambda}{\Delta s^2} \frac{\gamma \omega}{\omega^2 + \gamma^2}.\label{eq:drudelorentz}
\end{align}
Here $\Delta s$ is the separation between the system states. Depending on the
value of $n$, $J_\text{ExpCutoff}$ represents an Ohmic spectral density ($n=1$),
super-Ohmic spectral density ($n>1$) or sub-Ohmic spectral density ($n<1$). This
family of spectral densities is specified in terms of the dimensionless Kondo
parameter, $\xi$, and the cutoff frequency, $\omega_c$. The Drude-Lorentz
spectral density is another Ohmic spectral density but with a Lorentzian cutoff.
It is typically specified using a reorganization energy, $\lambda$, and the
characteristic bath time scale, $\gamma$.

There are a variety of approaches for simulating the dynamics of these systems
ranging from completely empirical to numerically exact. The implementations are
often challenging and hard to bring to a common interface. Because of the
typically strong system-environment couplings, perturbative methods of
calculation of dynamics are often not very accurate. However, they still might
provide useful starting points for understanding the dynamics. A broad set of
these exact and approximate methods are supported in QuantumDynamics.jl. They
can be roughly categorized as
\begin{enumerate}[noitemsep,topsep=0.25em]
    \item Empirical approaches
    \item Hierarchical Equations of Motion
    \item Path integral approaches
\end{enumerate}
It is often difficult to maintain consistency of the interface across these
different classes of approaches. However, within every category, the consistency
has been ensured. QuantumDynamics.jl does not support the rich gamut of
classical trajectory-based methods. Consequently, notable in its omission is the
ubiquitous surface hopping method~\cite{tullyMolecularDynamicsElectronic1990,
    tullyPerspectiveNonadiabaticDynamics2012, wangRecentProgressSurface2016}.
The NQCDynamics.jl package~\cite{gardnerNQCDynamicsJlJulia2022} implements
surface hopping both in its fewest switching form and in connection to
ring-polymer molecular dynamics. It also implements various other classical
trajectory-based approaches in a modular manner.

Within the group of empirical approaches, QuantumDynamics.jl
supports propagation of both Hermitian and non-Hermitian systems. It also
supports more rigorous approaches based on master equations such as BRME and
Lindblad master equation. HEOM~\cite{tanimuraTimeEvolutionQuantum1989,
    ishizakiQuantumDynamicsSystem2005, tanimuraReducedHierarchicalEquations2014,
    tanimuraNumericallyExactApproach2020} is implemented in its ``scaled''
form~\cite{shiEfficientHierarchicalLiouville2009}. While many other improvements
and extensions of HEOM exist in the literature, they have not yet been
implemented in QuantumDynamics.jl. These will be incorporated in future versions
as and when required.

The largest class of methods supported by QuantumDynamics.jl is in the path
integral approaches. In addition to the original
QuAPI~\cite{makriTensorPropagatorIterativeI1995,
    makriTensorPropagatorIterativeII1995, makriNumericalPathIntegral1995}, blip
decomposition of path integrals~\cite{makriBlipDecompositionPath2014,
    makriIterativeBlipsummedPath2017} (BSPI), the tensor network path
integral~\cite{boseTensorNetworkRepresentation2021} implementation of
time-evolving matrix product
operator~\cite{strathearnEfficientNonMarkovianQuantum2018} (TEMPO) approach and
the pairwise-connected tensor network path
integral~\cite{bosePairwiseConnectedTensor2022} (PC-TNPI) method are supported.
Quantum-classical path integral~\cite{lambertQuantumclassicalPathIntegralI2012,
    lambertQuantumclassicalPathIntegralII2012} (QCPI) using solvent-driven
references~\cite{banerjeeQuantumClassicalPathIntegral2013} in the harmonic
backreaction~\cite{wangQuantumclassicalPathIntegral2019} framework has been
implemented using the same interface. As elaborated in Sec.~\ref{sec:structure},
the code has been designed in a way that QCPI could be used with different
``backends'' corresponding to QuAPI or TEMPO.

Ideas of dynamical maps have been shown to be effective in understanding
non-Markovian evolution of systems~\cite{cerrilloNonMarkovianDynamicalMaps2014}.
The transfer tensor method~\cite{cerrilloNonMarkovianDynamicalMaps2014} (TTM)
allows construction of transfer tensors from dynamical maps, which for open
quantum systems are the forward-backward propagators augmented by the bath
influence, $\mathcal{E}(t) =
    \Tr_\text{bath}\left(\exp\left(-i\mathcal{L}t/\hbar\right)\right)$, where
$\mathcal{L}$ is the Liouvillean corresponding to the system-bath. These
transfer tensors can be further used to propagate the reduced density matrix of
the system beyond the memory length. This reduces the complexity of simulating
the time-evolution beyond memory length to multiplying matrices of the size of
the system and removes all storage requirements. TTM in QuantumDynamics.jl can
take advantage of the forward-backward augmented propagators obtained from other
path integral methods like QuAPI, TEMPO, PC-TNPI, and blips.

The small matrix decomposition of path
integral~\cite{makriSmallMatrixDisentanglement2020, makriSmallMatrixPath2020}
(SMatPI) is a rigorous QuAPI-based method which achieves a similar objective but
with more efficient implementations for extended memory
length~\cite{makriSmallMatrixPath2021a} and support for simulation of dynamics
under the influence of external fields~\cite{makriSmallMatrixPath2021}. It has
been noted by~\citet{makriSmallMatrixPath2020} that while TTM employs
time-translational invariance leading to generation of spurious memory, SMatPI
through a rigorous derivation based on QuAPI lifts this limitation.
QuantumDynamics.jl enables the use of tensor network-based methods like TEMPO
with TTM which allows inclusion of the possible spurious memory generated
without a significant increase in computational complexity.


All these methods with the exception of TTM have been implemented in a manner so
that they can simulate the dynamics of these systems in presence of external
time-dependent fields. One of the potential applications of such time-dependent
fields is simulation of dynamics in the presence of light described in a
semiclassical manner.

\subsection{Code Structure}\label{sec:structure}
QuantumDynamics.jl, being a Julia package, can be used on any operating system
and platform supported by the programming language. It has recently been
registered with the Julia package registry. Thus, the installation procedure is
relatively simple. After Julia has been setup, there are two ways of installing
QuantumDynamics.jl. The first way involves Julia's package manager
read-eval-print loop (REPL) interface:
\begin{minted}[breaklines, fontsize=\footnotesize, framesep=2mm, frame=lines]{julia}
julia> ]
pkg> add QuantumDynamics
\end{minted}
The alternate is to use the \verb|Pkg| module in Julia:
\begin{minted}[breaklines, fontsize=\footnotesize, framesep=2mm, frame=lines]{julia}
import Pkg
Pkg.add("QuantumDynamics")
\end{minted}
All the dependencies will automatically be installed. Julia comes with
implementations of OpenBlas built-in by default. However, depending on the
architecture, it may be preferable to install and use Intel's Math Kernel
Library (MKL), which can be installed as an additional package MKL.jl. If MKL is
used, it should be loaded before QuantumDynamics.jl in the source code.

In QuantumDynamics.jl, an attempt has been made to provide as flexible and
consistent an application programming interface (API) as possible across the
gamut of supported methods. This consistency is crucial in ensuring a successful
mix-and-match of various approaches. However, this is an extremely challenging
task given the different requirements and restrictions of various methods. In
this section, we discuss some of the crucial design choices present in this
package.

Each method has its own module. The empirical methods are completely grouped in
the \verb|Bare| module. Bloch-Redfield Master
Equation~\cite{redfieldTheoryRelaxationProcesses1957,
    blochGeneralizedTheoryRelaxation1957} and
HEOM~\cite{tanimuraNumericallyExactApproach2020} are supported in the
\verb|BlochRedfield| and \verb|HEOM| modules respectively.  The path
integral methods are more varied and have been afforded their individual
modules viz. \verb|QuAPI|~\cite{makriTensorPropagatorIterativeI1995,
    makriTensorPropagatorIterativeII1995},
\verb|Blip|~\cite{makriBlipDecompositionPath2014},
\verb|TEMPO|~\cite{strathearnEfficientNonMarkovianQuantum2018},
\verb|PCTNPI|~\cite{bosePairwiseConnectedTensor2022}, etc. All the path
integral methods, with the exception of quantum-classical path
integral~\cite{lambertQuantumclassicalPathIntegralI2012,
    lambertQuantumclassicalPathIntegralII2012}, builds on top of a time-series of
forward-backward propagators corresponding to the bare or isolated system.
Because of this decision, it becomes possible for QCPI to use any of the base
path integral methods as the engine to simulate the dynamics. For every sampled
phase space point of the solvent, the QCPI routine provides the underlying path
integral routine a sequence of solvent-driven reference
propagators~\cite{banerjeeQuantumClassicalPathIntegral2013}, and obtains as an
output the reduced density matrices after incorporation of the backreaction in
the harmonic approximation~\cite{wangQuantumclassicalPathIntegral2019}. The
toggle of whether the full memory needs to be incorporated or just the quantum
memory from the back reaction is necessary, is determined by the boolean
parameter, \verb|reference_propagator|. If \verb|reference_propagator| is false,
which is the default behavior, then the full influence functional is
incorporated, else only the quantum memory is incorporated. For any method, the
function for simulating the dynamics of an reduced density matrix is called
\verb|propagate|. Individual methods often have convergence parameters that
differ wildly from each other. All such parameters are grouped in
method-specific argument types all derived from \verb|Utilities.ExtraArgs|.

QCPI requires definition of a solvent, which is treated by classical
trajectories. This facility is provided by the abstract struct
\verb|Solvents.Solvent|, which can be inheritted from for different types of
solvents. Currently only a discrete harmonic bath is provided. There is scope
for providing wrappers around emerging Julia libraries for doing molecular
dynamics as more detailed solvents. Associated with each solvent is a
description of the corresponding phase space, and an iterator which generates
phase space points that are distributed according to the thermal Boltzmann
distribution.

Many of the empirical methods and HEOM require solution of differential
equations, which is done numerically using the
DifferentialEquations.jl~\cite{rackauckasDifferentialequationsJlPerformant2017}
package. It implements a variety of methods for solving differential equations.
The details that control differential equation solver like the method of
simulation, relative error and absolute error, are controlled through the
structure, \verb|Utilities.DiffEqArgs|. In QuantumDynamics.jl, the default
method of solution is an adaptive Runge-Kutta approach of orders 5
(4)~\cite{tsitourasRungeKuttaPairs2011}, though other methods can be easily used
by suitably changing the \verb|Utilities.DiffEqArgs| passed to the method. The
methods based on tensor network are built on the open-source
\textsc{ITensor}~\cite{fishmanITensorSoftwareLibrary2022,
    fishmanCodebaseReleaseITensor2022} library.

For the specification of the bath spectral densities, QuantumDynamics.jl
provides a SpectralDensities module. Currently we support
\verb|ExponentialCutoff| for Eq.~\ref{eq:expcutoff} and \verb|DrudeLorentz| for
Eq.~\ref{eq:drudelorentz}. Facilities are provided for reading in tabulated
spectral densities obtained as Fourier transforms of numerically simulated bath
response functions is also provided through \verb|SpectralDensityTable|. Utility
functions are provided for reading the tabulated data for both $J(\omega)$ and
$J(\omega)/\omega$.

Finally, TTM~\cite{cerrilloNonMarkovianDynamicalMaps2014} builds on propagators
from initial time, $t=0$, to final time. Thus, in addition to providing routines
for propagating a reduced density matrix, the various sub-modules for path
integral also provide \verb|build_augmented_propagator| functions that calculate
the time-series of propagators including the solvent effects using the
corresponding full path methods. As detailed in the numerical examples,
Sec.~\ref{sec:num_example}, these functions make it possible to use TTM to
propagate a system whose augmented propagators have been calculated using some
path integral method.

The full
\href{https://amartyabose.github.io/QuantumDynamics.jl/dev/}{documentation} of
the package also shows other examples along with detailed description of the
various arguments and parameters supported by these methods. It will remain
updated as the package continues to evolve and implement other methods.

\section{Numerical Examples}\label{sec:num_example}
\subsection{Empirical Approaches to Open Quantum Systems}\label{sec:empirical}
\subsubsection{Isolated Hermitian \& Non-Hermitian Systems}
The simplest case of propagation happens to be for an isolated system. The
dynamics is Markovian. QuantumDynamics.jl provides interface for simulating this
dynamics both for Hermitian and non-Hermitian systems defined by a Hamiltonian,
$\hat{H}$. The equation of motion for the density matrix,
\begin{align}
    i\hbar\partial_t\rho(t) = \hat{H}\rho(t) - \rho(t)\hat{H}^\dag,
\end{align}
works for both types of systems.

Consider two degenerate states that are described by the Hamiltonian,
\begin{align}
    \hat{H} = \begin{pmatrix}
                  0.0  & -1.0 \\
                  -1.0 & 0.0
              \end{pmatrix}
\end{align}
This is a Hermitian Hamiltonian. In addition, also consider a non-Hermitian
Hamiltonian where the two states are lossy with different rates:
\begin{align}
    \hat{H}_\text{nh} = \begin{pmatrix}
                            -0.1i & -1.0  \\
                            -1.0  & -0.5i
                        \end{pmatrix}
\end{align}

\begin{figure}
    \subfloat[Hermitian system, $\hat{H}$]{\includegraphics{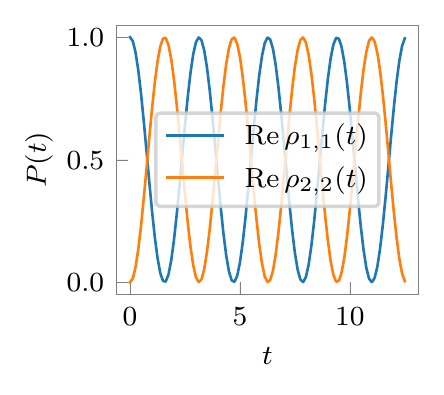}}
    ~\subfloat[Non-Hermitian system, $\hat{H}_\text{nh}$]{\includegraphics{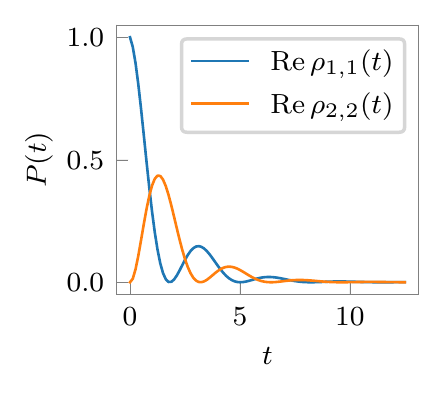}}

    \subfloat[Hermitian system, $\hat{H} + V(t)$]{\includegraphics{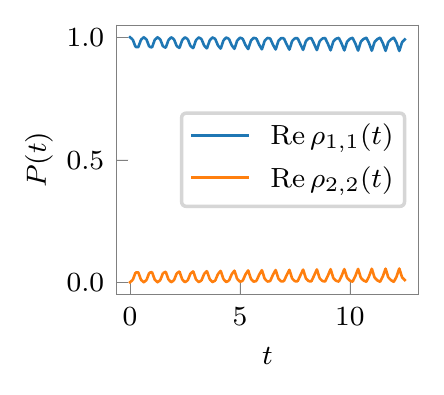}}
    ~\subfloat[Non-Hermitian system, $\hat{H}_\text{nh} + V(t)$]{\includegraphics{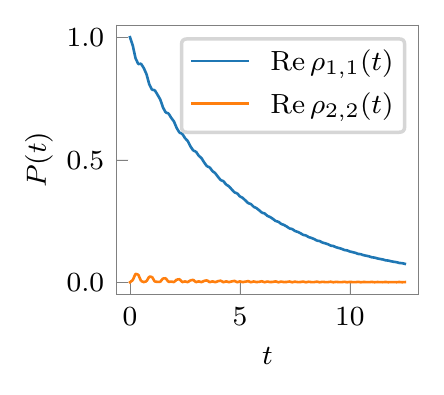}}
    \caption{Dynamics of the different elements of the density matrix.}\label{fig:isolated_dynamics}
\end{figure}
The Hamiltonian in either case can be defined as a $2\times 2$ complex matrix or
by using the convenience function \verb|Utilities.create_tls_hamiltonian|.
Currently, QuantumDynamics.jl also provides another convenience function for
creating a periodic or aperiodic nearest-neighbor Hamiltonian,
\verb|Utilities.create_nn_hamiltonian|.

The dynamics of a system under these two Hamiltonians starting with a density
matrix of
\begin{align}
    \rho(0) = \begin{pmatrix}
                  1.0 & 0.0 \\
                  0.0 & 0.0
              \end{pmatrix}.
\end{align}
is shown in Figs.~\ref{fig:isolated_dynamics}~(a) and (b). When a time-dependent
external field, $V(t) = 12\cos(10t)$, is coupled with the operator
$\hat\sigma_z$, the dynamics changes substantially. The dynamics under the
external field for the Hermitian and non-Hermitian systems are shown in
Fig.~\ref{fig:isolated_dynamics}~(c) and (d) respectively.

The code snippet for simulating the dynamics of the non-Hermitian system in
presence of the external field is as follows:
\begin{minted}[breaklines, fontsize=\footnotesize, framesep=2mm, frame=lines]{julia}
using QuantumDynamics

# define the Hamiltonian
H = [-0.1im -1.0; -1.0 -0.5im]
V(t) = 12 * cos(10.0 * t)
EF = Utilities.ExternalField(V, [1.0+0.0im 0.0; 0.0 -1.0])

# define the initial condition, the time step and the number of steps of simulation
ρ0 = [1.0+0.0im 0.0; 0.0 0.0]
dt = 0.125
ntimes = 100

# simulate the dynamics using the propagate method
times, ρs = Bare.propagate(; Hamiltonian=H, ρ0, dt, ntimes, external_fields=[EF])
\end{minted}
The other cases are also similar. Notice that the Hamiltonian is defined as a
simple $2\times2$ matrix. The design moves away from defining classes
hierarchies for these fundamental objects because that creates barriers when
using different hardwares. For example, with the current design, implementing
the same algorithm on a graphics processing unit (GPU) should be as simple as
using the array abstractions in a Julia library like
CUDA.jl~\cite{besardEffectiveExtensibleProgramming2018}. The external field,
\verb|Utilities.ExternalField|, is a \verb|struct| with a simple function of
time and the system operator that couples to the field. The same
\verb|Bare.propagate| function works for Hermitian or non-Hermitian systems with
or without external fields.

\subsubsection{Lindblad Master Equation}\label{sec:lindblad}
Consider a system interacting with a variety of environment degrees of freedom
as in Eq.~\ref{eq:sys_bath}. An empirical approach of incorporating effects of
these environments on the dynamics of the reduced density matrix (RDM) of the
quantum systems is through the use of the Lindblad Master Equation,
\begin{align}
    \dv{\rho(t)}{t} & = -\frac{i}{\hbar}\commutator{\hat{H}_0}{\rho(t)} + \sum_j \left( L_j\rho(t)L_j^\dag - \frac{1}{2}\anticommutator{L_j^\dag L_j}{\rho(t)} \right),
\end{align}
where $\hat{H}_0$ is the Hamiltonian of the system, $\rho(t)$ is the
time-evolved system RDM. The impact of the environment is empirically modeled
through the so-called Lindblad ``jump'' operators, $L_j$. Different processes
require different types of jump operators. A couple of examples are demonstrated
here.

\begin{figure}
    \centering
    \includegraphics{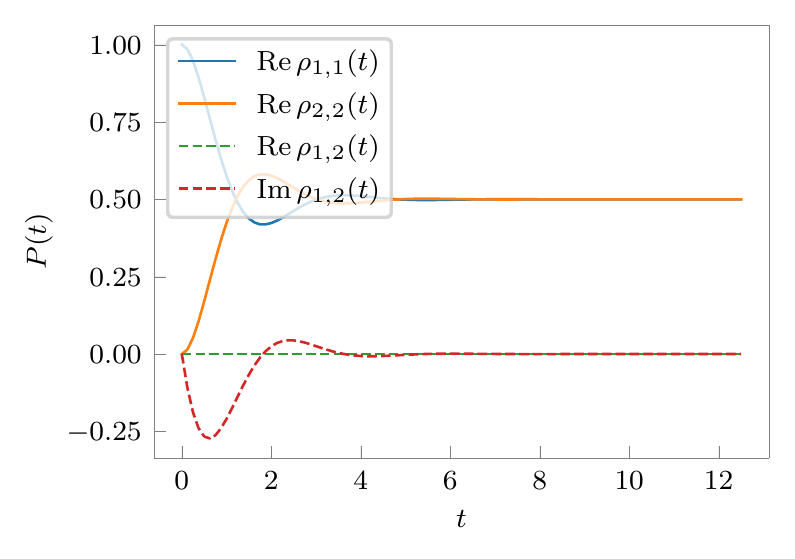}
    \caption{Simulation of dynamics corresponding to a typical spin-boson parameter.}\label{fig:lindblad_spin_boson}
\end{figure}

For mapping a spin-boson problem onto a system described with the Lindblad
master equation, we obtain a jump operator proportional to $\sigma_z$. The
strength of the system-bath coupling in a spin-boson parameter is related to
this proportionality constant. Consider a system Hamiltonian given by
$\hat{H}_0=-\sigma_x$, and a localized initial condition. The code to simulate
the dynamics using QuantumDynamics.jl is:
\begin{minted}[breaklines, fontsize=\footnotesize, framesep=2mm, frame=lines]{julia}
using QuantumDynamics

# define the Hamiltonian, the initial RDM, and the simulation details
H = Matrix{ComplexF64}([
    0.0 -1.0
    -1.0 0.0
])
ρ0 = [1.0+0.0im 0.0; 0.0 0.0]
dt = 0.125
ntimes = 100

# define the Lindbladian jump operator
L = [1.0+0.0im 0.0; 0.0 -1.0]

# call propagate with a list of jump operators to be applied
times, ρs = Bare.propagate(; Hamiltonian=H, ρ0, dt, ntimes, L=[L])
\end{minted}
The resultant dynamics is shown in Fig.~\ref{fig:lindblad_spin_boson}. One can
notice the features reminiscent of a typical spin-boson
parameters~\cite{makriTensorPropagatorIterativeI1995,
    makriTensorPropagatorIterativeII1995}. Also note that the only change from
the simulations of the isolated Hermitian and non-Hermitian systems is the
new argument, \verb|L|, containing a vector of jump operators $L_j$ that is
being passed in.

Now, consider a more involved example. We want to model an excitation transport
between a dimer of molecules while accounting for possibility of spontaneous
emission which will bring one molecule to the ground state without exciting the
other. This possibility does not allow for modeling the problem in the so-called
first excitation subspace. In the full Hilbert space, the system Hamiltonian is
taken to be
\begin{align}
    \hat{H}_0 & = 20.0\dyad{ee} + 10.0\dyad{ge} + 10.0\dyad{eg}\nonumber \\
              & - 1.0\dyad{eg}{ge} - 1.0\dyad{ge}{eg}.
\end{align}
The simulation is started from an initial condition of $\dyad{ge}$. For the
effects of the molecular vibrations moving the energies of the excited and the
ground states, we use jump operators proportional to $\sigma_z\otimes\mathbb{I}$
and $\mathbb{I}\otimes\sigma_z$. For capturing the spontaneous decay process, we
introduce jump operators proportional to $\sigma_m\otimes\mathbb{I}$ and
$\mathbb{I}\otimes\sigma_m$. The code for simulating this system is as follows:
\begin{figure}
    \centering
    \subfloat[Dynamics for \texttt{bo}=0.7071 and \texttt{se}=0.0.]{\includegraphics{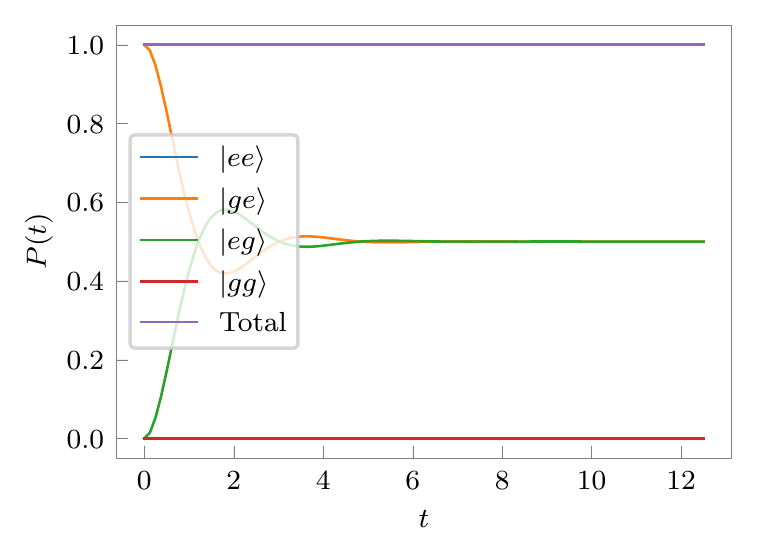}}

    \subfloat[Dynamics for \texttt{bo}=0.7071 and \texttt{se}=0.25.]{\includegraphics{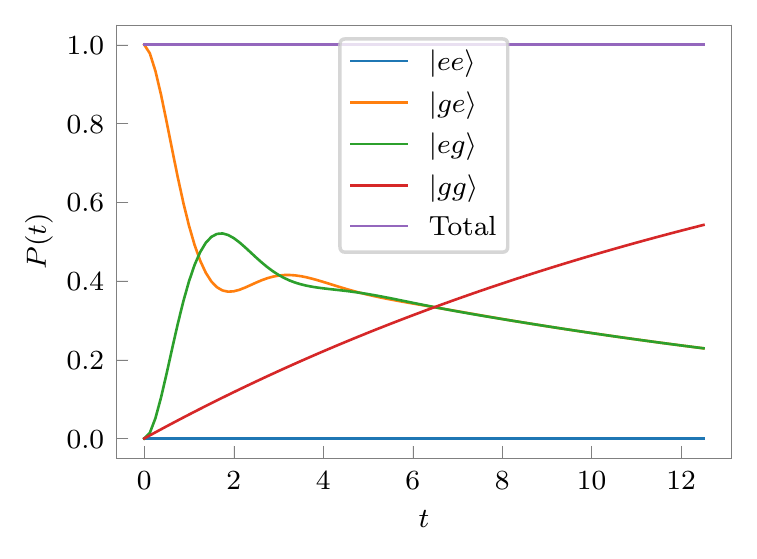}}
    \caption{Simulation of dynamics corresponding to an excitation transfer in a dimer with and without spontaneous emission.}\label{fig:excitation_lindblad}
\end{figure}

\begin{minted}[breaklines, fontsize=\footnotesize, framesep=2mm, frame=lines]{julia}
using QuantumDynamics

# define Hamiltonian, initial RDM and simulation parameters
H = Matrix{ComplexF64}([
    20.0 0.0 0.0 0.0
    0.0 10.0 -1.0 0.0
    0.0 -1.0 10.0 0.0
    0.0 0.0 0.0 0.0
])
ρ0 = Matrix{ComplexF64}([
    0.0 0.0 0.0 0.0
    0.0 1.0 0.0 0.0
    0.0 0.0 0.0 0.0
    0.0 0.0 0.0 0.0
])
dt = 0.125
ntimes = 100

# define the jump operators corresponding to the decohering effects of the individual Born-Oppenheimer surfaces.
# bo is the coupling strength
σz = bo * Matrix{ComplexF64}([
    1.0 0.0
    0.0 -1.0
])
id = Matrix{ComplexF64}([
    1.0 0.0
    0.0 1.0
])
L1 = kron(σz, id)
L2 = kron(id, σz)

# define the jump operators corresponding to spontaneous emission
# se is the coupling strength
σm = se * Matrix{ComplexF64}([
    0.0 0.0
    1.0 0.0
])
L3 = kron(σm, id)
L4 = kron(id, σm)

times, ρs = Bare.propagate(; Hamiltonian=H, ρ0, dt, ntimes, L=[L1, L2, L3, L4])
\end{minted}
where the proportionality constants have been given as \verb|bo| and \verb|se|
for the molecular vibrations and the spontaneous emission processes
respectively. Figure~\ref{fig:excitation_lindblad} demonstrates the dynamics
obtained using this code when spontaneous emission is switched off and on. We
see the expected conservation of the number of excitation when spontaneous
emission is turned off, and a gradual build up of population in the ground state
in presence of spontaneous emission.

\subsection{Perturbative \& Non-Perturbative Dynamics of Open Quantum Systems}
While QuantumDynamics.jl supports empirical methods as described in
Sec.~\ref{sec:empirical}, the primary focus is rigorous methods of simulation of
open quantum systems. Now, we turn our attention to the more numerical involved
methods. For these examples, we will specify the detailed characteristics of the
harmonic bath using spectral densities.

\subsubsection{Bloch-Redfield Master Equation}
The Bloch-Redfield master equation~\cite{blochGeneralizedTheoryRelaxation1957,
    redfieldTheoryRelaxationProcesses1957} (BRME) is one of the simplest and most
versatile approaches to perturbatively simulate the dynamics of open quantum
systems. It includes a perturbative description of the system-environment
interaction with the environment being described under the Born approximation.
Finally, an additional approximation of Markovian dynamics is invoked to obtain
BRME. While the combination of approximations involved often makes the method
unsuitable for strongly coupled solvents, it is still useful for understanding
the very rough timescales of dynamics. Combining BRME with ideas of polaron
transform is successful in extending its applicability to strongly coupled
non-perturbative solvents~\cite{silbeyVariationalCalculationDynamics1984,
    xuNoncanonicalDistributionNonequilibrium2016,
    xuPolaronEffectsPerformance2016, jangPartiallyPolarontransformedQuantum2022,
    leeAccuracySecondOrder2012}.

For the system-solvent Hamiltonian, Eq.~\ref{eq:sys_bath}, under the Born
approximation and Markovian limit of the environment, BRME can be expressed as
an equation of motion for the reduced density matrix in the eigen-basis of the
system Hamiltonian, $\hat{H}_0$:
\begin{align}
    \dv{\rho_{ab}}{t} & = -i\omega_{ab}\rho_{ab}(t) + \sum_{cd} R_{abcd}\rho_{cd}(t),
\end{align}
where $R_{abcd}$ is the Redfield tensor that captures the impact of the solvent
on the system in a perturbative manner:
\begin{align}
    R_{abcd} = -\frac{1}{2}\sum_{k=1}^{N_\text{env}} & \left(\delta_{bd}\sum_n \bra{a}\hat{s}_k\dyad{n}\hat{s}_k\ket{c} J_k(\omega_c - \omega_n)\right.\nonumber              \\
                                                     & - \mel{a}{\hat{s}_k}{c}\mel{d}{\hat{s}_k}{b} J_k(\omega_c - \omega_a)\nonumber                                         \\
                                                     & + \delta_{ac}\sum_n \bra{d}\hat{s}_k\dyad{n}\hat{s}_k\ket{b} J_k(\omega_d - \omega_n)\nonumber                         \\
                                                     & - \left.\vphantom{\sum_{k=1}^{N_\text{env}}}\mel{a}{\hat{s}_k}{c}\mel{d}{\hat{s}_k}{b} J_k(\omega_d - \omega_b)\right)
\end{align}

\begin{figure}
    \includegraphics{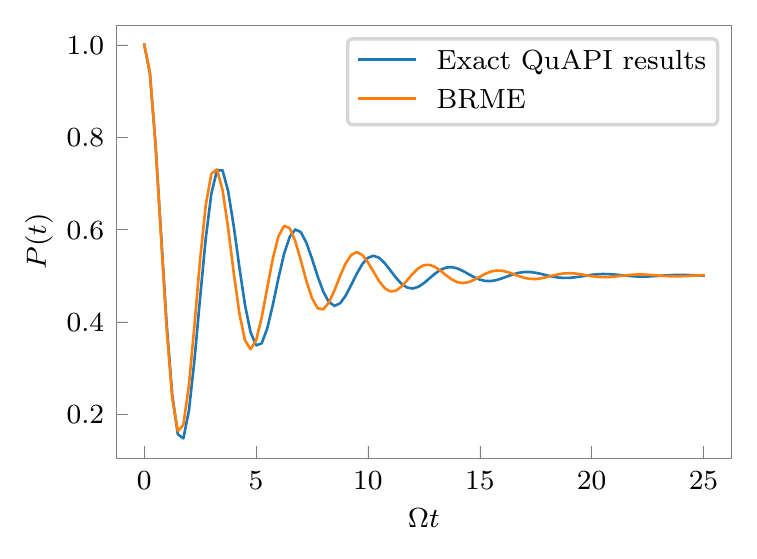}
    \caption{Comparison between BRME simulation and numerically exact QuAPI calculations. Discussion of QuAPI is given later in Sec.~\ref{sec:pi_methods}.}\label{fig:brme}
\end{figure}

A particular example of the results of BRME for a spin-boson system and its
comparison with exact quantum dynamical calculations using QuAPI is shown in
Fig.~\ref{fig:brme}. The code for simulating the BRME equations using
QuantumDynamics.jl for this particular case is as follows:
\begin{minted}[breaklines, fontsize=\footnotesize, framesep=2mm, frame=lines]{julia}
using QuantumDynamics

# define the system Hamiltonian
H = Matrix{ComplexF64}([
    0.0 -1.0
    -1.0 0.0
])

# specify the spectral density describing the bath and the inverse temperature
Jw = SpectralDensities.ExponentialCutoff(; ξ=0.1, ωc=7.5)
β = 5.0

ρ0 = Matrix{ComplexF64}([
    1.0 0.0
    0.0 0.0
])
dt = 0.25
ntimes = 100
time, ρs = BlochRedfield.propagate(; Hamiltonian=H, Jw=[Jw], β, ρ0, dt, ntimes, sys_ops=[[1.0+0.0im 0.0; 0.0 -1.0]])
\end{minted}

\subsubsection{Hierarchical Equations of Motion}
The hierarchical equations of motion
(HEOM)~\cite{tanimuraTimeEvolutionQuantum1989,
    tanimuraQuantumClassicalFokkerPlanck1991,
    tanimuraReducedHierarchicalEquations2014, tanimuraNumericallyExactApproach2020}
is one of the two foundational numerically exact, non-perturbative methods for
simulating the dynamics of an open quantum system interacting with a harmonic
bath. While originally formulated primarily for the Drude-Lorentz spectral
density, recent work has made it possible to use this method with more general
spectral densities~\cite{duanStudyExtendedHierarchy2017,
    popescuUsingChebychevExpansion2015, tianApplicationHierarchicalEquations2013,
    liuReducedQuantumDynamics2014}. Other developments have improved the numerical
stability of HEOM at lower
temperatures~\cite{dunnRemovingInstabilitiesHierarchical2019}.
QuantumDynamics.jl supports the scaled version of
HEOM~\cite{shiEfficientHierarchicalLiouville2009} for the Drude-Lorentz spectral
density. The more advanced approaches required to handle other spectral
densities will be incorporated in later versions of the package.

The general problem that HEOM solves is Eq.~\ref{eq:sys_bath}. However, for
HEOM, the system-environment interaction Hamiltonian is not exactly
Eq.~\ref{eq:quapi_bath}. It is given by
\begin{align}
    \hat{H}^{(b)}_\text{env} & = \sum_j \frac{p_{jb}^2}{2m_{jb}} + \frac{1}{2}m_{jb}\omega_{jb}^2 x_{jb}^2 - c_{jb} \hat{s}_b x_{jb}
\end{align}
Notice that the difference with Eq.~\ref{eq:quapi_bath} is that here, the square
is not completed. For the implementation of HEOM in QuantumDynamics.jl, the
baths need to be characterized by spectral densities having the Drude-Lorentz
form:
\begin{align}
    J_b(\omega) & = \frac{2 \lambda_b}{\Delta s_b^2} \frac{\gamma_b \omega}{\omega^2 + \gamma_b^2}.
\end{align}
The separation between the system states is $\Delta s_b$. For problems involving
exciton transport, the spectral density is specified using $\Delta s_b = 1$. For
application of HEOM, the correlation functions corresponding to the spectral
densities are written in a sum over poles
form~\cite{shiEfficientHierarchicalLiouville2009},
\begin{align}
    C_b(t) & = \sum_{m=0}^\infty c_{bm} \exp(-\nu_{bm}t),
\end{align}
where $\nu_{b0} = \gamma_b$ is the Drude decay constant and $\nu_{bm\ge 1} =
    2m\pi/\beta$ are the Matsubara frequencies. The coefficients $c_{bm}$ are given
by
\begin{align}
    c_{b0}      & = \gamma_b \frac{\lambda_b}{\Delta s_b^2} \left(\cot\left(\frac{\beta\hbar\gamma_b}{2}\right)-i\right), \\
    c_{bm\ge 1} & = \frac{4\lambda_b\gamma_b}{\beta\hbar\Delta s_b^2}\left(\frac{\nu_{bm}}{\nu_{bm}^2-\gamma_b^2}\right).
\end{align}

For such a system, the primary expression for HEOM is given as:
\begin{align}
    \dv{\rho_\mathbf{n}}{t} & = -\frac{i}{\hbar}\commutator{\hat{H}_0}{\rho_\mathbf{n}} + \sum_{b=1}^{N_\text{env}}\sum_{m=0}^M n_{bm}\nu_{bm}\rho_\mathbf{n}\nonumber                                                              \\
                            & - \sum_{b=1}^{N_\text{env}} \left( \frac{2\lambda_b}{\beta\hbar^2\gamma_b} - \sum_{m=0}^M\frac{c_{bm}}{\hbar\nu_{bm}} \right)\commutator{\hat{s}_b}{\commutator{\hat{s}_b}{\rho_\mathbf{n}}}\nonumber \\
                            & - i\sum_{b=1}^{N_\text{env}} \commutator{\hat{s}_b}{\sum_{m=0}^M \rho_{\mathbf{n}_{bm}^+}}\nonumber                                                                                                   \\
                            & - \frac{i}{\hbar}\sum_{b=1}^{N_\text{env}}\sum_{m=0}^M n_{bm}\left(c_{bm}\hat{s}_b\rho_{\mathbf{n}_{bm}^-} - c_{bm}^* \rho_{\mathbf{n}_{bm}^-}\hat{s}_b\right),\label{eq:heom}
\end{align}
where $\rho_\mathbf{n}$ represents the generalized density operators ---
when $\mathbf{n}=0,0,0,\ldots$, it is the reduced density operator, for all other
$\mathbf{n}$, it is an auxiliary density operator. The subscript vectors,
$\mathbf{n}$ are of length $N_\text{env} K (M+1)$, where $K$ is the depth of the
hierarchy. Each density matrix is assigned a depth of $L =
    \sum_{b=1}^{N_\text{env}}\sum_{m=0}^M n_{bm}$. The term in the second line
of Eq.~\ref{eq:heom} is the correction term is the Ishizaki-Tanimura scheme
of truncating the truncating the Matsubara terms by treating $m>M$ using a
Markovian approximation~\cite{ishizakiQuantumDynamicsSystem2005,
    tanimuraStochasticLiouvilleLangevin2006}.

The scaled version of HEOM~\cite{shiEfficientHierarchicalLiouville2009} rescales
the auxiliary density operators in a manner that allows truncation of the
hierarchy at a lower value of $K$.
\begin{align}
    \tilde\rho_\mathbf{n} & = \left(\prod_b\prod_m n_{bm}! |c_{bm}|^{n_{bm}}\right)^{-\frac{1}{2}}\rho_\mathbf{n},
\end{align}
which changes Eq.~\ref{eq:heom} to:
\begin{align}
    \dv{\tilde\rho_\mathbf{n}}{t} & = -\frac{i}{\hbar}\commutator{\hat{H}_0}{\tilde\rho_\mathbf{n}} + \sum_{b=1}^{N_\text{env}}\sum_{m=0}^M n_{bm}\nu_{bm}\tilde\rho_\mathbf{n}\nonumber                                                                      \\
                                  & - \sum_{b=1}^{N_\text{env}} \left( \frac{2\lambda_b}{\beta\hbar^2\gamma_b} - \sum_{m=0}^M\frac{c_{bm}}{\hbar\nu_{bm}} \right)\commutator{\hat{s}_b}{\commutator{\hat{s}_b}{\tilde\rho_\mathbf{n}}}\nonumber               \\
                                  & - i\sum_{b=1}^{N_\text{env}} \commutator{\hat{s}_b}{\sum_{m=0}^M \sqrt{(n_{bm}+1)|c_{bm}|}\tilde\rho_{\mathbf{n}_{bm}^+}}\nonumber                                                                                        \\
                                  & - \frac{i}{\hbar}\sum_{b=1}^{N_\text{env}}\sum_{m=0}^M \sqrt{\frac{n_{bm}}{|c_{bm}|}}\left(c_{bm}\hat{s}_b\tilde\rho_{\mathbf{n}_{bm}^-} - c_{bm}^* \tilde\rho_{\mathbf{n}_{bm}^-}\hat{s}_b\right).\label{eq:scaled_heom}
\end{align}
In this new version, Eq.~\ref{eq:scaled_heom}, the number of levels of hierarchy
required for convergence decreases significantly in comparison to the original
unscaled HEOM, Eq.~\ref{eq:heom}. This is the version that is used by default in
QuantumDynamics.jl. To use the unscaled version of HEOM, one needs to set
\verb|scaled| to \verb|false| while calling the \verb|HEOM.propagate| function.

\begin{figure}
    \centering
    \subfloat[$T=\SI{77}{\kelvin}$]{\includegraphics{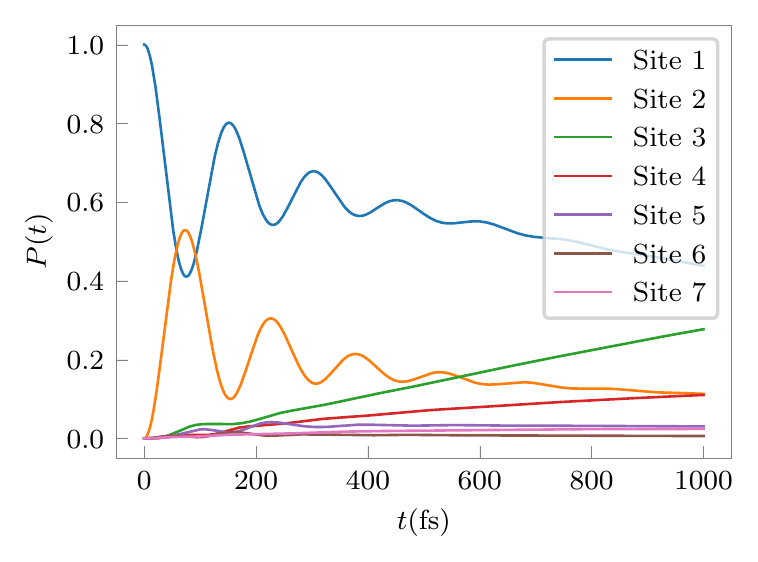}}

    \subfloat[$T=\SI{300}{\kelvin}$]{\includegraphics{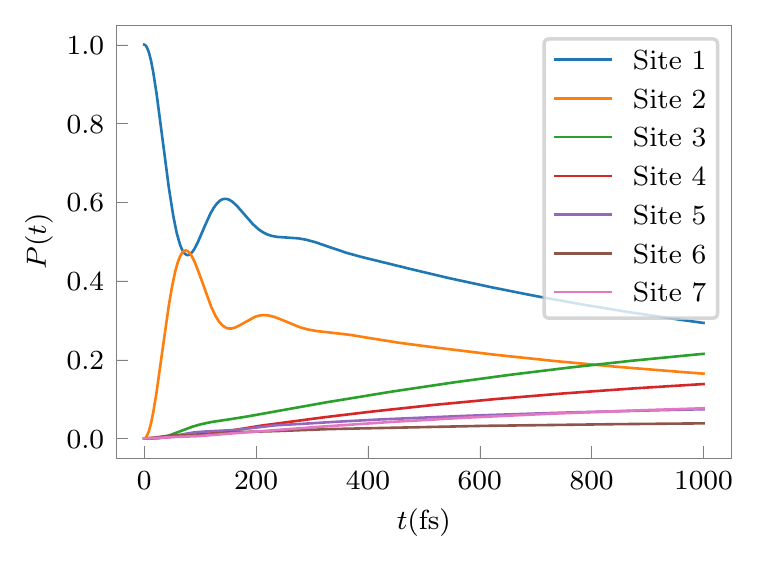}}
    \caption{Excitation population dynamics in the 7-state FMO model introduced by~\citet{ishizakiTheoreticalExaminationQuantum2009}.}\label{fig:ishizaki_fleming_heom}
\end{figure}

As a demonstration of the HEOM module in the code, we simulate the dynamics of
the chromophoric excitation in the famous 7-state model for the
Fenna-Matthews-Olson (FMO) complex~\cite{adolphsHowProteinsTrigger2006,
    ishizakiTheoreticalExaminationQuantum2009} at $T=\SIlist{77; 300}{\kelvin}$.
The code snippet for this part is quite self-explanatory:
\begin{minted}[breaklines, fontsize=\footnotesize, framesep=2mm, frame=lines]{julia}
using QuantumDynamics

# unit conversion
invcm2au = 4.55633e-6
au2fs = 0.02418884254

function FMO(num_modes, Lmax, β)
    # set up the system Hamiltonian
    H = Matrix{ComplexF64}([
        12410 -87.7   5.5  -5.9   6.7 -13.7  -9.9
        -87.7 12530  30.8   8.2   0.7  11.8   4.3
            5.5  30.8 12210 -53.5  -2.2  -9.6   6.0
            -5.9   8.2 -53.5 12320 -70.7 -17.0 -63.3
            6.7   0.7  -2.2 -70.7 12480  81.1  -1.3
        -13.7  11.8  -9.6 -17.0  81.1 12630  39.7
        -9.9    4.3   6.0 -63.3  -1.3  39.7 12440
    ]) * invcm2au
    nsteps = 500                  # number of steps of simulations
    dt = 1000 / au2fs / nsteps    # dt for simulation

    # initial density matrix with excitation localized on BChl unit 1
    ρ0 = Matrix{ComplexF64}(zeros(7, 7))
    ρ0[1, 1] = 1

    # create spectral densities on each of the BChl sites
    λs = [35.0, 35.0, 35.0, 35.0, 35.0, 35.0, 35.0] * invcm2au
    γs = 1 ./ ([50.0, 50.0, 50.0, 50.0, 50.0, 50.0, 50.0] / au2fs)
    Jw = Vector{SpectralDensities.DrudeLorentz}()
    sys_ops = Vector{Matrix{ComplexF64}}()
    for (j, (λ, γ)) in enumerate(zip(λs, γs))
        push!(Jw, SpectralDensities.DrudeLorentz(; λ, γ, Δs=1.0))
        op = zeros(7, 7)
        op[j, j] = 1.0
        push!(sys_ops, op)
    end

    # simulate the time evolution of ρ0 using HEOM
    t, ρ = HEOM.propagate(; Hamiltonian=H, ρ0=ρ0, Jw, β, ntimes=nsteps, dt, sys_ops, num_modes, Lmax)
    t .*= au2fs
    t, ρ
end
\end{minted}
Here, we use the \verb|propagate| function under the \verb|HEOM| submodule. It
takes a list of spectral densities, \verb|Jw|, along with the corresponding
system operators that couple to a particular bath, \verb|sys_ops|. The number
of Matsubara modes that are required to converge the results is \verb|num_modes|
and \verb|Lmax| is the number of auxiliary density operators considered in the
calculation. At both temperatures, well converged results were obtained with
\verb|num_modes| $=2$ and \verb|Lmax| $=3$. The dynamics obtained using this
code is shown in Fig.~\ref{fig:ishizaki_fleming_heom} which matches the original
results reported by~\citet{ishizakiTheoreticalExaminationQuantum2009}.

\begin{figure}
    \includegraphics{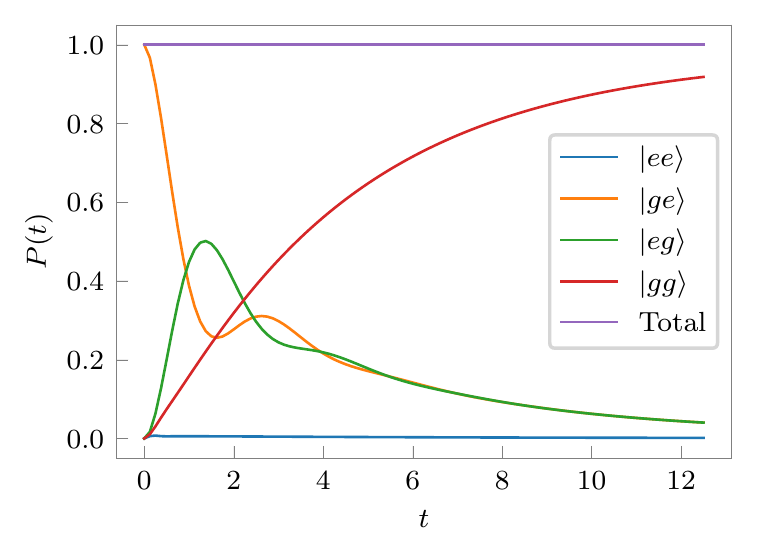}
    \caption{Dynamics of an excitonic dimer with multiple non-commuting baths.}\label{fig:spontaneous_emission_HEOM}
\end{figure}

As a final example of HEOM, let us consider the case of spontaneous emission
that was modelled empirically using Lindblad Master Equation,
Fig.~\ref{fig:excitation_lindblad}, in Sec.~\ref{sec:lindblad}. Spontaneous
emission happens because of the presence of an environment or bath that is able
to couple the molecular excited state to the molecular ground state, thereby
reducing the excited state lifetime to some finite value. Once again, the ground
and excited states will include the corresponding vibrations and the changes in
energy that they bring about. The resultant dynamics is shown in
Fig.~\ref{fig:spontaneous_emission_HEOM}. The bath which enables a spontaneous
excitation or relaxation of the molecular eigenstate is acts through the system
$\hat\sigma_x$ operator, whereas the baths representing the vibrational motion
on the Born-Oppenheimer surfaces act through $\hat\sigma_z$. HEOM is able to
handle both a ``diagonal'' and an ``off-diagonal'' bath on the same footing
without an increase in computational complexity.

\begin{minted}[breaklines, fontsize=\footnotesize, framesep=2mm, frame=lines]{julia}
# define Hamiltonian, initial RDM and simulation parameters
H = Matrix{ComplexF64}([
    20.0 0.0 0.0 0.0
    0.0 10.0 -1.0 0.0
    0.0 -1.0 10.0 0.0
    0.0 0.0 0.0 0.0
])
ρ0 = Matrix{ComplexF64}([
    0.0 0.0 0.0 0.0
    0.0 1.0 0.0 0.0
    0.0 0.0 0.0 0.0
    0.0 0.0 0.0 0.0
])
dt = 0.125
ntimes = 100

# define the spectral density for the vibrational bath
Jw = Vector{SpectralDensities.DrudeLorentz}()
svec = Vector{Matrix{ComplexF64}}()
σz = Matrix{ComplexF64}([
    1.0 0.0
    0.0 -1.0
])
id = Matrix{ComplexF64}([
    1.0 0.0
    0.0 1.0
])
jw1 = SpectralDensities.DrudeLorentz(; λ=bo, γ=5.0, Δs=1.0)
bo1 = kron(σz, id)
bo2 = kron(id, σz)

# define the bath that cause a change in the excitation state of a monomer
σx = Matrix{ComplexF64}([
    0.0 1.0
    1.0 0.0
])
jw3 = SpectralDensities.DrudeLorentz(; λ=se, γ=5.0, Δs=1.0)
se3 = kron(σx, id) + kron(id, σx)

times, ρs = HEOM.propagate(; Hamiltonian=H, ρ0=ρ0, Jw=[jw1, jw1, jw3], β, ntimes, dt, sys_ops=[bo1, bo2, se3], num_modes, Lmax)
\end{minted}

\subsubsection{Path Integral Methods}\label{sec:pi_methods}
Path integral approaches form the other numerically exact family of
computational methods for simulating the dynamics of open quantum systems as
described by Eqs.~\ref{eq:sys_bath} and~\ref{eq:quapi_bath}. Since the original
papers~\cite{makriTensorPropagatorIterativeI1995,
    makriTensorPropagatorIterativeII1995, makriNumericalPathIntegral1995}
significant developments~\cite{makriBlipDecompositionPath2014,
    makriIterativeBlipsummedPath2017, strathearnEfficientNonMarkovianQuantum2018,
    makriSmallMatrixPath2020, makriSmallMatrixDisentanglement2020,
    makriSmallMatrixModular2021, makriSmallMatrixPath2021,
    makriSmallMatrixPath2021a, boseTensorNetworkRepresentation2021,
    boseMultisiteDecompositionTensor2022, bosePairwiseConnectedTensor2022} have
led to the proliferation of methods based on the foundations of path
integrals with Feynman-Vernon influence
functional~\cite{feynmanTheoryGeneralQuantum1963}.

Starting from an initial state given as a direct product of the system reduced
density matrix and the thermal distribution of the environment, the dynamics of
the system after $N$ time-steps can be expressed as a path integral,
\begin{widetext}
    \begin{align}
        \mel{s_N^+}{\rho(N\Delta t)}{s_N^-} & = \sum_{s_0^\pm}\sum_{s_1^\pm}\ldots\sum_{s_{N-1}^\pm} \bra{s_N^+}\hat{U}\dyad{s_{N-1}^+}\hat{U}\ket{s_{N-2}^+}\ldots\nonumber                     \\
                                            & \times\bra{s_1^+}\hat{U}\dyad{s_0^+}\rho(0)\dyad{s_0^-}\hat{U}^\dag\ket{s_1^-}\ldots\bra{s_{N-1}^-}\hat{U}\ket{s_N^-} F[\{s^\pm_j\}] \label{eq:pi} \\
        \text{where }F[\{s^\pm_j\}]         & = \exp\left(-\frac{1}{\hbar}\sum_{k=0}^{N}(s_k^+-s_k^-)\sum_{k'=0}^{k}(\eta_{kk'}s_{k'}^+ - \eta^*_{kk'}s_{k'}^-)\right).\label{eq:fvif}
    \end{align}
\end{widetext}
Here, $\hat{U}$ is the bare system propagator, and $F$ is the Feynman-Vernon
influence functional corresponding to the forward-backward path $s^\pm_j$. The
influence functional for a system coupled to multiple environments is given as a
product of the influence functionals corresponding to the individual
environments. The cost of simulations do not increase as long as all the
operators that the environments couple to commute with each other. The bath
response function is discretized into
$\eta$-coefficients~\cite{makriTensorPropagatorIterativeI1995}. The
non-Markovian nature of the dynamics is brought in by the dependence of the
influence functional on the full path of the system. However, in condensed
phases, the memory decays away with the time difference of the interacting
points. Thus, after a full-memory simulation of $L$ time steps, which is a
convergence parameter, one can use an iterative algorithm to propagate the
reduced density matrix further out in time. The summand of the right-hand side
of Eq.~\ref{eq:pi} can be thought of as a tensor indexed by the forward-backward
system paths called the path amplitude tensor.

Various approaches have been used to reduce the computational complexity of the
problem which na\"ively grows as $\mathcal{O}(d^{2L})$ where $d$ is the system
dimensionality and $L$ is the memory length. This is the original QuAPI algorithm.
Other approaches attempt to decrease this exponentially growing computational
and storage requirements. The blip
decomposition~\cite{makriBlipDecompositionPath2014,
    makriIterativeBlipsummedPath2017} of path integral uses the fact that the
influence functional, Eq.~\ref{eq:fvif}, depends on the value of $\Delta s = s^+
    - s^-$ for the latter point. That means that for all the paths with no
time-point where $\Delta s = 0$ the influence functional is one.  Thus this set
of paths can be summed up in a Markovian manner. In fact, any segment of path
that consists solely of points with $\Delta s = 0$, or ``sojourns'', can be
summed up through iterative matrix-vector multiplications thereby reducing the
effective number of paths that need to be considered.

Recently tensor networks have been used in a variety of ways to reduce the
complexity of these path integral calculations. Most prominent of these is the
time-evolved matrix product operators
approach~\cite{strathearnEfficientNonMarkovianQuantum2018} (TEMPO) which uses a
matrix product state to give a compact represent the path amplitude tensor
utilizing the decaying correlation between indices with large separation. Under
the tensor network path integral~\cite{boseTensorNetworkRepresentation2021}
(TNPI) implementation of the TEMPO algorithm, it has been shown that the
influence functional for multiple baths can be analytically represented in the
form of an optimal matrix product operator. Additionally, PC-TNPI is a new
tensor network that has been designed to manifestly capture the symmetries
present in the influence functional~\cite{bosePairwiseConnectedTensor2022}.

There are four basic modules of path integral simulations that are supported ---
\verb|QuAPI| implementing ideas in
Refs.~\cite{makriTensorPropagatorIterativeI1995,
    makriTensorPropagatorIterativeII1995}, \verb|Blip| implementing
Ref.~\cite{makriBlipDecompositionPath2014}, \verb|TEMPO| implemeting
Ref.~\cite{strathearnEfficientNonMarkovianQuantum2018}, and \verb|PCTNPI|
implementing Ref.~\cite{bosePairwiseConnectedTensor2022}. In principle iterative
propagation of reduced density matrices beyond the memory time is possible in
all of these methods, however based on our experience, of these methods TEMPO
gives the greatest ability to access long memory lengths and large systems. Thus
iterative propagation is implemented only in \verb|TEMPO| and in base
\verb|QuAPI|. All the modules support creation of augmented propagators which
are the effective propagators of the system in presence of the solvent.

First, we demonstrate the QuantumDynamics.jl code both for the most fundamental
path integral method, QuAPI~\cite{makriTensorPropagatorIterativeI1995,
    makriTensorPropagatorIterativeII1995} and for
TEMPO~\cite{strathearnEfficientNonMarkovianQuantum2018}. Consider the symmetric
system, $\hat{H}_0=-\hbar\Omega\sigma_x$ coupled with a bath of harmonic
oscillators characterized by an Ohmic spectral density where $\xi$ is the
dimensionless Kondo parameter and $\omega_c$ is the cutoff frequency. The
simulation with any of the methods will have the following outline:
\begin{minted}[breaklines, fontsize=\footnotesize, framesep=2mm, frame=lines]{julia}
# define the system Hamiltonian
H = Matrix{ComplexF64}([
    0.0 -1.0
    -1.0 0.0
])

# calculate the sequence of bare propagators with a given time-step
barefbU = Propagators.calculate_bare_propagators(; Hamiltonian=H, dt, ntimes)

# define the initial condition and the spectral density
ρ0 = [1.0+0.0im 0.0; 0.0 0.0]
Jw = SpectralDensities.ExponentialCutoff(; ξ, ωc)

# use a given method to propagate the initial state in presence of the environment at an inverse temperature, β.
# method is one of QuAPI.propagate or TEMPO.propagate
t, ρs = method(; fbU=barefbU, Jw=[Jw], β, ρ0, dt, ntimes, L, svec)
\end{minted}
\verb|method| is currently one of \verb|QuAPI.propagate| or
\verb|TEMPO.propagate|. These propagate methods take custom extra arguments
which specify how to tune the algorithms in specific ways to improve the
performance.

\begin{figure}
    \subfloat[$\xi=0.1$, $\omega_c=7.5\Omega$, $\hbar\Omega\beta=5$, $L=6$, $\Delta t=0.25$. Parameters from Ref.~\cite{makriTensorPropagatorIterativeI1995}. Run with \texttt{method}=\texttt{QuAPI.propagate}.]{\includegraphics{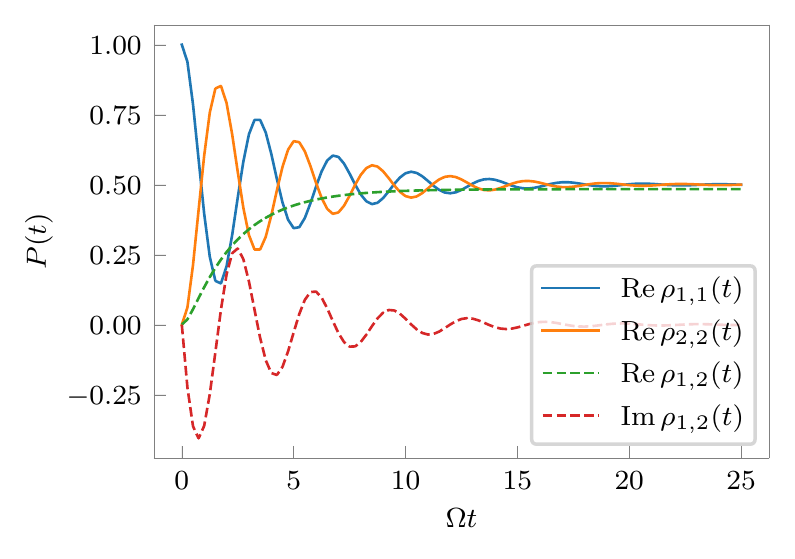}}\label{fig:quapi_eg}

    \subfloat[$\xi=2.0$, $\omega_c=\Omega$, $\hbar\Omega\beta=1$, $L=150$, $\Delta t = 0.125$. Parameters from Ref.~\cite{makriSmallMatrixPath2021, waltersIterativeQuantumclassicalPath2016}. Run with \texttt{method}=\texttt{TEMPO.propagate}.]{\includegraphics{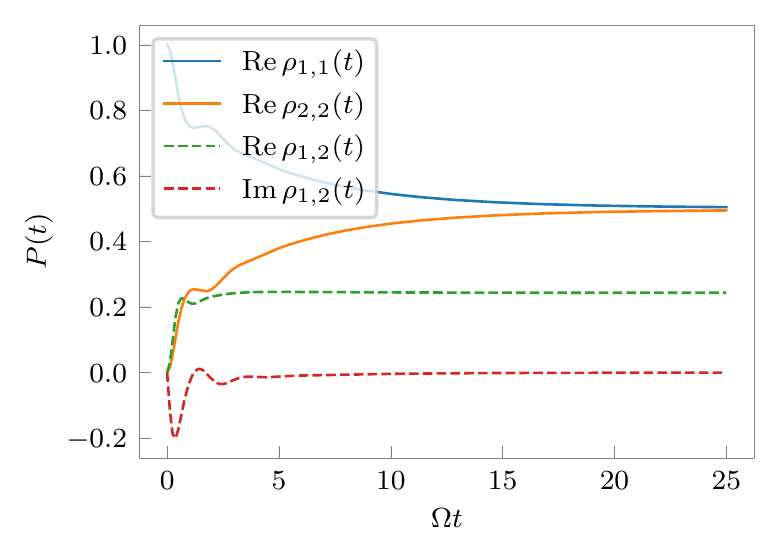}}\label{fig:tempo_eg}
    \caption{Example of dynamics using QuAPI-related methods simulated using QuantumDynamics.jl.}\label{fig:pi_eg1}
\end{figure}
As an illustration, we demonstrate two different parameters using base QuAPI
(Fig.~\ref{fig:pi_eg1}~(a)) and using TEMPO (Fig.~\ref{fig:pi_eg1}~(b)). For the
example simulated using QuAPI, we use a parameter that was introduced in
Ref.~\cite{makriTensorPropagatorIterativeI1995}. For the example that we
simulated using TEMPO, we chose a parameter that was originally simulated using
quantum-classical path integral up to a short
time~\cite{waltersIterativeQuantumclassicalPath2016} and more recently using
SMatPI till equilibration~\cite{makriSmallMatrixPath2021}. For this case, the
bath is localized around the initial system state.

\begin{figure}
    \hspace{-1.2cm}
    \subfloat[$\lambda=\SI{20}{\per\cm}$]{\includegraphics{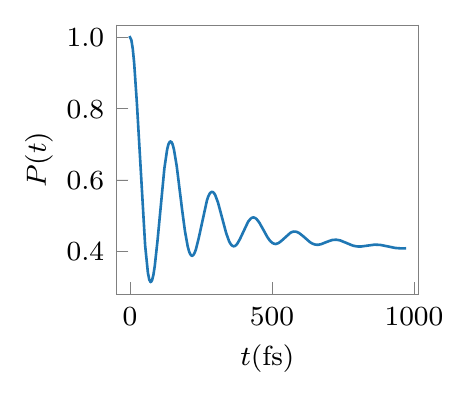}}
    ~\subfloat[$\lambda=\SI{100}{\per\cm}$]{\includegraphics{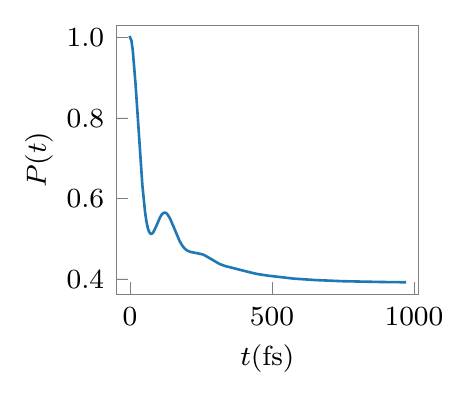}}
    \caption{Simulation of a excitation energy transfer dimer with parameters obtained from Ref.~\cite{ishizakiUnifiedTreatmentQuantum2009}. $\epsilon=\SI{100}{\per\cm}$, $\Omega=\SI{-100}{\per\cm}$.}\label{fig:ishizaki_fleming_dimer}
\end{figure}

For problems where the iterative portion of the dynamics is significantly longer
than the full-memory portion, the cost of the iteration, which is proportional
to the number of paths, adds up. One way of solving this is to use
TTM~\cite{cerrilloNonMarkovianDynamicalMaps2014} to reduce the cost to a
``convolution'' of these transfer tensors and the augmented propagators. TTM
uses the other base path integral methods to generate the propagators for some
number of time-steps and then uses them calculate the propagators further out in
time. We demonstrate the use of TTM using the strongly excitation energy
transfer (EET) dimer from Ref.~\cite{ishizakiUnifiedTreatmentQuantum2009}. The
structure of a code using TTM is shown below:
\begin{minted}[breaklines, fontsize=\footnotesize, framesep=2mm, frame=lines]{julia}
# define the EET dimer Hamiltonian
H = Matrix{ComplexF64}([
    50 100.0
    100.0 -50
] .* invcm2au)
# using a time-step, dt, calculate the bare forward-backward propagators for ntimes time-steps
fbU = Propagators.calculate_bare_propagators(; Hamiltonian=H, ntimes, dt)

# define the spectral densities acting on the different monomers
J1 = SpectralDensities.DrudeLorentz(; λ=λ * invcm2au, γ=γ * invcm2au, Δs=1.0)
J2 = SpectralDensities.DrudeLorentz(; λ=λ * invcm2au, γ=γ * invcm2au, Δs=1.0)
svec = [1.0 0.0; 0.0 1.0]
β = 1052.0

ρ0 = [1.0+0.0im 0.0; 0.0 0.0]
# TTM.propagate takes a particular path_integral_routine and the corresponding extraargs.
times, ρs = TTM.propagate(; fbU=fbU, ρ0=ρ0, Jw=[J1, J2], β, ntimes, dt, svec, rmax=rmax, extraargs=TEMPO.TEMPOArgs(; cutoff=1e-13, maxdim=10000), path_integral_routine = TEMPO.build_augmented_propagator, verbose=true)
\end{minted}
These calculations were done with full memory simulations of 75 steps with a
time-step of $\Delta t=\SI{4.84}{\fs}$. The spectral density used is the
Drude-Lorentz spectral density, Eq.~\ref{eq:drudelorentz}, with $\gamma =
    \SI{53.08}{\per\cm}$. The results are shown for two different reorganization
energies, $\lambda$, are shown in Fig.~\ref{fig:ishizaki_fleming_dimer}.

Finally, the last major method supported by QuantumDynamics.jl is
QCPI~\cite{lambertQuantumclassicalPathIntegralI2012,
    lambertQuantumclassicalPathIntegralII2012} with reference
propagators~\cite{banerjeeQuantumClassicalPathIntegral2013} and harmonic
backreaction~\cite{wangQuantumclassicalPathIntegral2019}. The incorporation of
classical trajectories not only allows for larger time-steps but also reduces
the effective memory that needs to be accounted for through path integrals by
incorporating the classical part of the memory
completely~\cite{banerjeeQuantumClassicalPathIntegral2013}. A simulation that
only incorporates the classical part of the memory is called the ensemble
average classical path (EACP) simulation. With reference propagators one can do
this simulation in a Markovian manner. Currently the support is only for a
harmonic bath though the infrastructure is built in such a manner that it is
trivial to extend it to include anharmonic solvents in the reference propagators
either by solving the equations of motion using
DifferentialEquations.jl~\cite{rackauckasDifferentialequationsJlPerformant2017},
or couple it with a molecular dynamics frameworks like
\href{https://github.com/JuliaMolSim/Molly.jl}{Molly.jl} and the Atomic
Simulation Engine~\cite{hjorthlarsenAtomicSimulationEnvironment2017} (ASE). Due
to the modular nature of QuantumDynamics.jl, these different classical
trajectory backends will work in a plug-and-play manner.

\begin{figure}
    \includegraphics{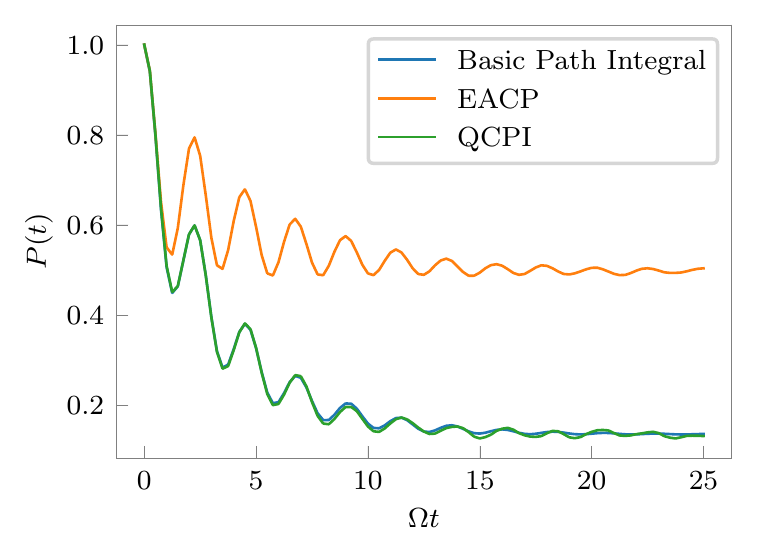}
    \caption{Comparison between QuAPI and QCPI runs for the parameters shown in the code. 10000 initial conditions were used for the EACP and QCPI calculations.}\label{fig:qcpi}
\end{figure}

Below is a code snippet which does both the EACP calculation and a full QCPI
calculation on a sample spin-boson parameter.
\begin{minted}[breaklines, fontsize=\footnotesize, framesep=2mm, frame=lines]{julia}
# specify the system Hamiltonian
H0 = Matrix{ComplexF64}([
    1.0 -1.0
    -1.0 -1.0
])

# specify the spectral density and the inverse temperature
Jw = SpectralDensities.ExponentialCutoff(; ξ=0.1, ωc=7.5)
β = 5.0

ρ0 = Matrix{ComplexF64}([
    1.0 0.0
    0.0 0.0
])

dt = 0.25
ntimes = 100

# discretize the spectral density and create a harmonic bath solvent
# for an atomistic solvent, here we would use the actual description based on an appropriate force field or ab initio DFT calculation
ω, c = SpectralDensities.discretize(Jw, 100)
hb = Solvents.HarmonicBath(β, ω, c, [1.0, -1.0], num_points)

# calculate EACP dynamics
EACP_fbU = Propagators.calculate_average_reference_propagators(; Hamiltonian=H0, solvent=hb, classical_dt=dt / 100, dt, ntimes)
times_EACP, ρs_EACP = Utilities.apply_propagator(; propagators=EACP_fbU, ρ0, ntimes, dt)

# simulate QCPI
times_QCPI, ρs_QCPI = QCPI.propagate(; Hamiltonian=H0, Jw, solvent=hb, ρ0, classical_dt=dt / 100, dt, ntimes, kmax=3, extraargs=QuAPI.QuAPIArgs(), path_integral_routine=QuAPI.propagate)
\end{minted}
QuantumDynamics.jl does not enforce any parallelization over the Monte Carlo
runs or binning and calculation of error statistics. That is left to the end
user to implement in a manner suited to the problem being studied. It will be quite
simple to spread \verb|QCPI.propagate| calls over multiple nodes and aggregate
across them using message-passing interface. The dynamics obtained with 10000
initial conditions is demonstrated in Fig.~\ref{fig:qcpi}.

\section{Conclusion}\label{sec:conclusions}
In this paper, we have introduced a new package called QuantumDynamics.jl for
simulations of non-adiabatic processes using the Feynman-Vernon influence
functional~\cite{feynmanTheoryGeneralQuantum1963}. The Julia programming
language has been emerging as a promising candidate for modern high-level
high-performance scientific computing, with a growing base of packages for
computational chemistry and physics. Being written in Julia, allows
QuantumDynamics.jl to take advantage of packages like DifferentialEquations.jl
for solving differential equations. This also allows us to avoid the
``two-language'' problem where the performance critical parts need to be
implemented in some lower-level high-performance language.

Simulating the dynamics of quantum systems interacting with environments is
often very difficult if done in a numerically exact manner. The exact methods
are quite involved from a theoretical perspective while being challenging to
implement in code. They are built on top of a variety of deep insights into the
structure of dynamics in these systems. Very few open-source packages exist that
aim to make these methods accessible to non-specialists while providing for a
platform to the specialists that encourages explorations and further theoretical
development. Inspired by the objectives behind
PySCF~\cite{sunPySCFPythonbasedSimulations2018}, QuantumDynamics.jl was designed
to address this particular problem. It joins the recently growing ranks of
computational packages for chemistry in the Julia programming
language~\cite{aroeiraFermiJlModern2022, gardnerNQCDynamicsJlJulia2022,
    herbstDFTKJulianApproach2021}.

QuantumDynamics.jl already supports a variety of methods. On the empirical and
perturbative end, methods like propagation of non-Hermitian Hamiltonians,
Lindblad master equation, and the perturbative Bloch-Redfield master equation
are all built on top of the backend provided DifferentialEquations.jl. BRME can
later be extended using polaron and variational polaron transformed approaches
to increase the applicability of the perturbative ideas. In terms of numerically
exact approaches, both HEOM-based and QuAPI-based methods are supported. In
HEOM, we have already implemented the unscaled and scaled version. Use of matrix
product states and other approaches of generalizing it to account for
non-Drude-Lorentz spectral densities will be implemented in the near future.

The largest set of methods implemented in QuantumDynamics.jl fall in the
category of path integral- or QuAPI-based approaches. The base methods of QuAPI,
blip decomposition, TEMPO, and PC-TNPI are all supported. QuAPI and TEMPO
support propagation of density matrices, while blips and PC-TNPI are currently
only capable of producing augmented forward-backward propagators. While this
does not hamper the usability of these methods in conjunction with TTM, this
deficiency will be remedied in a future version. Probably the single most useful
sub-module of the path integral methods is TEMPO. Given its ability to handle
comparatively large systems with long memories makes it exceptionally powerful.
The TNPI-based implementation allows use of multiple baths in an optimal
manner. The compatibility of all of these methods with TTM is a very useful
feature of QuantumDynamics.jl.

The goal is to provide the community with a platform that is fit for exploration
and method development in addition to a repository of methods that can directly
be used for accurate simulations of quantum dynamics. There are many other
developments that are yet to be incorporated in QuantumDynamics.jl. A notable
example is the recently developed multisite decomposition of the tensor network
path integral~\cite{boseMultisiteDecompositionTensor2022} (MS-TNPI), which
combines ideas from time-dependent density matrix renormalization
group~\cite{whiteRealTimeEvolutionUsing2004,
    schollwockDensitymatrixRenormalizationGroup2005,
    schollwockDensitymatrixRenormalizationGroup2011,
    schollwockDensitymatrixRenormalizationGroup2011a,
    paeckelTimeevolutionMethodsMatrixproduct2019} with the Feynman-Vernon influence
functional in order to make simulations of extended open quantum systems
feasible~\cite{boseEffectTemperatureGradient2022, boseTensorNetworkPath2022}.
While we will introduce some methods like
MS-TNPI~\cite{boseMultisiteDecompositionTensor2022} in the near future and
continue to develop into this package, we hope that QuantumDynamics.jl becomes a
toolbox for the community with others actively using and developing it as well.

\bibliography{library}

\begin{thebibliography}{91}%
\makeatletter
\providecommand \@ifxundefined [1]{%
 \@ifx{#1\undefined}
}%
\providecommand \@ifnum [1]{%
 \ifnum #1\expandafter \@firstoftwo
 \else \expandafter \@secondoftwo
 \fi
}%
\providecommand \@ifx [1]{%
 \ifx #1\expandafter \@firstoftwo
 \else \expandafter \@secondoftwo
 \fi
}%
\providecommand \natexlab [1]{#1}%
\providecommand \enquote  [1]{``#1''}%
\providecommand \bibnamefont  [1]{#1}%
\providecommand \bibfnamefont [1]{#1}%
\providecommand \citenamefont [1]{#1}%
\providecommand \href@noop [0]{\@secondoftwo}%
\providecommand \href [0]{\begingroup \@sanitize@url \@href}%
\providecommand \@href[1]{\@@startlink{#1}\@@href}%
\providecommand \@@href[1]{\endgroup#1\@@endlink}%
\providecommand \@sanitize@url [0]{\catcode `\\12\catcode `\$12\catcode
  `\&12\catcode `\#12\catcode `\^12\catcode `\_12\catcode `\%12\relax}%
\providecommand \@@startlink[1]{}%
\providecommand \@@endlink[0]{}%
\providecommand \url  [0]{\begingroup\@sanitize@url \@url }%
\providecommand \@url [1]{\endgroup\@href {#1}{\urlprefix }}%
\providecommand \urlprefix  [0]{URL }%
\providecommand \Eprint [0]{\href }%
\providecommand \doibase [0]{https://doi.org/}%
\providecommand \selectlanguage [0]{\@gobble}%
\providecommand \bibinfo  [0]{\@secondoftwo}%
\providecommand \bibfield  [0]{\@secondoftwo}%
\providecommand \translation [1]{[#1]}%
\providecommand \BibitemOpen [0]{}%
\providecommand \bibitemStop [0]{}%
\providecommand \bibitemNoStop [0]{.\EOS\space}%
\providecommand \EOS [0]{\spacefactor3000\relax}%
\providecommand \BibitemShut  [1]{\csname bibitem#1\endcsname}%
\let\auto@bib@innerbib\@empty
\bibitem [{\citenamefont {Bloch}(1957)}]{blochGeneralizedTheoryRelaxation1957}%
  \BibitemOpen
  \bibfield  {author} {\bibinfo {author} {\bibfnamefont {F.}~\bibnamefont
  {Bloch}},\ }\bibfield  {title} {\bibinfo {title} {Generalized {{Theory}} of
  {{Relaxation}}},\ }\href {https://doi.org/10.1103/PhysRev.105.1206}
  {\bibfield  {journal} {\bibinfo  {journal} {Phys. Rev.}\ }\textbf {\bibinfo
  {volume} {105}},\ \bibinfo {pages} {1206} (\bibinfo {year}
  {1957})}\BibitemShut {NoStop}%
\bibitem [{\citenamefont
  {Redfield}(1957)}]{redfieldTheoryRelaxationProcesses1957}%
  \BibitemOpen
  \bibfield  {author} {\bibinfo {author} {\bibfnamefont {A.~G.}\ \bibnamefont
  {Redfield}},\ }\bibfield  {title} {\bibinfo {title} {On the {{Theory}} of
  {{Relaxation Processes}}},\ }\href@noop {} {\bibfield  {journal} {\bibinfo
  {journal} {IBM J. Res. Dev.}\ }\textbf {\bibinfo {volume} {1}},\ \bibinfo
  {pages} {19} (\bibinfo {year} {1957})}\BibitemShut {NoStop}%
\bibitem [{\citenamefont {Makri}\ and\ \citenamefont
  {Makarov}(1995{\natexlab{a}})}]{makriTensorPropagatorIterativeI1995}%
  \BibitemOpen
  \bibfield  {author} {\bibinfo {author} {\bibfnamefont {N.}~\bibnamefont
  {Makri}}\ and\ \bibinfo {author} {\bibfnamefont {D.~E.}\ \bibnamefont
  {Makarov}},\ }\bibfield  {title} {\bibinfo {title} {Tensor propagator for
  iterative quantum time evolution of reduced density matrices. {{I}}.
  {{Theory}}},\ }\href {https://doi.org/10.1063/1.469508} {\bibfield  {journal}
  {\bibinfo  {journal} {The Journal of Chemical Physics}\ }\textbf {\bibinfo
  {volume} {102}},\ \bibinfo {pages} {4600} (\bibinfo {year}
  {1995}{\natexlab{a}})}\BibitemShut {NoStop}%
\bibitem [{\citenamefont {Makri}\ and\ \citenamefont
  {Makarov}(1995{\natexlab{b}})}]{makriTensorPropagatorIterativeII1995}%
  \BibitemOpen
  \bibfield  {author} {\bibinfo {author} {\bibfnamefont {N.}~\bibnamefont
  {Makri}}\ and\ \bibinfo {author} {\bibfnamefont {D.~E.}\ \bibnamefont
  {Makarov}},\ }\bibfield  {title} {\bibinfo {title} {Tensor propagator for
  iterative quantum time evolution of reduced density matrices. {{II}}.
  {{Numerical}} methodology},\ }\href {https://doi.org/10.1063/1.469509}
  {\bibfield  {journal} {\bibinfo  {journal} {The Journal of Chemical Physics}\
  }\textbf {\bibinfo {volume} {102}},\ \bibinfo {pages} {4611} (\bibinfo {year}
  {1995}{\natexlab{b}})}\BibitemShut {NoStop}%
\bibitem [{\citenamefont {Makri}(1995)}]{makriNumericalPathIntegral1995}%
  \BibitemOpen
  \bibfield  {author} {\bibinfo {author} {\bibfnamefont {N.}~\bibnamefont
  {Makri}},\ }\bibfield  {title} {\bibinfo {title} {Numerical path integral
  techniques for long time dynamics of quantum dissipative systems},\ }\href
  {https://doi.org/10.1063/1.531046} {\bibfield  {journal} {\bibinfo  {journal}
  {Journal of Mathematical Physics}\ }\textbf {\bibinfo {volume} {36}},\
  \bibinfo {pages} {2430} (\bibinfo {year} {1995})}\BibitemShut {NoStop}%
\bibitem [{\citenamefont {Tanimura}\ and\ \citenamefont
  {Kubo}(1989)}]{tanimuraTimeEvolutionQuantum1989}%
  \BibitemOpen
  \bibfield  {author} {\bibinfo {author} {\bibfnamefont {Y.}~\bibnamefont
  {Tanimura}}\ and\ \bibinfo {author} {\bibfnamefont {R.}~\bibnamefont
  {Kubo}},\ }\bibfield  {title} {\bibinfo {title} {Time {{Evolution}} of a
  {{Quantum System}} in {{Contact}} with a {{Nearly Gaussian-Markoffian Noise
  Bath}}},\ }\href {https://doi.org/10.1143/JPSJ.58.101} {\bibfield  {journal}
  {\bibinfo  {journal} {Journal of the Physical Society of Japan}\ }\textbf
  {\bibinfo {volume} {58}},\ \bibinfo {pages} {101} (\bibinfo {year}
  {1989})}\BibitemShut {NoStop}%
\bibitem [{\citenamefont {Ishizaki}\ and\ \citenamefont
  {Tanimura}(2005)}]{ishizakiQuantumDynamicsSystem2005}%
  \BibitemOpen
  \bibfield  {author} {\bibinfo {author} {\bibfnamefont {A.}~\bibnamefont
  {Ishizaki}}\ and\ \bibinfo {author} {\bibfnamefont {Y.}~\bibnamefont
  {Tanimura}},\ }\bibfield  {title} {\bibinfo {title} {Quantum {{Dynamics}} of
  {{System Strongly Coupled}} to {{Low-Temperature Colored Noise Bath}}:
  {{Reduced Hierarchy Equations Approach}}},\ }\href
  {https://doi.org/10.1143/JPSJ.74.3131} {\bibfield  {journal} {\bibinfo
  {journal} {J. Phys. Soc. Jpn.}\ }\textbf {\bibinfo {volume} {74}},\ \bibinfo
  {pages} {3131} (\bibinfo {year} {2005})}\BibitemShut {NoStop}%
\bibitem [{\citenamefont
  {Tanimura}(2006)}]{tanimuraStochasticLiouvilleLangevin2006}%
  \BibitemOpen
  \bibfield  {author} {\bibinfo {author} {\bibfnamefont {Y.}~\bibnamefont
  {Tanimura}},\ }\bibfield  {title} {\bibinfo {title} {Stochastic
  {{Liouville}}, {{Langevin}}, {{Fokker}}\textendash{{Planck}}, and {{Master
  Equation Approaches}} to {{Quantum Dissipative Systems}}},\ }\href
  {https://doi.org/10.1143/JPSJ.75.082001} {\bibfield  {journal} {\bibinfo
  {journal} {J. Phys. Soc. Jpn.}\ }\textbf {\bibinfo {volume} {75}},\ \bibinfo
  {pages} {082001} (\bibinfo {year} {2006})}\BibitemShut {NoStop}%
\bibitem [{\citenamefont {Feynman}\ and\ \citenamefont
  {Vernon}(1963)}]{feynmanTheoryGeneralQuantum1963}%
  \BibitemOpen
  \bibfield  {author} {\bibinfo {author} {\bibfnamefont {R.~P.}\ \bibnamefont
  {Feynman}}\ and\ \bibinfo {author} {\bibfnamefont {F.~L.}\ \bibnamefont
  {Vernon}},\ }\bibfield  {title} {\bibinfo {title} {The theory of a general
  quantum system interacting with a linear dissipative system},\ }\href
  {https://doi.org/10.1016/0003-4916(63)90068-x} {\bibfield  {journal}
  {\bibinfo  {journal} {Annals of Physics}\ }\textbf {\bibinfo {volume} {24}},\
  \bibinfo {pages} {118} (\bibinfo {year} {1963})}\BibitemShut {NoStop}%
\bibitem [{\citenamefont {Makri}(2014)}]{makriBlipDecompositionPath2014}%
  \BibitemOpen
  \bibfield  {author} {\bibinfo {author} {\bibfnamefont {N.}~\bibnamefont
  {Makri}},\ }\bibfield  {title} {\bibinfo {title} {Blip decomposition of the
  path integral: {{Exponential}} acceleration of real-time calculations on
  quantum dissipative systems},\ }\href {https://doi.org/10.1063/1.4896736}
  {\bibfield  {journal} {\bibinfo  {journal} {The Journal of Chemical Physics}\
  }\textbf {\bibinfo {volume} {141}},\ \bibinfo {pages} {134117} (\bibinfo
  {year} {2014})}\BibitemShut {NoStop}%
\bibitem [{\citenamefont {Makri}(2017)}]{makriIterativeBlipsummedPath2017}%
  \BibitemOpen
  \bibfield  {author} {\bibinfo {author} {\bibfnamefont {N.}~\bibnamefont
  {Makri}},\ }\bibfield  {title} {\bibinfo {title} {Iterative blip-summed path
  integral for quantum dynamics in strongly dissipative environments},\ }\href
  {https://doi.org/10.1063/1.4979197} {\bibfield  {journal} {\bibinfo
  {journal} {The Journal of Chemical Physics}\ }\textbf {\bibinfo {volume}
  {146}},\ \bibinfo {pages} {134101} (\bibinfo {year} {2017})}\BibitemShut
  {NoStop}%
\bibitem [{\citenamefont
  {Bose}(2022{\natexlab{a}})}]{bosePairwiseConnectedTensor2022}%
  \BibitemOpen
  \bibfield  {author} {\bibinfo {author} {\bibfnamefont {A.}~\bibnamefont
  {Bose}},\ }\bibfield  {title} {\bibinfo {title} {Pairwise connected tensor
  network representation of path integrals},\ }\href
  {https://doi.org/10.1103/PhysRevB.105.024309} {\bibfield  {journal} {\bibinfo
   {journal} {Physical Review B}\ }\textbf {\bibinfo {volume} {105}},\ \bibinfo
  {pages} {024309} (\bibinfo {year} {2022}{\natexlab{a}})}\BibitemShut
  {NoStop}%
\bibitem [{\citenamefont {Bose}\ and\ \citenamefont
  {Walters}(2022{\natexlab{a}})}]{boseMultisiteDecompositionTensor2022}%
  \BibitemOpen
  \bibfield  {author} {\bibinfo {author} {\bibfnamefont {A.}~\bibnamefont
  {Bose}}\ and\ \bibinfo {author} {\bibfnamefont {P.~L.}\ \bibnamefont
  {Walters}},\ }\bibfield  {title} {\bibinfo {title} {A multisite decomposition
  of the tensor network path integrals},\ }\href
  {https://doi.org/10.1063/5.0073234} {\bibfield  {journal} {\bibinfo
  {journal} {The Journal of Chemical Physics}\ }\textbf {\bibinfo {volume}
  {156}},\ \bibinfo {pages} {024101} (\bibinfo {year}
  {2022}{\natexlab{a}})}\BibitemShut {NoStop}%
\bibitem [{\citenamefont {Makri}(2018)}]{makriModularPathIntegral2018}%
  \BibitemOpen
  \bibfield  {author} {\bibinfo {author} {\bibfnamefont {N.}~\bibnamefont
  {Makri}},\ }\bibfield  {title} {\bibinfo {title} {Modular path integral
  methodology for real-time quantum dynamics},\ }\href
  {https://doi.org/10.1063/1.5058223} {\bibfield  {journal} {\bibinfo
  {journal} {The Journal of Chemical Physics}\ }\textbf {\bibinfo {volume}
  {149}},\ \bibinfo {pages} {214108} (\bibinfo {year} {2018})}\BibitemShut
  {NoStop}%
\bibitem [{\citenamefont {Bose}\ and\ \citenamefont
  {Walters}(2021)}]{boseTensorNetworkRepresentation2021}%
  \BibitemOpen
  \bibfield  {author} {\bibinfo {author} {\bibfnamefont {A.}~\bibnamefont
  {Bose}}\ and\ \bibinfo {author} {\bibfnamefont {P.~L.}\ \bibnamefont
  {Walters}},\ }\bibfield  {title} {\bibinfo {title} {A tensor network
  representation of path integrals: {{Implementation}} and analysis},\
  }\href@noop {} {\bibfield  {journal} {\bibinfo  {journal} {arXiv pre-print
  server arXiv:2106.12523}\ } (\bibinfo {year} {2021})},\ \Eprint
  {https://arxiv.org/abs/2106.12523} {arxiv:2106.12523} \BibitemShut {NoStop}%
\bibitem [{\citenamefont {J{\o}rgensen}\ and\ \citenamefont
  {Pollock}(2019)}]{jorgensenExploitingCausalTensor2019}%
  \BibitemOpen
  \bibfield  {author} {\bibinfo {author} {\bibfnamefont {M.~R.}\ \bibnamefont
  {J{\o}rgensen}}\ and\ \bibinfo {author} {\bibfnamefont {F.~A.}\ \bibnamefont
  {Pollock}},\ }\bibfield  {title} {\bibinfo {title} {Exploiting the {{Causal
  Tensor Network Structure}} of {{Quantum Processes}} to {{Efficiently Simulate
  Non-Markovian Path Integrals}}},\ }\href
  {https://doi.org/10.1103/physrevlett.123.240602} {\bibfield  {journal}
  {\bibinfo  {journal} {Physical Review Letters}\ }\textbf {\bibinfo {volume}
  {123}},\ \bibinfo {pages} {240602} (\bibinfo {year} {2019})}\BibitemShut
  {NoStop}%
\bibitem [{\citenamefont
  {Makri}(2020{\natexlab{a}})}]{makriSmallMatrixDisentanglement2020}%
  \BibitemOpen
  \bibfield  {author} {\bibinfo {author} {\bibfnamefont {N.}~\bibnamefont
  {Makri}},\ }\bibfield  {title} {\bibinfo {title} {Small matrix
  disentanglement of the path integral: {{Overcoming}} the exponential tensor
  scaling with memory length},\ }\href {https://doi.org/10.1063/1.5139473}
  {\bibfield  {journal} {\bibinfo  {journal} {The Journal of Chemical Physics}\
  }\textbf {\bibinfo {volume} {152}},\ \bibinfo {pages} {41104} (\bibinfo
  {year} {2020}{\natexlab{a}})}\BibitemShut {NoStop}%
\bibitem [{\citenamefont
  {Makri}(2020{\natexlab{b}})}]{makriSmallMatrixPath2020}%
  \BibitemOpen
  \bibfield  {author} {\bibinfo {author} {\bibfnamefont {N.}~\bibnamefont
  {Makri}},\ }\bibfield  {title} {\bibinfo {title} {Small {{Matrix Path
  Integral}} for {{System-Bath Dynamics}}},\ }\href
  {https://doi.org/10.1021/acs.jctc.0c00039} {\bibfield  {journal} {\bibinfo
  {journal} {Journal of Chemical Theory and Computation}\ }\textbf {\bibinfo
  {volume} {16}},\ \bibinfo {pages} {4038} (\bibinfo {year}
  {2020}{\natexlab{b}})}\BibitemShut {NoStop}%
\bibitem [{\citenamefont
  {Makri}(2021{\natexlab{a}})}]{makriSmallMatrixModular2021}%
  \BibitemOpen
  \bibfield  {author} {\bibinfo {author} {\bibfnamefont {N.}~\bibnamefont
  {Makri}},\ }\bibfield  {title} {\bibinfo {title} {Small matrix modular path
  integral: Iterative quantum dynamics in space and time},\ }\href
  {https://doi.org/10.1039/D1CP01483H} {\bibfield  {journal} {\bibinfo
  {journal} {Physical Chemistry Chemical Physics}\ }\textbf {\bibinfo {volume}
  {23}},\ \bibinfo {pages} {12537} (\bibinfo {year}
  {2021}{\natexlab{a}})}\BibitemShut {NoStop}%
\bibitem [{\citenamefont
  {Makri}(2021{\natexlab{b}})}]{makriSmallMatrixPath2021a}%
  \BibitemOpen
  \bibfield  {author} {\bibinfo {author} {\bibfnamefont {N.}~\bibnamefont
  {Makri}},\ }\bibfield  {title} {\bibinfo {title} {Small matrix path integral
  for driven dissipative dynamics},\ }\href
  {https://doi.org/10.1021/acs.jpca.1c08230} {\bibfield  {journal} {\bibinfo
  {journal} {The Journal of Physical Chemistry A}\ }\textbf {\bibinfo {volume}
  {125}},\ \bibinfo {pages} {10500} (\bibinfo {year} {2021}{\natexlab{b}})},\
  \Eprint {https://arxiv.org/abs/https://doi.org/10.1021/acs.jpca.1c08230}
  {https://doi.org/10.1021/acs.jpca.1c08230} \BibitemShut {NoStop}%
\bibitem [{\citenamefont {Strathearn}\ \emph {et~al.}(2018)\citenamefont
  {Strathearn}, \citenamefont {Kirton}, \citenamefont {Kilda}, \citenamefont
  {Keeling},\ and\ \citenamefont
  {Lovett}}]{strathearnEfficientNonMarkovianQuantum2018}%
  \BibitemOpen
  \bibfield  {author} {\bibinfo {author} {\bibfnamefont {A.}~\bibnamefont
  {Strathearn}}, \bibinfo {author} {\bibfnamefont {P.}~\bibnamefont {Kirton}},
  \bibinfo {author} {\bibfnamefont {D.}~\bibnamefont {Kilda}}, \bibinfo
  {author} {\bibfnamefont {J.}~\bibnamefont {Keeling}},\ and\ \bibinfo {author}
  {\bibfnamefont {B.~W.}\ \bibnamefont {Lovett}},\ }\bibfield  {title}
  {\bibinfo {title} {Efficient non-{{Markovian}} quantum dynamics using
  time-evolving matrix product operators},\ }\bibfield  {journal} {\bibinfo
  {journal} {Nature Communications}\ }\textbf {\bibinfo {volume} {9}},\ \href
  {https://doi.org/10.1038/s41467-018-05617-3} {10.1038/s41467-018-05617-3}
  (\bibinfo {year} {2018})\BibitemShut {NoStop}%
\bibitem [{\citenamefont {Hu}\ \emph {et~al.}(2011)\citenamefont {Hu},
  \citenamefont {Luo}, \citenamefont {Jiang}, \citenamefont {Xu},\ and\
  \citenamefont {Yan}}]{huPadeSpectrumDecompositions2011}%
  \BibitemOpen
  \bibfield  {author} {\bibinfo {author} {\bibfnamefont {J.}~\bibnamefont
  {Hu}}, \bibinfo {author} {\bibfnamefont {M.}~\bibnamefont {Luo}}, \bibinfo
  {author} {\bibfnamefont {F.}~\bibnamefont {Jiang}}, \bibinfo {author}
  {\bibfnamefont {R.-X.}\ \bibnamefont {Xu}},\ and\ \bibinfo {author}
  {\bibfnamefont {Y.}~\bibnamefont {Yan}},\ }\bibfield  {title} {\bibinfo
  {title} {Pad\'e spectrum decompositions of quantum distribution functions and
  optimal hierarchical equations of motion construction for quantum open
  systems},\ }\href {https://doi.org/10.1063/1.3602466} {\bibfield  {journal}
  {\bibinfo  {journal} {J. Chem. Phys.}\ }\textbf {\bibinfo {volume} {134}},\
  \bibinfo {pages} {244106} (\bibinfo {year} {2011})}\BibitemShut {NoStop}%
\bibitem [{\citenamefont {Shi}\ \emph {et~al.}(2009)\citenamefont {Shi},
  \citenamefont {Chen}, \citenamefont {Nan}, \citenamefont {Xu},\ and\
  \citenamefont {Yan}}]{shiEfficientHierarchicalLiouville2009}%
  \BibitemOpen
  \bibfield  {author} {\bibinfo {author} {\bibfnamefont {Q.}~\bibnamefont
  {Shi}}, \bibinfo {author} {\bibfnamefont {L.}~\bibnamefont {Chen}}, \bibinfo
  {author} {\bibfnamefont {G.}~\bibnamefont {Nan}}, \bibinfo {author}
  {\bibfnamefont {R.-X.}\ \bibnamefont {Xu}},\ and\ \bibinfo {author}
  {\bibfnamefont {Y.}~\bibnamefont {Yan}},\ }\bibfield  {title} {\bibinfo
  {title} {Efficient hierarchical {{Liouville}} space propagator to quantum
  dissipative dynamics},\ }\href {https://doi.org/10.1063/1.3077918} {\bibfield
   {journal} {\bibinfo  {journal} {J. Chem. Phys.}\ }\textbf {\bibinfo {volume}
  {130}},\ \bibinfo {pages} {084105} (\bibinfo {year} {2009})}\BibitemShut
  {NoStop}%
\bibitem [{\citenamefont {Shi}\ \emph {et~al.}(2018)\citenamefont {Shi},
  \citenamefont {Xu}, \citenamefont {Yan},\ and\ \citenamefont
  {Xu}}]{shiEfficientPropagationHierarchical2018}%
  \BibitemOpen
  \bibfield  {author} {\bibinfo {author} {\bibfnamefont {Q.}~\bibnamefont
  {Shi}}, \bibinfo {author} {\bibfnamefont {Y.}~\bibnamefont {Xu}}, \bibinfo
  {author} {\bibfnamefont {Y.}~\bibnamefont {Yan}},\ and\ \bibinfo {author}
  {\bibfnamefont {M.}~\bibnamefont {Xu}},\ }\bibfield  {title} {\bibinfo
  {title} {Efficient propagation of the hierarchical equations of motion using
  the matrix product state method},\ }\href {https://doi.org/10.1063/1.5026753}
  {\bibfield  {journal} {\bibinfo  {journal} {The Journal of Chemical Physics}\
  }\textbf {\bibinfo {volume} {148}},\ \bibinfo {pages} {174102} (\bibinfo
  {year} {2018})}\BibitemShut {NoStop}%
\bibitem [{\citenamefont {Yan}\ \emph {et~al.}(2021)\citenamefont {Yan},
  \citenamefont {Xu}, \citenamefont {Li},\ and\ \citenamefont
  {Shi}}]{yanEfficientPropagationHierarchical2021}%
  \BibitemOpen
  \bibfield  {author} {\bibinfo {author} {\bibfnamefont {Y.}~\bibnamefont
  {Yan}}, \bibinfo {author} {\bibfnamefont {M.}~\bibnamefont {Xu}}, \bibinfo
  {author} {\bibfnamefont {T.}~\bibnamefont {Li}},\ and\ \bibinfo {author}
  {\bibfnamefont {Q.}~\bibnamefont {Shi}},\ }\bibfield  {title} {\bibinfo
  {title} {Efficient propagation of the hierarchical equations of motion using
  the {{Tucker}} and hierarchical {{Tucker}} tensors},\ }\href
  {https://doi.org/10.1063/5.0050720} {\bibfield  {journal} {\bibinfo
  {journal} {The Journal of Chemical Physics}\ }\textbf {\bibinfo {volume}
  {154}},\ \bibinfo {pages} {194104} (\bibinfo {year} {2021})}\BibitemShut
  {NoStop}%
\bibitem [{\citenamefont {Ikeda}\ and\ \citenamefont
  {Scholes}(2020)}]{ikedaGeneralizationHierarchicalEquations2020}%
  \BibitemOpen
  \bibfield  {author} {\bibinfo {author} {\bibfnamefont {T.}~\bibnamefont
  {Ikeda}}\ and\ \bibinfo {author} {\bibfnamefont {G.~D.}\ \bibnamefont
  {Scholes}},\ }\bibfield  {title} {\bibinfo {title} {Generalization of the
  hierarchical equations of motion theory for efficient calculations with
  arbitrary correlation functions},\ }\href {https://doi.org/10.1063/5.0007327}
  {\bibfield  {journal} {\bibinfo  {journal} {The Journal of Chemical Physics}\
  }\textbf {\bibinfo {volume} {152}},\ \bibinfo {pages} {204101} (\bibinfo
  {year} {2020})}\BibitemShut {NoStop}%
\bibitem [{\citenamefont {Apr{\`a}}\ \emph {et~al.}(2020)\citenamefont
  {Apr{\`a}}, \citenamefont {Bylaska}, \citenamefont {{de Jong}}, \citenamefont
  {Govind}, \citenamefont {Kowalski}, \citenamefont {Straatsma}, \citenamefont
  {Valiev}, \citenamefont {{van Dam}}, \citenamefont {Alexeev}, \citenamefont
  {Anchell}, \citenamefont {Anisimov}, \citenamefont {Aquino}, \citenamefont
  {{Atta-Fynn}}, \citenamefont {Autschbach}, \citenamefont {Bauman},
  \citenamefont {Becca}, \citenamefont {Bernholdt}, \citenamefont
  {{Bhaskaran-Nair}}, \citenamefont {Bogatko}, \citenamefont {Borowski},
  \citenamefont {Boschen}, \citenamefont {Brabec}, \citenamefont {Bruner},
  \citenamefont {Cau{\"e}t}, \citenamefont {Chen}, \citenamefont {Chuev},
  \citenamefont {Cramer}, \citenamefont {Daily}, \citenamefont {Deegan},
  \citenamefont {Dunning}, \citenamefont {Dupuis}, \citenamefont {Dyall},
  \citenamefont {Fann}, \citenamefont {Fischer}, \citenamefont {Fonari},
  \citenamefont {Fr{\"u}chtl}, \citenamefont {Gagliardi}, \citenamefont
  {Garza}, \citenamefont {Gawande}, \citenamefont {Ghosh}, \citenamefont
  {Glaesemann}, \citenamefont {G{\"o}tz}, \citenamefont {Hammond},
  \citenamefont {Helms}, \citenamefont {Hermes}, \citenamefont {Hirao},
  \citenamefont {Hirata}, \citenamefont {Jacquelin}, \citenamefont {Jensen},
  \citenamefont {Johnson}, \citenamefont {J{\'o}nsson}, \citenamefont
  {Kendall}, \citenamefont {Klemm}, \citenamefont {Kobayashi}, \citenamefont
  {Konkov}, \citenamefont {Krishnamoorthy}, \citenamefont {Krishnan},
  \citenamefont {Lin}, \citenamefont {Lins}, \citenamefont {Littlefield},
  \citenamefont {Logsdail}, \citenamefont {Lopata}, \citenamefont {Ma},
  \citenamefont {Marenich}, \citenamefont {{Martin del Campo}}, \citenamefont
  {{Mejia-Rodriguez}}, \citenamefont {Moore}, \citenamefont {Mullin},
  \citenamefont {Nakajima}, \citenamefont {Nascimento}, \citenamefont
  {Nichols}, \citenamefont {Nichols}, \citenamefont {Nieplocha}, \citenamefont
  {{Otero-de-la-Roza}}, \citenamefont {Palmer}, \citenamefont {Panyala},
  \citenamefont {Pirojsirikul}, \citenamefont {Peng}, \citenamefont {Peverati},
  \citenamefont {Pittner}, \citenamefont {Pollack}, \citenamefont {Richard},
  \citenamefont {Sadayappan}, \citenamefont {Schatz}, \citenamefont {Shelton},
  \citenamefont {Silverstein}, \citenamefont {Smith}, \citenamefont {Soares},
  \citenamefont {Song}, \citenamefont {Swart}, \citenamefont {Taylor},
  \citenamefont {Thomas}, \citenamefont {Tipparaju}, \citenamefont {Truhlar},
  \citenamefont {Tsemekhman}, \citenamefont {Van~Voorhis}, \citenamefont
  {{V{\'a}zquez-Mayagoitia}}, \citenamefont {Verma}, \citenamefont {Villa},
  \citenamefont {Vishnu}, \citenamefont {Vogiatzis}, \citenamefont {Wang},
  \citenamefont {Weare}, \citenamefont {Williamson}, \citenamefont {Windus},
  \citenamefont {Woli{\'n}ski}, \citenamefont {Wong}, \citenamefont {Wu},
  \citenamefont {Yang}, \citenamefont {Yu}, \citenamefont {Zacharias},
  \citenamefont {Zhang}, \citenamefont {Zhao},\ and\ \citenamefont
  {Harrison}}]{apraNWChemPresentFuture2020}%
  \BibitemOpen
  \bibfield  {author} {\bibinfo {author} {\bibfnamefont {E.}~\bibnamefont
  {Apr{\`a}}}, \bibinfo {author} {\bibfnamefont {E.~J.}\ \bibnamefont
  {Bylaska}}, \bibinfo {author} {\bibfnamefont {W.~A.}\ \bibnamefont {{de
  Jong}}}, \bibinfo {author} {\bibfnamefont {N.}~\bibnamefont {Govind}},
  \bibinfo {author} {\bibfnamefont {K.}~\bibnamefont {Kowalski}}, \bibinfo
  {author} {\bibfnamefont {T.~P.}\ \bibnamefont {Straatsma}}, \bibinfo {author}
  {\bibfnamefont {M.}~\bibnamefont {Valiev}}, \bibinfo {author} {\bibfnamefont
  {H.~J.~J.}\ \bibnamefont {{van Dam}}}, \bibinfo {author} {\bibfnamefont
  {Y.}~\bibnamefont {Alexeev}}, \bibinfo {author} {\bibfnamefont
  {J.}~\bibnamefont {Anchell}}, \bibinfo {author} {\bibfnamefont
  {V.}~\bibnamefont {Anisimov}}, \bibinfo {author} {\bibfnamefont {F.~W.}\
  \bibnamefont {Aquino}}, \bibinfo {author} {\bibfnamefont {R.}~\bibnamefont
  {{Atta-Fynn}}}, \bibinfo {author} {\bibfnamefont {J.}~\bibnamefont
  {Autschbach}}, \bibinfo {author} {\bibfnamefont {N.~P.}\ \bibnamefont
  {Bauman}}, \bibinfo {author} {\bibfnamefont {J.~C.}\ \bibnamefont {Becca}},
  \bibinfo {author} {\bibfnamefont {D.~E.}\ \bibnamefont {Bernholdt}}, \bibinfo
  {author} {\bibfnamefont {K.}~\bibnamefont {{Bhaskaran-Nair}}}, \bibinfo
  {author} {\bibfnamefont {S.}~\bibnamefont {Bogatko}}, \bibinfo {author}
  {\bibfnamefont {P.}~\bibnamefont {Borowski}}, \bibinfo {author}
  {\bibfnamefont {J.}~\bibnamefont {Boschen}}, \bibinfo {author} {\bibfnamefont
  {J.}~\bibnamefont {Brabec}}, \bibinfo {author} {\bibfnamefont
  {A.}~\bibnamefont {Bruner}}, \bibinfo {author} {\bibfnamefont
  {E.}~\bibnamefont {Cau{\"e}t}}, \bibinfo {author} {\bibfnamefont
  {Y.}~\bibnamefont {Chen}}, \bibinfo {author} {\bibfnamefont {G.~N.}\
  \bibnamefont {Chuev}}, \bibinfo {author} {\bibfnamefont {C.~J.}\ \bibnamefont
  {Cramer}}, \bibinfo {author} {\bibfnamefont {J.}~\bibnamefont {Daily}},
  \bibinfo {author} {\bibfnamefont {M.~J.~O.}\ \bibnamefont {Deegan}}, \bibinfo
  {author} {\bibfnamefont {T.~H.}\ \bibnamefont {Dunning}}, \bibinfo {author}
  {\bibfnamefont {M.}~\bibnamefont {Dupuis}}, \bibinfo {author} {\bibfnamefont
  {K.~G.}\ \bibnamefont {Dyall}}, \bibinfo {author} {\bibfnamefont {G.~I.}\
  \bibnamefont {Fann}}, \bibinfo {author} {\bibfnamefont {S.~A.}\ \bibnamefont
  {Fischer}}, \bibinfo {author} {\bibfnamefont {A.}~\bibnamefont {Fonari}},
  \bibinfo {author} {\bibfnamefont {H.}~\bibnamefont {Fr{\"u}chtl}}, \bibinfo
  {author} {\bibfnamefont {L.}~\bibnamefont {Gagliardi}}, \bibinfo {author}
  {\bibfnamefont {J.}~\bibnamefont {Garza}}, \bibinfo {author} {\bibfnamefont
  {N.}~\bibnamefont {Gawande}}, \bibinfo {author} {\bibfnamefont
  {S.}~\bibnamefont {Ghosh}}, \bibinfo {author} {\bibfnamefont
  {K.}~\bibnamefont {Glaesemann}}, \bibinfo {author} {\bibfnamefont {A.~W.}\
  \bibnamefont {G{\"o}tz}}, \bibinfo {author} {\bibfnamefont {J.}~\bibnamefont
  {Hammond}}, \bibinfo {author} {\bibfnamefont {V.}~\bibnamefont {Helms}},
  \bibinfo {author} {\bibfnamefont {E.~D.}\ \bibnamefont {Hermes}}, \bibinfo
  {author} {\bibfnamefont {K.}~\bibnamefont {Hirao}}, \bibinfo {author}
  {\bibfnamefont {S.}~\bibnamefont {Hirata}}, \bibinfo {author} {\bibfnamefont
  {M.}~\bibnamefont {Jacquelin}}, \bibinfo {author} {\bibfnamefont
  {L.}~\bibnamefont {Jensen}}, \bibinfo {author} {\bibfnamefont {B.~G.}\
  \bibnamefont {Johnson}}, \bibinfo {author} {\bibfnamefont {H.}~\bibnamefont
  {J{\'o}nsson}}, \bibinfo {author} {\bibfnamefont {R.~A.}\ \bibnamefont
  {Kendall}}, \bibinfo {author} {\bibfnamefont {M.}~\bibnamefont {Klemm}},
  \bibinfo {author} {\bibfnamefont {R.}~\bibnamefont {Kobayashi}}, \bibinfo
  {author} {\bibfnamefont {V.}~\bibnamefont {Konkov}}, \bibinfo {author}
  {\bibfnamefont {S.}~\bibnamefont {Krishnamoorthy}}, \bibinfo {author}
  {\bibfnamefont {M.}~\bibnamefont {Krishnan}}, \bibinfo {author}
  {\bibfnamefont {Z.}~\bibnamefont {Lin}}, \bibinfo {author} {\bibfnamefont
  {R.~D.}\ \bibnamefont {Lins}}, \bibinfo {author} {\bibfnamefont {R.~J.}\
  \bibnamefont {Littlefield}}, \bibinfo {author} {\bibfnamefont {A.~J.}\
  \bibnamefont {Logsdail}}, \bibinfo {author} {\bibfnamefont {K.}~\bibnamefont
  {Lopata}}, \bibinfo {author} {\bibfnamefont {W.}~\bibnamefont {Ma}}, \bibinfo
  {author} {\bibfnamefont {A.~V.}\ \bibnamefont {Marenich}}, \bibinfo {author}
  {\bibfnamefont {J.}~\bibnamefont {{Martin del Campo}}}, \bibinfo {author}
  {\bibfnamefont {D.}~\bibnamefont {{Mejia-Rodriguez}}}, \bibinfo {author}
  {\bibfnamefont {J.~E.}\ \bibnamefont {Moore}}, \bibinfo {author}
  {\bibfnamefont {J.~M.}\ \bibnamefont {Mullin}}, \bibinfo {author}
  {\bibfnamefont {T.}~\bibnamefont {Nakajima}}, \bibinfo {author}
  {\bibfnamefont {D.~R.}\ \bibnamefont {Nascimento}}, \bibinfo {author}
  {\bibfnamefont {J.~A.}\ \bibnamefont {Nichols}}, \bibinfo {author}
  {\bibfnamefont {P.~J.}\ \bibnamefont {Nichols}}, \bibinfo {author}
  {\bibfnamefont {J.}~\bibnamefont {Nieplocha}}, \bibinfo {author}
  {\bibfnamefont {A.}~\bibnamefont {{Otero-de-la-Roza}}}, \bibinfo {author}
  {\bibfnamefont {B.}~\bibnamefont {Palmer}}, \bibinfo {author} {\bibfnamefont
  {A.}~\bibnamefont {Panyala}}, \bibinfo {author} {\bibfnamefont
  {T.}~\bibnamefont {Pirojsirikul}}, \bibinfo {author} {\bibfnamefont
  {B.}~\bibnamefont {Peng}}, \bibinfo {author} {\bibfnamefont {R.}~\bibnamefont
  {Peverati}}, \bibinfo {author} {\bibfnamefont {J.}~\bibnamefont {Pittner}},
  \bibinfo {author} {\bibfnamefont {L.}~\bibnamefont {Pollack}}, \bibinfo
  {author} {\bibfnamefont {R.~M.}\ \bibnamefont {Richard}}, \bibinfo {author}
  {\bibfnamefont {P.}~\bibnamefont {Sadayappan}}, \bibinfo {author}
  {\bibfnamefont {G.~C.}\ \bibnamefont {Schatz}}, \bibinfo {author}
  {\bibfnamefont {W.~A.}\ \bibnamefont {Shelton}}, \bibinfo {author}
  {\bibfnamefont {D.~W.}\ \bibnamefont {Silverstein}}, \bibinfo {author}
  {\bibfnamefont {D.~M.~A.}\ \bibnamefont {Smith}}, \bibinfo {author}
  {\bibfnamefont {T.~A.}\ \bibnamefont {Soares}}, \bibinfo {author}
  {\bibfnamefont {D.}~\bibnamefont {Song}}, \bibinfo {author} {\bibfnamefont
  {M.}~\bibnamefont {Swart}}, \bibinfo {author} {\bibfnamefont {H.~L.}\
  \bibnamefont {Taylor}}, \bibinfo {author} {\bibfnamefont {G.~S.}\
  \bibnamefont {Thomas}}, \bibinfo {author} {\bibfnamefont {V.}~\bibnamefont
  {Tipparaju}}, \bibinfo {author} {\bibfnamefont {D.~G.}\ \bibnamefont
  {Truhlar}}, \bibinfo {author} {\bibfnamefont {K.}~\bibnamefont {Tsemekhman}},
  \bibinfo {author} {\bibfnamefont {T.}~\bibnamefont {Van~Voorhis}}, \bibinfo
  {author} {\bibfnamefont {{\'A}.}~\bibnamefont {{V{\'a}zquez-Mayagoitia}}},
  \bibinfo {author} {\bibfnamefont {P.}~\bibnamefont {Verma}}, \bibinfo
  {author} {\bibfnamefont {O.}~\bibnamefont {Villa}}, \bibinfo {author}
  {\bibfnamefont {A.}~\bibnamefont {Vishnu}}, \bibinfo {author} {\bibfnamefont
  {K.~D.}\ \bibnamefont {Vogiatzis}}, \bibinfo {author} {\bibfnamefont
  {D.}~\bibnamefont {Wang}}, \bibinfo {author} {\bibfnamefont {J.~H.}\
  \bibnamefont {Weare}}, \bibinfo {author} {\bibfnamefont {M.~J.}\ \bibnamefont
  {Williamson}}, \bibinfo {author} {\bibfnamefont {T.~L.}\ \bibnamefont
  {Windus}}, \bibinfo {author} {\bibfnamefont {K.}~\bibnamefont
  {Woli{\'n}ski}}, \bibinfo {author} {\bibfnamefont {A.~T.}\ \bibnamefont
  {Wong}}, \bibinfo {author} {\bibfnamefont {Q.}~\bibnamefont {Wu}}, \bibinfo
  {author} {\bibfnamefont {C.}~\bibnamefont {Yang}}, \bibinfo {author}
  {\bibfnamefont {Q.}~\bibnamefont {Yu}}, \bibinfo {author} {\bibfnamefont
  {M.}~\bibnamefont {Zacharias}}, \bibinfo {author} {\bibfnamefont
  {Z.}~\bibnamefont {Zhang}}, \bibinfo {author} {\bibfnamefont
  {Y.}~\bibnamefont {Zhao}},\ and\ \bibinfo {author} {\bibfnamefont {R.~J.}\
  \bibnamefont {Harrison}},\ }\bibfield  {title} {\bibinfo {title} {{{NWChem}}:
  {{Past}}, present, and future},\ }\href {https://doi.org/10.1063/5.0004997}
  {\bibfield  {journal} {\bibinfo  {journal} {J. Chem. Phys.}\ }\textbf
  {\bibinfo {volume} {152}},\ \bibinfo {pages} {184102} (\bibinfo {year}
  {2020})}\BibitemShut {NoStop}%
\bibitem [{\citenamefont {Frisch}\ \emph {et~al.}(2016)\citenamefont {Frisch},
  \citenamefont {Trucks}, \citenamefont {Schlegel}, \citenamefont {Scuseria},
  \citenamefont {Robb}, \citenamefont {Cheeseman}, \citenamefont {Scalmani},
  \citenamefont {Barone}, \citenamefont {Petersson}, \citenamefont {Nakatsuji},
  \citenamefont {Li}, \citenamefont {Caricato}, \citenamefont {Marenich},
  \citenamefont {Bloino}, \citenamefont {Janesko}, \citenamefont {Gomperts},
  \citenamefont {Mennucci}, \citenamefont {Hratchian}, \citenamefont {Ortiz},
  \citenamefont {Izmaylov}, \citenamefont {Sonnenberg}, \citenamefont
  {{Williams}}, \citenamefont {Ding}, \citenamefont {Lipparini}, \citenamefont
  {Egidi}, \citenamefont {Goings}, \citenamefont {Peng}, \citenamefont
  {Petrone}, \citenamefont {Henderson}, \citenamefont {Ranasinghe},
  \citenamefont {Zakrzewski}, \citenamefont {Gao}, \citenamefont {Rega},
  \citenamefont {Zheng}, \citenamefont {Liang}, \citenamefont {Hada},
  \citenamefont {Ehara}, \citenamefont {Toyota}, \citenamefont {Fukuda},
  \citenamefont {Hasegawa}, \citenamefont {Ishida}, \citenamefont {Nakajima},
  \citenamefont {Honda}, \citenamefont {Kitao}, \citenamefont {Nakai},
  \citenamefont {Vreven}, \citenamefont {Throssell}, \citenamefont
  {Montgomery~Jr.}, \citenamefont {Peralta}, \citenamefont {Ogliaro},
  \citenamefont {Bearpark}, \citenamefont {Heyd}, \citenamefont {Brothers},
  \citenamefont {Kudin}, \citenamefont {Staroverov}, \citenamefont {Keith},
  \citenamefont {Kobayashi}, \citenamefont {Normand}, \citenamefont
  {Raghavachari}, \citenamefont {Rendell}, \citenamefont {Burant},
  \citenamefont {Iyengar}, \citenamefont {Tomasi}, \citenamefont {Cossi},
  \citenamefont {Millam}, \citenamefont {Klene}, \citenamefont {Adamo},
  \citenamefont {Cammi}, \citenamefont {Ochterski}, \citenamefont {Martin},
  \citenamefont {Morokuma}, \citenamefont {Farkas}, \citenamefont {Foresman},\
  and\ \citenamefont {Fox}}]{frischGaussian16Rev2016}%
  \BibitemOpen
  \bibfield  {author} {\bibinfo {author} {\bibfnamefont {M.~J.}\ \bibnamefont
  {Frisch}}, \bibinfo {author} {\bibfnamefont {G.~W.}\ \bibnamefont {Trucks}},
  \bibinfo {author} {\bibfnamefont {H.~B.}\ \bibnamefont {Schlegel}}, \bibinfo
  {author} {\bibfnamefont {G.~E.}\ \bibnamefont {Scuseria}}, \bibinfo {author}
  {\bibfnamefont {M.~A.}\ \bibnamefont {Robb}}, \bibinfo {author}
  {\bibfnamefont {J.~R.}\ \bibnamefont {Cheeseman}}, \bibinfo {author}
  {\bibfnamefont {G.}~\bibnamefont {Scalmani}}, \bibinfo {author}
  {\bibfnamefont {V.}~\bibnamefont {Barone}}, \bibinfo {author} {\bibfnamefont
  {G.~A.}\ \bibnamefont {Petersson}}, \bibinfo {author} {\bibfnamefont
  {H.}~\bibnamefont {Nakatsuji}}, \bibinfo {author} {\bibfnamefont
  {X.}~\bibnamefont {Li}}, \bibinfo {author} {\bibfnamefont {M.}~\bibnamefont
  {Caricato}}, \bibinfo {author} {\bibfnamefont {A.~V.}\ \bibnamefont
  {Marenich}}, \bibinfo {author} {\bibfnamefont {J.}~\bibnamefont {Bloino}},
  \bibinfo {author} {\bibfnamefont {B.~G.}\ \bibnamefont {Janesko}}, \bibinfo
  {author} {\bibfnamefont {R.}~\bibnamefont {Gomperts}}, \bibinfo {author}
  {\bibfnamefont {B.}~\bibnamefont {Mennucci}}, \bibinfo {author}
  {\bibfnamefont {H.~P.}\ \bibnamefont {Hratchian}}, \bibinfo {author}
  {\bibfnamefont {J.~V.}\ \bibnamefont {Ortiz}}, \bibinfo {author}
  {\bibfnamefont {A.~F.}\ \bibnamefont {Izmaylov}}, \bibinfo {author}
  {\bibfnamefont {J.~L.}\ \bibnamefont {Sonnenberg}}, \bibinfo {author}
  {\bibnamefont {{Williams}}}, \bibinfo {author} {\bibfnamefont
  {F.}~\bibnamefont {Ding}}, \bibinfo {author} {\bibfnamefont {F.}~\bibnamefont
  {Lipparini}}, \bibinfo {author} {\bibfnamefont {F.}~\bibnamefont {Egidi}},
  \bibinfo {author} {\bibfnamefont {J.}~\bibnamefont {Goings}}, \bibinfo
  {author} {\bibfnamefont {B.}~\bibnamefont {Peng}}, \bibinfo {author}
  {\bibfnamefont {A.}~\bibnamefont {Petrone}}, \bibinfo {author} {\bibfnamefont
  {T.}~\bibnamefont {Henderson}}, \bibinfo {author} {\bibfnamefont
  {D.}~\bibnamefont {Ranasinghe}}, \bibinfo {author} {\bibfnamefont {V.~G.}\
  \bibnamefont {Zakrzewski}}, \bibinfo {author} {\bibfnamefont
  {J.}~\bibnamefont {Gao}}, \bibinfo {author} {\bibfnamefont {N.}~\bibnamefont
  {Rega}}, \bibinfo {author} {\bibfnamefont {G.}~\bibnamefont {Zheng}},
  \bibinfo {author} {\bibfnamefont {W.}~\bibnamefont {Liang}}, \bibinfo
  {author} {\bibfnamefont {M.}~\bibnamefont {Hada}}, \bibinfo {author}
  {\bibfnamefont {M.}~\bibnamefont {Ehara}}, \bibinfo {author} {\bibfnamefont
  {K.}~\bibnamefont {Toyota}}, \bibinfo {author} {\bibfnamefont
  {R.}~\bibnamefont {Fukuda}}, \bibinfo {author} {\bibfnamefont
  {J.}~\bibnamefont {Hasegawa}}, \bibinfo {author} {\bibfnamefont
  {M.}~\bibnamefont {Ishida}}, \bibinfo {author} {\bibfnamefont
  {T.}~\bibnamefont {Nakajima}}, \bibinfo {author} {\bibfnamefont
  {Y.}~\bibnamefont {Honda}}, \bibinfo {author} {\bibfnamefont
  {O.}~\bibnamefont {Kitao}}, \bibinfo {author} {\bibfnamefont
  {H.}~\bibnamefont {Nakai}}, \bibinfo {author} {\bibfnamefont
  {T.}~\bibnamefont {Vreven}}, \bibinfo {author} {\bibfnamefont
  {K.}~\bibnamefont {Throssell}}, \bibinfo {author} {\bibfnamefont {J.~A.}\
  \bibnamefont {Montgomery~Jr.}}, \bibinfo {author} {\bibfnamefont {J.~E.}\
  \bibnamefont {Peralta}}, \bibinfo {author} {\bibfnamefont {F.}~\bibnamefont
  {Ogliaro}}, \bibinfo {author} {\bibfnamefont {M.~J.}\ \bibnamefont
  {Bearpark}}, \bibinfo {author} {\bibfnamefont {J.~J.}\ \bibnamefont {Heyd}},
  \bibinfo {author} {\bibfnamefont {E.~N.}\ \bibnamefont {Brothers}}, \bibinfo
  {author} {\bibfnamefont {K.~N.}\ \bibnamefont {Kudin}}, \bibinfo {author}
  {\bibfnamefont {V.~N.}\ \bibnamefont {Staroverov}}, \bibinfo {author}
  {\bibfnamefont {T.~A.}\ \bibnamefont {Keith}}, \bibinfo {author}
  {\bibfnamefont {R.}~\bibnamefont {Kobayashi}}, \bibinfo {author}
  {\bibfnamefont {J.}~\bibnamefont {Normand}}, \bibinfo {author} {\bibfnamefont
  {K.}~\bibnamefont {Raghavachari}}, \bibinfo {author} {\bibfnamefont {A.~P.}\
  \bibnamefont {Rendell}}, \bibinfo {author} {\bibfnamefont {J.~C.}\
  \bibnamefont {Burant}}, \bibinfo {author} {\bibfnamefont {S.~S.}\
  \bibnamefont {Iyengar}}, \bibinfo {author} {\bibfnamefont {J.}~\bibnamefont
  {Tomasi}}, \bibinfo {author} {\bibfnamefont {M.}~\bibnamefont {Cossi}},
  \bibinfo {author} {\bibfnamefont {J.~M.}\ \bibnamefont {Millam}}, \bibinfo
  {author} {\bibfnamefont {M.}~\bibnamefont {Klene}}, \bibinfo {author}
  {\bibfnamefont {C.}~\bibnamefont {Adamo}}, \bibinfo {author} {\bibfnamefont
  {R.}~\bibnamefont {Cammi}}, \bibinfo {author} {\bibfnamefont {J.~W.}\
  \bibnamefont {Ochterski}}, \bibinfo {author} {\bibfnamefont {R.~L.}\
  \bibnamefont {Martin}}, \bibinfo {author} {\bibfnamefont {K.}~\bibnamefont
  {Morokuma}}, \bibinfo {author} {\bibfnamefont {O.}~\bibnamefont {Farkas}},
  \bibinfo {author} {\bibfnamefont {J.~B.}\ \bibnamefont {Foresman}},\ and\
  \bibinfo {author} {\bibfnamefont {D.~J.}\ \bibnamefont {Fox}},\ }\href@noop
  {} {\bibinfo {title} {Gaussian 16 {{Rev}}. {{C}}.01}} (\bibinfo {year}
  {2016})\BibitemShut {NoStop}%
\bibitem [{\citenamefont {Smith}\ \emph {et~al.}(2020)\citenamefont {Smith},
  \citenamefont {Burns}, \citenamefont {Simmonett}, \citenamefont {Parrish},
  \citenamefont {Schieber}, \citenamefont {Galvelis}, \citenamefont {Kraus},
  \citenamefont {Kruse}, \citenamefont {Di~Remigio}, \citenamefont {Alenaizan},
  \citenamefont {James}, \citenamefont {Lehtola}, \citenamefont {Misiewicz},
  \citenamefont {Scheurer}, \citenamefont {Shaw}, \citenamefont {Schriber},
  \citenamefont {Xie}, \citenamefont {Glick}, \citenamefont {Sirianni},
  \citenamefont {O'Brien}, \citenamefont {Waldrop}, \citenamefont {Kumar},
  \citenamefont {Hohenstein}, \citenamefont {Pritchard}, \citenamefont
  {Brooks}, \citenamefont {Schaefer}, \citenamefont {Sokolov}, \citenamefont
  {Patkowski}, \citenamefont {DePrince}, \citenamefont {Bozkaya}, \citenamefont
  {King}, \citenamefont {Evangelista}, \citenamefont {Turney}, \citenamefont
  {Crawford},\ and\ \citenamefont
  {Sherrill}}]{smithPSI4OpensourceSoftware2020}%
  \BibitemOpen
  \bibfield  {author} {\bibinfo {author} {\bibfnamefont {D.~G.~A.}\
  \bibnamefont {Smith}}, \bibinfo {author} {\bibfnamefont {L.~A.}\ \bibnamefont
  {Burns}}, \bibinfo {author} {\bibfnamefont {A.~C.}\ \bibnamefont
  {Simmonett}}, \bibinfo {author} {\bibfnamefont {R.~M.}\ \bibnamefont
  {Parrish}}, \bibinfo {author} {\bibfnamefont {M.~C.}\ \bibnamefont
  {Schieber}}, \bibinfo {author} {\bibfnamefont {R.}~\bibnamefont {Galvelis}},
  \bibinfo {author} {\bibfnamefont {P.}~\bibnamefont {Kraus}}, \bibinfo
  {author} {\bibfnamefont {H.}~\bibnamefont {Kruse}}, \bibinfo {author}
  {\bibfnamefont {R.}~\bibnamefont {Di~Remigio}}, \bibinfo {author}
  {\bibfnamefont {A.}~\bibnamefont {Alenaizan}}, \bibinfo {author}
  {\bibfnamefont {A.~M.}\ \bibnamefont {James}}, \bibinfo {author}
  {\bibfnamefont {S.}~\bibnamefont {Lehtola}}, \bibinfo {author} {\bibfnamefont
  {J.~P.}\ \bibnamefont {Misiewicz}}, \bibinfo {author} {\bibfnamefont
  {M.}~\bibnamefont {Scheurer}}, \bibinfo {author} {\bibfnamefont {R.~A.}\
  \bibnamefont {Shaw}}, \bibinfo {author} {\bibfnamefont {J.~B.}\ \bibnamefont
  {Schriber}}, \bibinfo {author} {\bibfnamefont {Y.}~\bibnamefont {Xie}},
  \bibinfo {author} {\bibfnamefont {Z.~L.}\ \bibnamefont {Glick}}, \bibinfo
  {author} {\bibfnamefont {D.~A.}\ \bibnamefont {Sirianni}}, \bibinfo {author}
  {\bibfnamefont {J.~S.}\ \bibnamefont {O'Brien}}, \bibinfo {author}
  {\bibfnamefont {J.~M.}\ \bibnamefont {Waldrop}}, \bibinfo {author}
  {\bibfnamefont {A.}~\bibnamefont {Kumar}}, \bibinfo {author} {\bibfnamefont
  {E.~G.}\ \bibnamefont {Hohenstein}}, \bibinfo {author} {\bibfnamefont
  {B.~P.}\ \bibnamefont {Pritchard}}, \bibinfo {author} {\bibfnamefont {B.~R.}\
  \bibnamefont {Brooks}}, \bibinfo {author} {\bibfnamefont {H.~F.}\
  \bibnamefont {Schaefer}}, \bibinfo {author} {\bibfnamefont {A.~Y.}\
  \bibnamefont {Sokolov}}, \bibinfo {author} {\bibfnamefont {K.}~\bibnamefont
  {Patkowski}}, \bibinfo {author} {\bibfnamefont {A.~E.}\ \bibnamefont
  {DePrince}}, \bibinfo {author} {\bibfnamefont {U.}~\bibnamefont {Bozkaya}},
  \bibinfo {author} {\bibfnamefont {R.~A.}\ \bibnamefont {King}}, \bibinfo
  {author} {\bibfnamefont {F.~A.}\ \bibnamefont {Evangelista}}, \bibinfo
  {author} {\bibfnamefont {J.~M.}\ \bibnamefont {Turney}}, \bibinfo {author}
  {\bibfnamefont {T.~D.}\ \bibnamefont {Crawford}},\ and\ \bibinfo {author}
  {\bibfnamefont {C.~D.}\ \bibnamefont {Sherrill}},\ }\bibfield  {title}
  {\bibinfo {title} {{{PSI4}} 1.4: {{Open-source}} software for high-throughput
  quantum chemistry},\ }\href {https://doi.org/10.1063/5.0006002} {\bibfield
  {journal} {\bibinfo  {journal} {J. Chem. Phys.}\ }\textbf {\bibinfo {volume}
  {152}},\ \bibinfo {pages} {184108} (\bibinfo {year} {2020})}\BibitemShut
  {NoStop}%
\bibitem [{\citenamefont {Sun}\ \emph {et~al.}(2018)\citenamefont {Sun},
  \citenamefont {Berkelbach}, \citenamefont {Blunt}, \citenamefont {Booth},
  \citenamefont {Guo}, \citenamefont {Li}, \citenamefont {Liu}, \citenamefont
  {McClain}, \citenamefont {Sayfutyarova}, \citenamefont {Sharma},
  \citenamefont {Wouters},\ and\ \citenamefont
  {Chan}}]{sunPySCFPythonbasedSimulations2018}%
  \BibitemOpen
  \bibfield  {author} {\bibinfo {author} {\bibfnamefont {Q.}~\bibnamefont
  {Sun}}, \bibinfo {author} {\bibfnamefont {T.~C.}\ \bibnamefont {Berkelbach}},
  \bibinfo {author} {\bibfnamefont {N.~S.}\ \bibnamefont {Blunt}}, \bibinfo
  {author} {\bibfnamefont {G.~H.}\ \bibnamefont {Booth}}, \bibinfo {author}
  {\bibfnamefont {S.}~\bibnamefont {Guo}}, \bibinfo {author} {\bibfnamefont
  {Z.}~\bibnamefont {Li}}, \bibinfo {author} {\bibfnamefont {J.}~\bibnamefont
  {Liu}}, \bibinfo {author} {\bibfnamefont {J.~D.}\ \bibnamefont {McClain}},
  \bibinfo {author} {\bibfnamefont {E.~R.}\ \bibnamefont {Sayfutyarova}},
  \bibinfo {author} {\bibfnamefont {S.}~\bibnamefont {Sharma}}, \bibinfo
  {author} {\bibfnamefont {S.}~\bibnamefont {Wouters}},\ and\ \bibinfo {author}
  {\bibfnamefont {G.~K.-L.}\ \bibnamefont {Chan}},\ }\bibfield  {title}
  {\bibinfo {title} {{{PySCF}}: The {{Python-based}} simulations of chemistry
  framework},\ }\href {https://doi.org/10.1002/wcms.1340} {\bibfield  {journal}
  {\bibinfo  {journal} {WIREs Computational Molecular Science}\ }\textbf
  {\bibinfo {volume} {8}},\ \bibinfo {pages} {e1340} (\bibinfo {year}
  {2018})}\BibitemShut {NoStop}%
\bibitem [{\citenamefont {Balasubramani}\ \emph {et~al.}(2020)\citenamefont
  {Balasubramani}, \citenamefont {Chen}, \citenamefont {Coriani}, \citenamefont
  {Diedenhofen}, \citenamefont {Frank}, \citenamefont {Franzke}, \citenamefont
  {Furche}, \citenamefont {Grotjahn}, \citenamefont {Harding}, \citenamefont
  {H{\"a}ttig}, \citenamefont {Hellweg}, \citenamefont {{Helmich-Paris}},
  \citenamefont {Holzer}, \citenamefont {Huniar}, \citenamefont {Kaupp},
  \citenamefont {Marefat~Khah}, \citenamefont {Karbalaei~Khani}, \citenamefont
  {M{\"u}ller}, \citenamefont {Mack}, \citenamefont {Nguyen}, \citenamefont
  {Parker}, \citenamefont {Perlt}, \citenamefont {Rappoport}, \citenamefont
  {Reiter}, \citenamefont {Roy}, \citenamefont {R{\"u}ckert}, \citenamefont
  {Schmitz}, \citenamefont {Sierka}, \citenamefont {Tapavicza}, \citenamefont
  {Tew}, \citenamefont {{van W{\"u}llen}}, \citenamefont {Voora}, \citenamefont
  {Weigend}, \citenamefont {Wody{\'n}ski},\ and\ \citenamefont
  {Yu}}]{balasubramaniTURBOMOLEModularProgram2020}%
  \BibitemOpen
  \bibfield  {author} {\bibinfo {author} {\bibfnamefont {S.~G.}\ \bibnamefont
  {Balasubramani}}, \bibinfo {author} {\bibfnamefont {G.~P.}\ \bibnamefont
  {Chen}}, \bibinfo {author} {\bibfnamefont {S.}~\bibnamefont {Coriani}},
  \bibinfo {author} {\bibfnamefont {M.}~\bibnamefont {Diedenhofen}}, \bibinfo
  {author} {\bibfnamefont {M.~S.}\ \bibnamefont {Frank}}, \bibinfo {author}
  {\bibfnamefont {Y.~J.}\ \bibnamefont {Franzke}}, \bibinfo {author}
  {\bibfnamefont {F.}~\bibnamefont {Furche}}, \bibinfo {author} {\bibfnamefont
  {R.}~\bibnamefont {Grotjahn}}, \bibinfo {author} {\bibfnamefont {M.~E.}\
  \bibnamefont {Harding}}, \bibinfo {author} {\bibfnamefont {C.}~\bibnamefont
  {H{\"a}ttig}}, \bibinfo {author} {\bibfnamefont {A.}~\bibnamefont {Hellweg}},
  \bibinfo {author} {\bibfnamefont {B.}~\bibnamefont {{Helmich-Paris}}},
  \bibinfo {author} {\bibfnamefont {C.}~\bibnamefont {Holzer}}, \bibinfo
  {author} {\bibfnamefont {U.}~\bibnamefont {Huniar}}, \bibinfo {author}
  {\bibfnamefont {M.}~\bibnamefont {Kaupp}}, \bibinfo {author} {\bibfnamefont
  {A.}~\bibnamefont {Marefat~Khah}}, \bibinfo {author} {\bibfnamefont
  {S.}~\bibnamefont {Karbalaei~Khani}}, \bibinfo {author} {\bibfnamefont
  {T.}~\bibnamefont {M{\"u}ller}}, \bibinfo {author} {\bibfnamefont
  {F.}~\bibnamefont {Mack}}, \bibinfo {author} {\bibfnamefont {B.~D.}\
  \bibnamefont {Nguyen}}, \bibinfo {author} {\bibfnamefont {S.~M.}\
  \bibnamefont {Parker}}, \bibinfo {author} {\bibfnamefont {E.}~\bibnamefont
  {Perlt}}, \bibinfo {author} {\bibfnamefont {D.}~\bibnamefont {Rappoport}},
  \bibinfo {author} {\bibfnamefont {K.}~\bibnamefont {Reiter}}, \bibinfo
  {author} {\bibfnamefont {S.}~\bibnamefont {Roy}}, \bibinfo {author}
  {\bibfnamefont {M.}~\bibnamefont {R{\"u}ckert}}, \bibinfo {author}
  {\bibfnamefont {G.}~\bibnamefont {Schmitz}}, \bibinfo {author} {\bibfnamefont
  {M.}~\bibnamefont {Sierka}}, \bibinfo {author} {\bibfnamefont
  {E.}~\bibnamefont {Tapavicza}}, \bibinfo {author} {\bibfnamefont {D.~P.}\
  \bibnamefont {Tew}}, \bibinfo {author} {\bibfnamefont {C.}~\bibnamefont {{van
  W{\"u}llen}}}, \bibinfo {author} {\bibfnamefont {V.~K.}\ \bibnamefont
  {Voora}}, \bibinfo {author} {\bibfnamefont {F.}~\bibnamefont {Weigend}},
  \bibinfo {author} {\bibfnamefont {A.}~\bibnamefont {Wody{\'n}ski}},\ and\
  \bibinfo {author} {\bibfnamefont {J.~M.}\ \bibnamefont {Yu}},\ }\bibfield
  {title} {\bibinfo {title} {{{TURBOMOLE}}: {{Modular}} program suite for ab
  initio quantum-chemical and condensed-matter simulations},\ }\href
  {https://doi.org/10.1063/5.0004635} {\bibfield  {journal} {\bibinfo
  {journal} {J. Chem. Phys.}\ }\textbf {\bibinfo {volume} {152}},\ \bibinfo
  {pages} {184107} (\bibinfo {year} {2020})}\BibitemShut {NoStop}%
\bibitem [{\citenamefont {K{\"u}hne}\ \emph {et~al.}(2020)\citenamefont
  {K{\"u}hne}, \citenamefont {Iannuzzi}, \citenamefont {Del~Ben}, \citenamefont
  {Rybkin}, \citenamefont {Seewald}, \citenamefont {Stein}, \citenamefont
  {Laino}, \citenamefont {Khaliullin}, \citenamefont {Sch{\"u}tt},
  \citenamefont {Schiffmann}, \citenamefont {Golze}, \citenamefont {Wilhelm},
  \citenamefont {Chulkov}, \citenamefont {{Bani-Hashemian}}, \citenamefont
  {Weber}, \citenamefont {Bor{\v s}tnik}, \citenamefont {Taillefumier},
  \citenamefont {Jakobovits}, \citenamefont {Lazzaro}, \citenamefont {Pabst},
  \citenamefont {M{\"u}ller}, \citenamefont {Schade}, \citenamefont {Guidon},
  \citenamefont {Andermatt}, \citenamefont {Holmberg}, \citenamefont
  {Schenter}, \citenamefont {Hehn}, \citenamefont {Bussy}, \citenamefont
  {Belleflamme}, \citenamefont {Tabacchi}, \citenamefont {Gl{\"o}{\ss}},
  \citenamefont {Lass}, \citenamefont {Bethune}, \citenamefont {Mundy},
  \citenamefont {Plessl}, \citenamefont {Watkins}, \citenamefont
  {Vandevondele}, \citenamefont {Krack},\ and\ \citenamefont
  {Hutter}}]{kuhneCP2KElectronicStructure2020}%
  \BibitemOpen
  \bibfield  {author} {\bibinfo {author} {\bibfnamefont {T.~D.}\ \bibnamefont
  {K{\"u}hne}}, \bibinfo {author} {\bibfnamefont {M.}~\bibnamefont {Iannuzzi}},
  \bibinfo {author} {\bibfnamefont {M.}~\bibnamefont {Del~Ben}}, \bibinfo
  {author} {\bibfnamefont {V.~V.}\ \bibnamefont {Rybkin}}, \bibinfo {author}
  {\bibfnamefont {P.}~\bibnamefont {Seewald}}, \bibinfo {author} {\bibfnamefont
  {F.}~\bibnamefont {Stein}}, \bibinfo {author} {\bibfnamefont
  {T.}~\bibnamefont {Laino}}, \bibinfo {author} {\bibfnamefont {R.~Z.}\
  \bibnamefont {Khaliullin}}, \bibinfo {author} {\bibfnamefont
  {O.}~\bibnamefont {Sch{\"u}tt}}, \bibinfo {author} {\bibfnamefont
  {F.}~\bibnamefont {Schiffmann}}, \bibinfo {author} {\bibfnamefont
  {D.}~\bibnamefont {Golze}}, \bibinfo {author} {\bibfnamefont
  {J.}~\bibnamefont {Wilhelm}}, \bibinfo {author} {\bibfnamefont
  {S.}~\bibnamefont {Chulkov}}, \bibinfo {author} {\bibfnamefont {M.~H.}\
  \bibnamefont {{Bani-Hashemian}}}, \bibinfo {author} {\bibfnamefont
  {V.}~\bibnamefont {Weber}}, \bibinfo {author} {\bibfnamefont
  {U.}~\bibnamefont {Bor{\v s}tnik}}, \bibinfo {author} {\bibfnamefont
  {M.}~\bibnamefont {Taillefumier}}, \bibinfo {author} {\bibfnamefont {A.~S.}\
  \bibnamefont {Jakobovits}}, \bibinfo {author} {\bibfnamefont
  {A.}~\bibnamefont {Lazzaro}}, \bibinfo {author} {\bibfnamefont
  {H.}~\bibnamefont {Pabst}}, \bibinfo {author} {\bibfnamefont
  {T.}~\bibnamefont {M{\"u}ller}}, \bibinfo {author} {\bibfnamefont
  {R.}~\bibnamefont {Schade}}, \bibinfo {author} {\bibfnamefont
  {M.}~\bibnamefont {Guidon}}, \bibinfo {author} {\bibfnamefont
  {S.}~\bibnamefont {Andermatt}}, \bibinfo {author} {\bibfnamefont
  {N.}~\bibnamefont {Holmberg}}, \bibinfo {author} {\bibfnamefont {G.~K.}\
  \bibnamefont {Schenter}}, \bibinfo {author} {\bibfnamefont {A.}~\bibnamefont
  {Hehn}}, \bibinfo {author} {\bibfnamefont {A.}~\bibnamefont {Bussy}},
  \bibinfo {author} {\bibfnamefont {F.}~\bibnamefont {Belleflamme}}, \bibinfo
  {author} {\bibfnamefont {G.}~\bibnamefont {Tabacchi}}, \bibinfo {author}
  {\bibfnamefont {A.}~\bibnamefont {Gl{\"o}{\ss}}}, \bibinfo {author}
  {\bibfnamefont {M.}~\bibnamefont {Lass}}, \bibinfo {author} {\bibfnamefont
  {I.}~\bibnamefont {Bethune}}, \bibinfo {author} {\bibfnamefont {C.~J.}\
  \bibnamefont {Mundy}}, \bibinfo {author} {\bibfnamefont {C.}~\bibnamefont
  {Plessl}}, \bibinfo {author} {\bibfnamefont {M.}~\bibnamefont {Watkins}},
  \bibinfo {author} {\bibfnamefont {J.}~\bibnamefont {Vandevondele}}, \bibinfo
  {author} {\bibfnamefont {M.}~\bibnamefont {Krack}},\ and\ \bibinfo {author}
  {\bibfnamefont {J.}~\bibnamefont {Hutter}},\ }\bibfield  {title} {\bibinfo
  {title} {{{CP2K}}: {{An}} electronic structure and molecular dynamics
  software package - {{Quickstep}}: {{Efficient}} and accurate electronic
  structure calculations},\ }\href {https://doi.org/10.1063/5.0007045}
  {\bibfield  {journal} {\bibinfo  {journal} {The Journal of Chemical Physics}\
  }\textbf {\bibinfo {volume} {152}},\ \bibinfo {pages} {194103} (\bibinfo
  {year} {2020})}\BibitemShut {NoStop}%
\bibitem [{\citenamefont {Giannozzi}\ \emph {et~al.}(2009)\citenamefont
  {Giannozzi}, \citenamefont {Baroni}, \citenamefont {Bonini}, \citenamefont
  {Calandra}, \citenamefont {Car}, \citenamefont {Cavazzoni}, \citenamefont
  {Ceresoli}, \citenamefont {Chiarotti}, \citenamefont {Cococcioni},
  \citenamefont {Dabo}, \citenamefont {Dal~Corso}, \citenamefont {{de
  Gironcoli}}, \citenamefont {Fabris}, \citenamefont {Fratesi}, \citenamefont
  {Gebauer}, \citenamefont {Gerstmann}, \citenamefont {Gougoussis},
  \citenamefont {Kokalj}, \citenamefont {Lazzeri}, \citenamefont
  {{Martin-Samos}}, \citenamefont {Marzari}, \citenamefont {Mauri},
  \citenamefont {Mazzarello}, \citenamefont {Paolini}, \citenamefont
  {Pasquarello}, \citenamefont {Paulatto}, \citenamefont {Sbraccia},
  \citenamefont {Scandolo}, \citenamefont {Sclauzero}, \citenamefont
  {Seitsonen}, \citenamefont {Smogunov}, \citenamefont {Umari},\ and\
  \citenamefont {Wentzcovitch}}]{giannozziQUANTUMESPRESSOModular2009}%
  \BibitemOpen
  \bibfield  {author} {\bibinfo {author} {\bibfnamefont {P.}~\bibnamefont
  {Giannozzi}}, \bibinfo {author} {\bibfnamefont {S.}~\bibnamefont {Baroni}},
  \bibinfo {author} {\bibfnamefont {N.}~\bibnamefont {Bonini}}, \bibinfo
  {author} {\bibfnamefont {M.}~\bibnamefont {Calandra}}, \bibinfo {author}
  {\bibfnamefont {R.}~\bibnamefont {Car}}, \bibinfo {author} {\bibfnamefont
  {C.}~\bibnamefont {Cavazzoni}}, \bibinfo {author} {\bibfnamefont
  {D.}~\bibnamefont {Ceresoli}}, \bibinfo {author} {\bibfnamefont {G.~L.}\
  \bibnamefont {Chiarotti}}, \bibinfo {author} {\bibfnamefont {M.}~\bibnamefont
  {Cococcioni}}, \bibinfo {author} {\bibfnamefont {I.}~\bibnamefont {Dabo}},
  \bibinfo {author} {\bibfnamefont {A.}~\bibnamefont {Dal~Corso}}, \bibinfo
  {author} {\bibfnamefont {S.}~\bibnamefont {{de Gironcoli}}}, \bibinfo
  {author} {\bibfnamefont {S.}~\bibnamefont {Fabris}}, \bibinfo {author}
  {\bibfnamefont {G.}~\bibnamefont {Fratesi}}, \bibinfo {author} {\bibfnamefont
  {R.}~\bibnamefont {Gebauer}}, \bibinfo {author} {\bibfnamefont
  {U.}~\bibnamefont {Gerstmann}}, \bibinfo {author} {\bibfnamefont
  {C.}~\bibnamefont {Gougoussis}}, \bibinfo {author} {\bibfnamefont
  {A.}~\bibnamefont {Kokalj}}, \bibinfo {author} {\bibfnamefont
  {M.}~\bibnamefont {Lazzeri}}, \bibinfo {author} {\bibfnamefont
  {L.}~\bibnamefont {{Martin-Samos}}}, \bibinfo {author} {\bibfnamefont
  {N.}~\bibnamefont {Marzari}}, \bibinfo {author} {\bibfnamefont
  {F.}~\bibnamefont {Mauri}}, \bibinfo {author} {\bibfnamefont
  {R.}~\bibnamefont {Mazzarello}}, \bibinfo {author} {\bibfnamefont
  {S.}~\bibnamefont {Paolini}}, \bibinfo {author} {\bibfnamefont
  {A.}~\bibnamefont {Pasquarello}}, \bibinfo {author} {\bibfnamefont
  {L.}~\bibnamefont {Paulatto}}, \bibinfo {author} {\bibfnamefont
  {C.}~\bibnamefont {Sbraccia}}, \bibinfo {author} {\bibfnamefont
  {S.}~\bibnamefont {Scandolo}}, \bibinfo {author} {\bibfnamefont
  {G.}~\bibnamefont {Sclauzero}}, \bibinfo {author} {\bibfnamefont {A.~P.}\
  \bibnamefont {Seitsonen}}, \bibinfo {author} {\bibfnamefont {A.}~\bibnamefont
  {Smogunov}}, \bibinfo {author} {\bibfnamefont {P.}~\bibnamefont {Umari}},\
  and\ \bibinfo {author} {\bibfnamefont {R.~M.}\ \bibnamefont {Wentzcovitch}},\
  }\bibfield  {title} {\bibinfo {title} {{{QUANTUM ESPRESSO}}: A modular and
  open-source software project for quantum simulations of materials},\ }\href
  {https://doi.org/10.1088/0953-8984/21/39/395502} {\bibfield  {journal}
  {\bibinfo  {journal} {Journal of Physics: Condensed Matter}\ }\textbf
  {\bibinfo {volume} {21}},\ \bibinfo {pages} {395502} (\bibinfo {year}
  {2009})}\BibitemShut {NoStop}%
\bibitem [{\citenamefont {Giannozzi}\ \emph {et~al.}(2017)\citenamefont
  {Giannozzi}, \citenamefont {Andreussi}, \citenamefont {Brumme}, \citenamefont
  {Bunau}, \citenamefont {Buongiorno~Nardelli}, \citenamefont {Calandra},
  \citenamefont {Car}, \citenamefont {Cavazzoni}, \citenamefont {Ceresoli},
  \citenamefont {Cococcioni}, \citenamefont {Colonna}, \citenamefont
  {Carnimeo}, \citenamefont {Dal~Corso}, \citenamefont {{de Gironcoli}},
  \citenamefont {Delugas}, \citenamefont {DiStasio}, \citenamefont {Ferretti},
  \citenamefont {Floris}, \citenamefont {Fratesi}, \citenamefont {Fugallo},
  \citenamefont {Gebauer}, \citenamefont {Gerstmann}, \citenamefont {Giustino},
  \citenamefont {Gorni}, \citenamefont {Jia}, \citenamefont {Kawamura},
  \citenamefont {Ko}, \citenamefont {Kokalj}, \citenamefont {K{\"u}{\c
  c}{\"u}kbenli}, \citenamefont {Lazzeri}, \citenamefont {Marsili},
  \citenamefont {Marzari}, \citenamefont {Mauri}, \citenamefont {Nguyen},
  \citenamefont {Nguyen}, \citenamefont {{Otero-de-la-Roza}}, \citenamefont
  {Paulatto}, \citenamefont {Ponc{\'e}}, \citenamefont {Rocca}, \citenamefont
  {Sabatini}, \citenamefont {Santra}, \citenamefont {Schlipf}, \citenamefont
  {Seitsonen}, \citenamefont {Smogunov}, \citenamefont {Timrov}, \citenamefont
  {Thonhauser}, \citenamefont {Umari}, \citenamefont {Vast}, \citenamefont
  {Wu},\ and\ \citenamefont
  {Baroni}}]{giannozziAdvancedCapabilitiesMaterials2017}%
  \BibitemOpen
  \bibfield  {author} {\bibinfo {author} {\bibfnamefont {P.}~\bibnamefont
  {Giannozzi}}, \bibinfo {author} {\bibfnamefont {O.}~\bibnamefont
  {Andreussi}}, \bibinfo {author} {\bibfnamefont {T.}~\bibnamefont {Brumme}},
  \bibinfo {author} {\bibfnamefont {O.}~\bibnamefont {Bunau}}, \bibinfo
  {author} {\bibfnamefont {M.}~\bibnamefont {Buongiorno~Nardelli}}, \bibinfo
  {author} {\bibfnamefont {M.}~\bibnamefont {Calandra}}, \bibinfo {author}
  {\bibfnamefont {R.}~\bibnamefont {Car}}, \bibinfo {author} {\bibfnamefont
  {C.}~\bibnamefont {Cavazzoni}}, \bibinfo {author} {\bibfnamefont
  {D.}~\bibnamefont {Ceresoli}}, \bibinfo {author} {\bibfnamefont
  {M.}~\bibnamefont {Cococcioni}}, \bibinfo {author} {\bibfnamefont
  {N.}~\bibnamefont {Colonna}}, \bibinfo {author} {\bibfnamefont
  {I.}~\bibnamefont {Carnimeo}}, \bibinfo {author} {\bibfnamefont
  {A.}~\bibnamefont {Dal~Corso}}, \bibinfo {author} {\bibfnamefont
  {S.}~\bibnamefont {{de Gironcoli}}}, \bibinfo {author} {\bibfnamefont
  {P.}~\bibnamefont {Delugas}}, \bibinfo {author} {\bibfnamefont {R.~A.}\
  \bibnamefont {DiStasio}}, \bibinfo {author} {\bibfnamefont {A.}~\bibnamefont
  {Ferretti}}, \bibinfo {author} {\bibfnamefont {A.}~\bibnamefont {Floris}},
  \bibinfo {author} {\bibfnamefont {G.}~\bibnamefont {Fratesi}}, \bibinfo
  {author} {\bibfnamefont {G.}~\bibnamefont {Fugallo}}, \bibinfo {author}
  {\bibfnamefont {R.}~\bibnamefont {Gebauer}}, \bibinfo {author} {\bibfnamefont
  {U.}~\bibnamefont {Gerstmann}}, \bibinfo {author} {\bibfnamefont
  {F.}~\bibnamefont {Giustino}}, \bibinfo {author} {\bibfnamefont
  {T.}~\bibnamefont {Gorni}}, \bibinfo {author} {\bibfnamefont
  {J.}~\bibnamefont {Jia}}, \bibinfo {author} {\bibfnamefont {M.}~\bibnamefont
  {Kawamura}}, \bibinfo {author} {\bibfnamefont {H.-Y.}\ \bibnamefont {Ko}},
  \bibinfo {author} {\bibfnamefont {A.}~\bibnamefont {Kokalj}}, \bibinfo
  {author} {\bibfnamefont {E.}~\bibnamefont {K{\"u}{\c c}{\"u}kbenli}},
  \bibinfo {author} {\bibfnamefont {M.}~\bibnamefont {Lazzeri}}, \bibinfo
  {author} {\bibfnamefont {M.}~\bibnamefont {Marsili}}, \bibinfo {author}
  {\bibfnamefont {N.}~\bibnamefont {Marzari}}, \bibinfo {author} {\bibfnamefont
  {F.}~\bibnamefont {Mauri}}, \bibinfo {author} {\bibfnamefont {N.~L.}\
  \bibnamefont {Nguyen}}, \bibinfo {author} {\bibfnamefont {H.-V.}\
  \bibnamefont {Nguyen}}, \bibinfo {author} {\bibfnamefont {A.}~\bibnamefont
  {{Otero-de-la-Roza}}}, \bibinfo {author} {\bibfnamefont {L.}~\bibnamefont
  {Paulatto}}, \bibinfo {author} {\bibfnamefont {S.}~\bibnamefont {Ponc{\'e}}},
  \bibinfo {author} {\bibfnamefont {D.}~\bibnamefont {Rocca}}, \bibinfo
  {author} {\bibfnamefont {R.}~\bibnamefont {Sabatini}}, \bibinfo {author}
  {\bibfnamefont {B.}~\bibnamefont {Santra}}, \bibinfo {author} {\bibfnamefont
  {M.}~\bibnamefont {Schlipf}}, \bibinfo {author} {\bibfnamefont {A.~P.}\
  \bibnamefont {Seitsonen}}, \bibinfo {author} {\bibfnamefont {A.}~\bibnamefont
  {Smogunov}}, \bibinfo {author} {\bibfnamefont {I.}~\bibnamefont {Timrov}},
  \bibinfo {author} {\bibfnamefont {T.}~\bibnamefont {Thonhauser}}, \bibinfo
  {author} {\bibfnamefont {P.}~\bibnamefont {Umari}}, \bibinfo {author}
  {\bibfnamefont {N.}~\bibnamefont {Vast}}, \bibinfo {author} {\bibfnamefont
  {X.}~\bibnamefont {Wu}},\ and\ \bibinfo {author} {\bibfnamefont
  {S.}~\bibnamefont {Baroni}},\ }\bibfield  {title} {\bibinfo {title} {Advanced
  capabilities for materials modelling with {{Quantum ESPRESSO}}},\ }\href
  {https://doi.org/10.1088/1361-648X/aa8f79} {\bibfield  {journal} {\bibinfo
  {journal} {Journal of Physics: Condensed Matter}\ }\textbf {\bibinfo {volume}
  {29}},\ \bibinfo {pages} {465901} (\bibinfo {year} {2017})}\BibitemShut
  {NoStop}%
\bibitem [{\citenamefont {Kapil}\ \emph {et~al.}(2019)\citenamefont {Kapil},
  \citenamefont {Rossi}, \citenamefont {Marsalek}, \citenamefont {Petraglia},
  \citenamefont {Litman}, \citenamefont {Spura}, \citenamefont {Cheng},
  \citenamefont {Cuzzocrea}, \citenamefont {Mei{\ss}ner}, \citenamefont
  {Wilkins}, \citenamefont {Helfrecht}, \citenamefont {Juda}, \citenamefont
  {Bienvenue}, \citenamefont {Fang}, \citenamefont {Kessler}, \citenamefont
  {Poltavsky}, \citenamefont {Vandenbrande}, \citenamefont {Wieme},
  \citenamefont {Corminboeuf}, \citenamefont {K{\"u}hne}, \citenamefont
  {Manolopoulos}, \citenamefont {Markland}, \citenamefont {Richardson},
  \citenamefont {Tkatchenko}, \citenamefont {Tribello}, \citenamefont
  {Van~Speybroeck},\ and\ \citenamefont
  {Ceriotti}}]{kapilIPIUniversalForce2019}%
  \BibitemOpen
  \bibfield  {author} {\bibinfo {author} {\bibfnamefont {V.}~\bibnamefont
  {Kapil}}, \bibinfo {author} {\bibfnamefont {M.}~\bibnamefont {Rossi}},
  \bibinfo {author} {\bibfnamefont {O.}~\bibnamefont {Marsalek}}, \bibinfo
  {author} {\bibfnamefont {R.}~\bibnamefont {Petraglia}}, \bibinfo {author}
  {\bibfnamefont {Y.}~\bibnamefont {Litman}}, \bibinfo {author} {\bibfnamefont
  {T.}~\bibnamefont {Spura}}, \bibinfo {author} {\bibfnamefont
  {B.}~\bibnamefont {Cheng}}, \bibinfo {author} {\bibfnamefont
  {A.}~\bibnamefont {Cuzzocrea}}, \bibinfo {author} {\bibfnamefont {R.~H.}\
  \bibnamefont {Mei{\ss}ner}}, \bibinfo {author} {\bibfnamefont {D.~M.}\
  \bibnamefont {Wilkins}}, \bibinfo {author} {\bibfnamefont {B.~A.}\
  \bibnamefont {Helfrecht}}, \bibinfo {author} {\bibfnamefont {P.}~\bibnamefont
  {Juda}}, \bibinfo {author} {\bibfnamefont {S.~P.}\ \bibnamefont {Bienvenue}},
  \bibinfo {author} {\bibfnamefont {W.}~\bibnamefont {Fang}}, \bibinfo {author}
  {\bibfnamefont {J.}~\bibnamefont {Kessler}}, \bibinfo {author} {\bibfnamefont
  {I.}~\bibnamefont {Poltavsky}}, \bibinfo {author} {\bibfnamefont
  {S.}~\bibnamefont {Vandenbrande}}, \bibinfo {author} {\bibfnamefont
  {J.}~\bibnamefont {Wieme}}, \bibinfo {author} {\bibfnamefont
  {C.}~\bibnamefont {Corminboeuf}}, \bibinfo {author} {\bibfnamefont {T.~D.}\
  \bibnamefont {K{\"u}hne}}, \bibinfo {author} {\bibfnamefont {D.~E.}\
  \bibnamefont {Manolopoulos}}, \bibinfo {author} {\bibfnamefont {T.~E.}\
  \bibnamefont {Markland}}, \bibinfo {author} {\bibfnamefont {J.~O.}\
  \bibnamefont {Richardson}}, \bibinfo {author} {\bibfnamefont
  {A.}~\bibnamefont {Tkatchenko}}, \bibinfo {author} {\bibfnamefont {G.~A.}\
  \bibnamefont {Tribello}}, \bibinfo {author} {\bibfnamefont {V.}~\bibnamefont
  {Van~Speybroeck}},\ and\ \bibinfo {author} {\bibfnamefont {M.}~\bibnamefont
  {Ceriotti}},\ }\bibfield  {title} {\bibinfo {title} {I-{{PI}} 2.0: {{A}}
  universal force engine for advanced molecular simulations},\ }\href
  {https://doi.org/10.1016/j.cpc.2018.09.020} {\bibfield  {journal} {\bibinfo
  {journal} {Computer Physics Communications}\ }\textbf {\bibinfo {volume}
  {236}},\ \bibinfo {pages} {214} (\bibinfo {year} {2019})}\BibitemShut
  {NoStop}%
\bibitem [{\citenamefont {Kreisbeck}\ and\ \citenamefont
  {Kramer}()}]{kreisbeckExcitonDynamicsLab}%
  \BibitemOpen
  \bibfield  {author} {\bibinfo {author} {\bibfnamefont {C.}~\bibnamefont
  {Kreisbeck}}\ and\ \bibinfo {author} {\bibfnamefont {T.}~\bibnamefont
  {Kramer}},\ }\bibfield  {title} {\bibinfo {title} {Exciton dynamics lab for
  light-harvesting complexes ({{GPU-HEOM}}), 2013},\ }\href@noop {} {\bibinfo
  {journal} {See nanohub. org for electronic tool, doi: https://doi.
  org/10.4231/D3RB6W248}\ }\BibitemShut {NoStop}%
\bibitem [{\citenamefont {Str{\"u}mpfer}\ and\ \citenamefont
  {Schulten}(2012)}]{strumpferOpenQuantumDynamics2012}%
  \BibitemOpen
\bibfield  {journal} {  }\bibfield  {author} {\bibinfo {author} {\bibfnamefont
  {J.}~\bibnamefont {Str{\"u}mpfer}}\ and\ \bibinfo {author} {\bibfnamefont
  {K.}~\bibnamefont {Schulten}},\ }\bibfield  {title} {\bibinfo {title} {Open
  {{Quantum Dynamics Calculations}} with the {{Hierarchy Equations}} of
  {{Motion}} on {{Parallel Computers}}},\ }\href
  {https://doi.org/10.1021/ct3003833} {\bibfield  {journal} {\bibinfo
  {journal} {J. Chem. Theory Comput.}\ }\textbf {\bibinfo {volume} {8}},\
  \bibinfo {pages} {2808} (\bibinfo {year} {2012})}\BibitemShut {NoStop}%
\bibitem [{\citenamefont {Tsuchimoto}\ and\ \citenamefont
  {Tanimura}(2015)}]{tsuchimotoSpinsDynamicsDissipative2015}%
  \BibitemOpen
  \bibfield  {author} {\bibinfo {author} {\bibfnamefont {M.}~\bibnamefont
  {Tsuchimoto}}\ and\ \bibinfo {author} {\bibfnamefont {Y.}~\bibnamefont
  {Tanimura}},\ }\bibfield  {title} {\bibinfo {title} {Spins {{Dynamics}} in a
  {{Dissipative Environment}}: {{Hierarchal Equations}} of {{Motion Approach
  Using}} a {{Graphics Processing Unit}} ({{GPU}})},\ }\href
  {https://doi.org/10.1021/acs.jctc.5b00488} {\bibfield  {journal} {\bibinfo
  {journal} {J. Chem. Theory Comput.}\ }\textbf {\bibinfo {volume} {11}},\
  \bibinfo {pages} {3859} (\bibinfo {year} {2015})}\BibitemShut {NoStop}%
\bibitem [{\citenamefont {Temen}\ \emph {et~al.}(2020)\citenamefont {Temen},
  \citenamefont {Jain},\ and\ \citenamefont
  {Akimov}}]{temenHierarchicalEquationsMotion2020}%
  \BibitemOpen
  \bibfield  {author} {\bibinfo {author} {\bibfnamefont {S.}~\bibnamefont
  {Temen}}, \bibinfo {author} {\bibfnamefont {A.}~\bibnamefont {Jain}},\ and\
  \bibinfo {author} {\bibfnamefont {A.~V.}\ \bibnamefont {Akimov}},\ }\bibfield
   {title} {\bibinfo {title} {Hierarchical equations of motion in the {{Libra}}
  software package},\ }\href {https://doi.org/10.1002/qua.26373} {\bibfield
  {journal} {\bibinfo  {journal} {International Journal of Quantum Chemistry}\
  }\textbf {\bibinfo {volume} {120}},\ \bibinfo {pages} {e26373} (\bibinfo
  {year} {2020})}\BibitemShut {NoStop}%
\bibitem [{\citenamefont {Johansson}\ \emph {et~al.}(2013)\citenamefont
  {Johansson}, \citenamefont {Nation},\ and\ \citenamefont
  {Nori}}]{johanssonQuTiPPythonFramework2013}%
  \BibitemOpen
  \bibfield  {author} {\bibinfo {author} {\bibfnamefont {J.}~\bibnamefont
  {Johansson}}, \bibinfo {author} {\bibfnamefont {P.}~\bibnamefont {Nation}},\
  and\ \bibinfo {author} {\bibfnamefont {F.}~\bibnamefont {Nori}},\ }\bibfield
  {title} {\bibinfo {title} {{{QuTiP}} 2: {{A Python}} framework for the
  dynamics of open quantum systems},\ }\href
  {https://doi.org/10.1016/j.cpc.2012.11.019} {\bibfield  {journal} {\bibinfo
  {journal} {Computer Physics Communications}\ }\textbf {\bibinfo {volume}
  {184}},\ \bibinfo {pages} {1234} (\bibinfo {year} {2013})}\BibitemShut
  {NoStop}%
\bibitem [{\citenamefont
  {Akimov}(2016)}]{akimovLibraOpensourceMethodology2016}%
  \BibitemOpen
  \bibfield  {author} {\bibinfo {author} {\bibfnamefont {A.~V.}\ \bibnamefont
  {Akimov}},\ }\bibfield  {title} {\bibinfo {title} {Libra: {{An}} open-source
  ``methodology discovery'' library for quantum and classical dynamics
  simulations},\ }\href {https://doi.org/10.1002/jcc.24367} {\bibfield
  {journal} {\bibinfo  {journal} {Journal of Computational Chemistry}\ }\textbf
  {\bibinfo {volume} {37}},\ \bibinfo {pages} {1626} (\bibinfo {year}
  {2016})}\BibitemShut {NoStop}%
\bibitem [{\citenamefont {Gardner}\ \emph {et~al.}(2022)\citenamefont
  {Gardner}, \citenamefont {{Douglas-Gallardo}}, \citenamefont {Stark},
  \citenamefont {Westermayr}, \citenamefont {Janke}, \citenamefont
  {Habershon},\ and\ \citenamefont {Maurer}}]{gardnerNQCDynamicsJlJulia2022}%
  \BibitemOpen
  \bibfield  {author} {\bibinfo {author} {\bibfnamefont {J.}~\bibnamefont
  {Gardner}}, \bibinfo {author} {\bibfnamefont {O.~A.}\ \bibnamefont
  {{Douglas-Gallardo}}}, \bibinfo {author} {\bibfnamefont {W.~G.}\ \bibnamefont
  {Stark}}, \bibinfo {author} {\bibfnamefont {J.}~\bibnamefont {Westermayr}},
  \bibinfo {author} {\bibfnamefont {S.~M.}\ \bibnamefont {Janke}}, \bibinfo
  {author} {\bibfnamefont {S.}~\bibnamefont {Habershon}},\ and\ \bibinfo
  {author} {\bibfnamefont {R.~J.}\ \bibnamefont {Maurer}},\ }\bibfield  {title}
  {\bibinfo {title} {{{NQCDynamics}}.jl: {{A Julia}} package for nonadiabatic
  quantum classical molecular dynamics in the condensed phase},\ }\href
  {https://doi.org/10.1063/5.0089436} {\bibfield  {journal} {\bibinfo
  {journal} {The Journal of Chemical Physics}\ }\textbf {\bibinfo {volume}
  {156}},\ \bibinfo {pages} {174801} (\bibinfo {year} {2022})},\ \Eprint
  {https://arxiv.org/abs/https://doi.org/10.1063/5.0089436}
  {https://doi.org/10.1063/5.0089436} \BibitemShut {NoStop}%
\bibitem [{\citenamefont {Bezanson}\ \emph {et~al.}(2017)\citenamefont
  {Bezanson}, \citenamefont {Edelman}, \citenamefont {Karpinski},\ and\
  \citenamefont {Shah}}]{bezansonJuliaFreshApproach2017}%
  \BibitemOpen
  \bibfield  {author} {\bibinfo {author} {\bibfnamefont {J.}~\bibnamefont
  {Bezanson}}, \bibinfo {author} {\bibfnamefont {A.}~\bibnamefont {Edelman}},
  \bibinfo {author} {\bibfnamefont {S.}~\bibnamefont {Karpinski}},\ and\
  \bibinfo {author} {\bibfnamefont {V.~B.}\ \bibnamefont {Shah}},\ }\bibfield
  {title} {\bibinfo {title} {Julia: {{A}} fresh approach to numerical
  computing},\ }\href {https://doi.org/10.1137/141000671} {\bibfield  {journal}
  {\bibinfo  {journal} {SIAM Review}\ }\textbf {\bibinfo {volume} {59}},\
  \bibinfo {pages} {65} (\bibinfo {year} {2017})}\BibitemShut {NoStop}%
\bibitem [{\citenamefont {Aroeira}\ \emph {et~al.}(2022)\citenamefont
  {Aroeira}, \citenamefont {Davis}, \citenamefont {Turney},\ and\ \citenamefont
  {Schaefer}}]{aroeiraFermiJlModern2022}%
  \BibitemOpen
  \bibfield  {author} {\bibinfo {author} {\bibfnamefont {G.~J.~R.}\
  \bibnamefont {Aroeira}}, \bibinfo {author} {\bibfnamefont {M.~M.}\
  \bibnamefont {Davis}}, \bibinfo {author} {\bibfnamefont {J.~M.}\ \bibnamefont
  {Turney}},\ and\ \bibinfo {author} {\bibfnamefont {H.~F.}\ \bibnamefont
  {Schaefer}},\ }\bibfield  {title} {\bibinfo {title} {Fermi.jl: {{A Modern
  Design}} for {{Quantum Chemistry}}},\ }\href
  {https://doi.org/10.1021/acs.jctc.1c00719} {\bibfield  {journal} {\bibinfo
  {journal} {Journal of Chemical Theory and Computation}\ }\textbf {\bibinfo
  {volume} {18}},\ \bibinfo {pages} {677} (\bibinfo {year} {2022})}\BibitemShut
  {NoStop}%
\bibitem [{\citenamefont {Herbst}\ \emph {et~al.}(2021)\citenamefont {Herbst},
  \citenamefont {Levitt},\ and\ \citenamefont
  {Canc{\`e}s}}]{herbstDFTKJulianApproach2021}%
  \BibitemOpen
  \bibfield  {author} {\bibinfo {author} {\bibfnamefont {M.~F.}\ \bibnamefont
  {Herbst}}, \bibinfo {author} {\bibfnamefont {A.}~\bibnamefont {Levitt}},\
  and\ \bibinfo {author} {\bibfnamefont {E.}~\bibnamefont {Canc{\`e}s}},\
  }\bibfield  {title} {\bibinfo {title} {{{DFTK}}: {{A Julian}} approach for
  simulating electrons in solids},\ }\href
  {https://doi.org/10.21105/jcon.00069} {\bibfield  {journal} {\bibinfo
  {journal} {Proc. JuliaCon Conf.}\ }\textbf {\bibinfo {volume} {3}},\ \bibinfo
  {pages} {69} (\bibinfo {year} {2021})}\BibitemShut {NoStop}%
\bibitem [{\citenamefont {Kr{\"a}mer}\ \emph {et~al.}(2018)\citenamefont
  {Kr{\"a}mer}, \citenamefont {Plankensteiner}, \citenamefont {Ostermann},\
  and\ \citenamefont {Ritsch}}]{kramerQuantumOpticsJlJulia2018}%
  \BibitemOpen
  \bibfield  {author} {\bibinfo {author} {\bibfnamefont {S.}~\bibnamefont
  {Kr{\"a}mer}}, \bibinfo {author} {\bibfnamefont {D.}~\bibnamefont
  {Plankensteiner}}, \bibinfo {author} {\bibfnamefont {L.}~\bibnamefont
  {Ostermann}},\ and\ \bibinfo {author} {\bibfnamefont {H.}~\bibnamefont
  {Ritsch}},\ }\bibfield  {title} {\bibinfo {title} {{{QuantumOptics}}.jl: {{A
  Julia}} framework for simulating open quantum systems},\ }\href
  {https://doi.org/10.1016/j.cpc.2018.02.004} {\bibfield  {journal} {\bibinfo
  {journal} {Computer Physics Communications}\ }\textbf {\bibinfo {volume}
  {227}},\ \bibinfo {pages} {109} (\bibinfo {year} {2018})}\BibitemShut
  {NoStop}%
\bibitem [{\citenamefont {Poole}\ \emph {et~al.}(2020)\citenamefont {Poole},
  \citenamefont {Galvez~Vallejo},\ and\ \citenamefont
  {Gordon}}]{pooleNewKidBlock2020}%
  \BibitemOpen
  \bibfield  {author} {\bibinfo {author} {\bibfnamefont {D.}~\bibnamefont
  {Poole}}, \bibinfo {author} {\bibfnamefont {J.~L.}\ \bibnamefont
  {Galvez~Vallejo}},\ and\ \bibinfo {author} {\bibfnamefont {M.~S.}\
  \bibnamefont {Gordon}},\ }\bibfield  {title} {\bibinfo {title} {A {{New Kid}}
  on the {{Block}}: {{Application}} of {{Julia}} to
  {{Hartree}}\textendash{{Fock Calculations}}},\ }\href
  {https://doi.org/10.1021/acs.jctc.0c00337} {\bibfield  {journal} {\bibinfo
  {journal} {J. Chem. Theory Comput.}\ }\textbf {\bibinfo {volume} {16}},\
  \bibinfo {pages} {5006} (\bibinfo {year} {2020})}\BibitemShut {NoStop}%
\bibitem [{\citenamefont {Poole}\ \emph {et~al.}(2022)\citenamefont {Poole},
  \citenamefont {Galvez~Vallejo},\ and\ \citenamefont
  {Gordon}}]{pooleTaskBasedApproachParallel2022}%
  \BibitemOpen
  \bibfield  {author} {\bibinfo {author} {\bibfnamefont {D.}~\bibnamefont
  {Poole}}, \bibinfo {author} {\bibfnamefont {J.~L.}\ \bibnamefont
  {Galvez~Vallejo}},\ and\ \bibinfo {author} {\bibfnamefont {M.~S.}\
  \bibnamefont {Gordon}},\ }\bibfield  {title} {\bibinfo {title} {A
  {{Task-Based Approach}} to {{Parallel Restricted Hartree}}\textendash{{Fock
  Calculations}}},\ }\href {https://doi.org/10.1021/acs.jctc.1c00820}
  {\bibfield  {journal} {\bibinfo  {journal} {J. Chem. Theory Comput.}\
  }\textbf {\bibinfo {volume} {18}},\ \bibinfo {pages} {2144} (\bibinfo {year}
  {2022})}\BibitemShut {NoStop}%
\bibitem [{\citenamefont {Makri}(1999)}]{makriLinearResponseApproximation1999}%
  \BibitemOpen
  \bibfield  {author} {\bibinfo {author} {\bibfnamefont {N.}~\bibnamefont
  {Makri}},\ }\bibfield  {title} {\bibinfo {title} {The {{Linear Response
  Approximation}} and {{Its Lowest Order Corrections}}: {{An Influence
  Functional Approach}}},\ }\href {https://doi.org/10.1021/jp9847540}
  {\bibfield  {journal} {\bibinfo  {journal} {The Journal of Physical Chemistry
  B}\ }\textbf {\bibinfo {volume} {103}},\ \bibinfo {pages} {2823} (\bibinfo
  {year} {1999})}\BibitemShut {NoStop}%
\bibitem [{\citenamefont {Allen}\ \emph {et~al.}(2016)\citenamefont {Allen},
  \citenamefont {Walters},\ and\ \citenamefont
  {Makri}}]{allenDirectComputationInfluence2016}%
  \BibitemOpen
  \bibfield  {author} {\bibinfo {author} {\bibfnamefont {T.~C.}\ \bibnamefont
  {Allen}}, \bibinfo {author} {\bibfnamefont {P.~L.}\ \bibnamefont {Walters}},\
  and\ \bibinfo {author} {\bibfnamefont {N.}~\bibnamefont {Makri}},\ }\bibfield
   {title} {\bibinfo {title} {Direct {{Computation}} of {{Influence Functional
  Coefficients}} from {{Numerical Correlation Functions}}},\ }\href
  {https://doi.org/10.1021/acs.jctc.6b00390} {\bibfield  {journal} {\bibinfo
  {journal} {Journal of Chemical Theory and Computation}\ }\textbf {\bibinfo
  {volume} {12}},\ \bibinfo {pages} {4169} (\bibinfo {year}
  {2016})}\BibitemShut {NoStop}%
\bibitem [{\citenamefont {Walters}\ \emph {et~al.}(2017)\citenamefont
  {Walters}, \citenamefont {Allen},\ and\ \citenamefont
  {Makri}}]{waltersDirectDeterminationDiscrete2017}%
  \BibitemOpen
  \bibfield  {author} {\bibinfo {author} {\bibfnamefont {P.~L.}\ \bibnamefont
  {Walters}}, \bibinfo {author} {\bibfnamefont {T.~C.}\ \bibnamefont {Allen}},\
  and\ \bibinfo {author} {\bibfnamefont {N.}~\bibnamefont {Makri}},\ }\bibfield
   {title} {\bibinfo {title} {Direct determination of discrete harmonic bath
  parameters from molecular dynamics simulations},\ }\href
  {https://doi.org/10.1002/jcc.24527} {\bibfield  {journal} {\bibinfo
  {journal} {Journal of Computational Chemistry}\ }\textbf {\bibinfo {volume}
  {38}},\ \bibinfo {pages} {110} (\bibinfo {year} {2017})}\BibitemShut
  {NoStop}%
\bibitem [{\citenamefont
  {Bose}(2022{\natexlab{b}})}]{boseZerocostCorrectionsInfluence2022}%
  \BibitemOpen
  \bibfield  {author} {\bibinfo {author} {\bibfnamefont {A.}~\bibnamefont
  {Bose}},\ }\bibfield  {title} {\bibinfo {title} {Zero-cost corrections to
  influence functional coefficients from bath response functions},\ }\href
  {https://doi.org/10.1063/5.0101396} {\bibfield  {journal} {\bibinfo
  {journal} {The Journal of Chemical Physics}\ }\textbf {\bibinfo {volume}
  {157}},\ \bibinfo {pages} {054107} (\bibinfo {year}
  {2022}{\natexlab{b}})}\BibitemShut {NoStop}%
\bibitem [{\citenamefont {Tully}(1990)}]{tullyMolecularDynamicsElectronic1990}%
  \BibitemOpen
  \bibfield  {author} {\bibinfo {author} {\bibfnamefont {J.~C.}\ \bibnamefont
  {Tully}},\ }\bibfield  {title} {\bibinfo {title} {Molecular dynamics with
  electronic transitions},\ }\href {https://doi.org/10.1063/1.459170}
  {\bibfield  {journal} {\bibinfo  {journal} {The Journal of Chemical Physics}\
  }\textbf {\bibinfo {volume} {93}},\ \bibinfo {pages} {1061} (\bibinfo {year}
  {1990})}\BibitemShut {NoStop}%
\bibitem [{\citenamefont
  {Tully}(2012)}]{tullyPerspectiveNonadiabaticDynamics2012}%
  \BibitemOpen
  \bibfield  {author} {\bibinfo {author} {\bibfnamefont {J.~C.}\ \bibnamefont
  {Tully}},\ }\bibfield  {title} {\bibinfo {title} {Perspective:
  {{Nonadiabatic}} dynamics theory},\ }\href
  {https://doi.org/10.1063/1.4757762} {\bibfield  {journal} {\bibinfo
  {journal} {The Journal of Chemical Physics}\ }\textbf {\bibinfo {volume}
  {137}},\ \bibinfo {pages} {22A301} (\bibinfo {year} {2012})}\BibitemShut
  {NoStop}%
\bibitem [{\citenamefont {Wang}\ \emph {et~al.}(2016)\citenamefont {Wang},
  \citenamefont {Akimov},\ and\ \citenamefont
  {Prezhdo}}]{wangRecentProgressSurface2016}%
  \BibitemOpen
  \bibfield  {author} {\bibinfo {author} {\bibfnamefont {L.}~\bibnamefont
  {Wang}}, \bibinfo {author} {\bibfnamefont {A.}~\bibnamefont {Akimov}},\ and\
  \bibinfo {author} {\bibfnamefont {O.~V.}\ \bibnamefont {Prezhdo}},\
  }\bibfield  {title} {\bibinfo {title} {Recent {{Progress}} in {{Surface
  Hopping}}: 2011\textendash 2015},\ }\href
  {https://doi.org/10.1021/acs.jpclett.6b00710} {\bibfield  {journal} {\bibinfo
   {journal} {J. Phys. Chem. Lett.}\ }\textbf {\bibinfo {volume} {7}},\
  \bibinfo {pages} {2100} (\bibinfo {year} {2016})}\BibitemShut {NoStop}%
\bibitem [{\citenamefont
  {Tanimura}(2014)}]{tanimuraReducedHierarchicalEquations2014}%
  \BibitemOpen
  \bibfield  {author} {\bibinfo {author} {\bibfnamefont {Y.}~\bibnamefont
  {Tanimura}},\ }\bibfield  {title} {\bibinfo {title} {Reduced hierarchical
  equations of motion in real and imaginary time: {{Correlated}} initial states
  and thermodynamic quantities},\ }\href {https://doi.org/10.1063/1.4890441}
  {\bibfield  {journal} {\bibinfo  {journal} {The Journal of Chemical Physics}\
  }\textbf {\bibinfo {volume} {141}},\ \bibinfo {pages} {044114} (\bibinfo
  {year} {2014})},\ \Eprint
  {https://arxiv.org/abs/https://doi.org/10.1063/1.4890441}
  {https://doi.org/10.1063/1.4890441} \BibitemShut {NoStop}%
\bibitem [{\citenamefont
  {Tanimura}(2020)}]{tanimuraNumericallyExactApproach2020}%
  \BibitemOpen
  \bibfield  {author} {\bibinfo {author} {\bibfnamefont {Y.}~\bibnamefont
  {Tanimura}},\ }\bibfield  {title} {\bibinfo {title} {Numerically ``exact''
  approach to open quantum dynamics: {{The}} hierarchical equations of motion
  ({{HEOM}})},\ }\href {https://doi.org/10.1063/5.0011599} {\bibfield
  {journal} {\bibinfo  {journal} {The Journal of Chemical Physics}\ }\textbf
  {\bibinfo {volume} {153}},\ \bibinfo {pages} {20901} (\bibinfo {year}
  {2020})}\BibitemShut {NoStop}%
\bibitem [{\citenamefont {Lambert}\ and\ \citenamefont
  {Makri}(2012{\natexlab{a}})}]{lambertQuantumclassicalPathIntegralI2012}%
  \BibitemOpen
  \bibfield  {author} {\bibinfo {author} {\bibfnamefont {R.}~\bibnamefont
  {Lambert}}\ and\ \bibinfo {author} {\bibfnamefont {N.}~\bibnamefont
  {Makri}},\ }\bibfield  {title} {\bibinfo {title} {Quantum-classical path
  integral. {{I}}. {{Classical}} memory and weak quantum nonlocality},\ }\href
  {https://doi.org/10.1063/1.4767931} {\bibfield  {journal} {\bibinfo
  {journal} {The Journal of Chemical Physics}\ }\textbf {\bibinfo {volume}
  {137}},\ \bibinfo {pages} {22A552} (\bibinfo {year}
  {2012}{\natexlab{a}})}\BibitemShut {NoStop}%
\bibitem [{\citenamefont {Lambert}\ and\ \citenamefont
  {Makri}(2012{\natexlab{b}})}]{lambertQuantumclassicalPathIntegralII2012}%
  \BibitemOpen
  \bibfield  {author} {\bibinfo {author} {\bibfnamefont {R.}~\bibnamefont
  {Lambert}}\ and\ \bibinfo {author} {\bibfnamefont {N.}~\bibnamefont
  {Makri}},\ }\bibfield  {title} {\bibinfo {title} {Quantum-classical path
  integral. {{II}}. {{Numerical}} methodology},\ }\href
  {https://doi.org/10.1063/1.4767980} {\bibfield  {journal} {\bibinfo
  {journal} {The Journal of Chemical Physics}\ }\textbf {\bibinfo {volume}
  {137}},\ \bibinfo {pages} {22A553} (\bibinfo {year}
  {2012}{\natexlab{b}})}\BibitemShut {NoStop}%
\bibitem [{\citenamefont {Banerjee}\ and\ \citenamefont
  {Makri}(2013)}]{banerjeeQuantumClassicalPathIntegral2013}%
  \BibitemOpen
  \bibfield  {author} {\bibinfo {author} {\bibfnamefont {T.}~\bibnamefont
  {Banerjee}}\ and\ \bibinfo {author} {\bibfnamefont {N.}~\bibnamefont
  {Makri}},\ }\bibfield  {title} {\bibinfo {title} {Quantum-{{Classical Path
  Integral}} with {{Self-Consistent Solvent-Driven Reference Propagators}}},\
  }\href {https://doi.org/10.1021/jp4043123} {\bibfield  {journal} {\bibinfo
  {journal} {The Journal of Physical Chemistry B}\ }\textbf {\bibinfo {volume}
  {117}},\ \bibinfo {pages} {13357} (\bibinfo {year} {2013})}\BibitemShut
  {NoStop}%
\bibitem [{\citenamefont {Wang}\ and\ \citenamefont
  {Makri}(2019)}]{wangQuantumclassicalPathIntegral2019}%
  \BibitemOpen
  \bibfield  {author} {\bibinfo {author} {\bibfnamefont {F.}~\bibnamefont
  {Wang}}\ and\ \bibinfo {author} {\bibfnamefont {N.}~\bibnamefont {Makri}},\
  }\bibfield  {title} {\bibinfo {title} {Quantum-classical path integral with a
  harmonic treatment of the back-reaction},\ }\href
  {https://doi.org/10.1063/1.5091725} {\bibfield  {journal} {\bibinfo
  {journal} {The Journal of Chemical Physics}\ }\textbf {\bibinfo {volume}
  {150}},\ \bibinfo {pages} {184102} (\bibinfo {year} {2019})}\BibitemShut
  {NoStop}%
\bibitem [{\citenamefont {Cerrillo}\ and\ \citenamefont
  {Cao}(2014)}]{cerrilloNonMarkovianDynamicalMaps2014}%
  \BibitemOpen
  \bibfield  {author} {\bibinfo {author} {\bibfnamefont {J.}~\bibnamefont
  {Cerrillo}}\ and\ \bibinfo {author} {\bibfnamefont {J.}~\bibnamefont {Cao}},\
  }\bibfield  {title} {\bibinfo {title} {Non-{{Markovian Dynamical Maps}}:
  {{Numerical Processing}} of {{Open Quantum Trajectories}}},\ }\href
  {https://doi.org/10.1103/PhysRevLett.112.110401} {\bibfield  {journal}
  {\bibinfo  {journal} {Phys. Rev. Lett.}\ }\textbf {\bibinfo {volume} {112}},\
  \bibinfo {pages} {110401} (\bibinfo {year} {2014})}\BibitemShut {NoStop}%
\bibitem [{\citenamefont
  {Makri}(2021{\natexlab{c}})}]{makriSmallMatrixPath2021}%
  \BibitemOpen
  \bibfield  {author} {\bibinfo {author} {\bibfnamefont {N.}~\bibnamefont
  {Makri}},\ }\bibfield  {title} {\bibinfo {title} {Small {{Matrix Path
  Integral}} with {{Extended Memory}}},\ }\href
  {https://doi.org/10.1021/acs.jctc.0c00987} {\bibfield  {journal} {\bibinfo
  {journal} {Journal of Chemical Theory and Computation}\ }\textbf {\bibinfo
  {volume} {17}},\ \bibinfo {pages} {1} (\bibinfo {year}
  {2021}{\natexlab{c}})}\BibitemShut {NoStop}%
\bibitem [{\citenamefont {Rackauckas}\ and\ \citenamefont
  {Nie}(2017)}]{rackauckasDifferentialequationsJlPerformant2017}%
  \BibitemOpen
  \bibfield  {author} {\bibinfo {author} {\bibfnamefont {C.}~\bibnamefont
  {Rackauckas}}\ and\ \bibinfo {author} {\bibfnamefont {Q.}~\bibnamefont
  {Nie}},\ }\bibfield  {title} {\bibinfo {title}
  {Differentialequations.jl\textendash a performant and feature-rich ecosystem
  for solving differential equations in julia},\ }\href@noop {} {\bibfield
  {journal} {\bibinfo  {journal} {Journal of Open Research Software}\ }\textbf
  {\bibinfo {volume} {5}},\ \bibinfo {pages} {15} (\bibinfo {year}
  {2017})}\BibitemShut {NoStop}%
\bibitem [{\citenamefont {Tsitouras}(2011)}]{tsitourasRungeKuttaPairs2011}%
  \BibitemOpen
  \bibfield  {author} {\bibinfo {author} {\bibfnamefont {{\relax
  Ch}.}~\bibnamefont {Tsitouras}},\ }\bibfield  {title} {\bibinfo {title}
  {Runge\textendash{{Kutta}} pairs of order 5(4) satisfying only the first
  column simplifying assumption},\ }\href
  {https://doi.org/10.1016/j.camwa.2011.06.002} {\bibfield  {journal} {\bibinfo
   {journal} {Computers \& Mathematics with Applications}\ }\textbf {\bibinfo
  {volume} {62}},\ \bibinfo {pages} {770} (\bibinfo {year} {2011})}\BibitemShut
  {NoStop}%
\bibitem [{\citenamefont {Fishman}\ \emph
  {et~al.}(2022{\natexlab{a}})\citenamefont {Fishman}, \citenamefont {White},\
  and\ \citenamefont {Stoudenmire}}]{fishmanITensorSoftwareLibrary2022}%
  \BibitemOpen
  \bibfield  {author} {\bibinfo {author} {\bibfnamefont {M.}~\bibnamefont
  {Fishman}}, \bibinfo {author} {\bibfnamefont {S.}~\bibnamefont {White}},\
  and\ \bibinfo {author} {\bibfnamefont {E.}~\bibnamefont {Stoudenmire}},\
  }\bibfield  {title} {\bibinfo {title} {The {{ITensor Software Library}} for
  {{Tensor Network Calculations}}},\ }\href
  {https://doi.org/10.21468/SciPostPhysCodeb.4} {\bibfield  {journal} {\bibinfo
   {journal} {SciPost Phys. Codebases}\ ,\ \bibinfo {pages} {4}} (\bibinfo
  {year} {2022}{\natexlab{a}})}\BibitemShut {NoStop}%
\bibitem [{\citenamefont {Fishman}\ \emph
  {et~al.}(2022{\natexlab{b}})\citenamefont {Fishman}, \citenamefont {White},\
  and\ \citenamefont {Stoudenmire}}]{fishmanCodebaseReleaseITensor2022}%
  \BibitemOpen
  \bibfield  {author} {\bibinfo {author} {\bibfnamefont {M.}~\bibnamefont
  {Fishman}}, \bibinfo {author} {\bibfnamefont {S.}~\bibnamefont {White}},\
  and\ \bibinfo {author} {\bibfnamefont {E.}~\bibnamefont {Stoudenmire}},\
  }\bibfield  {title} {\bibinfo {title} {Codebase release 0.3 for
  {{ITensor}}},\ }\href {https://doi.org/10.21468/SciPostPhysCodeb.4-r0.3}
  {\bibfield  {journal} {\bibinfo  {journal} {SciPost Phys. Codebases}\ ,\
  \bibinfo {pages} {4}} (\bibinfo {year} {2022}{\natexlab{b}})}\BibitemShut
  {NoStop}%
\bibitem [{\citenamefont {Besard}\ \emph {et~al.}(2018)\citenamefont {Besard},
  \citenamefont {Foket},\ and\ \citenamefont
  {De~Sutter}}]{besardEffectiveExtensibleProgramming2018}%
  \BibitemOpen
  \bibfield  {author} {\bibinfo {author} {\bibfnamefont {T.}~\bibnamefont
  {Besard}}, \bibinfo {author} {\bibfnamefont {C.}~\bibnamefont {Foket}},\ and\
  \bibinfo {author} {\bibfnamefont {B.}~\bibnamefont {De~Sutter}},\ }\bibfield
  {title} {\bibinfo {title} {Effective extensible programming: {{Unleashing
  Julia}} on {{GPUs}}},\ }\bibfield  {journal} {\bibinfo  {journal} {IEEE
  Transactions on Parallel and Distributed Systems}\ }\href
  {https://doi.org/10.1109/TPDS.2018.2872064} {10.1109/TPDS.2018.2872064}
  (\bibinfo {year} {2018}),\ \Eprint {https://arxiv.org/abs/1712.03112}
  {arxiv:1712.03112 [cs.PL]} \BibitemShut {NoStop}%
\bibitem [{\citenamefont {Silbey}\ and\ \citenamefont
  {Harris}(1984)}]{silbeyVariationalCalculationDynamics1984}%
  \BibitemOpen
  \bibfield  {author} {\bibinfo {author} {\bibfnamefont {R.}~\bibnamefont
  {Silbey}}\ and\ \bibinfo {author} {\bibfnamefont {R.~A.}\ \bibnamefont
  {Harris}},\ }\bibfield  {title} {\bibinfo {title} {Variational calculation of
  the dynamics of a two level system interacting with a bath},\ }\href
  {https://doi.org/10.1063/1.447055} {\bibfield  {journal} {\bibinfo  {journal}
  {The Journal of Chemical Physics}\ }\textbf {\bibinfo {volume} {80}},\
  \bibinfo {pages} {2615} (\bibinfo {year} {1984})}\BibitemShut {NoStop}%
\bibitem [{\citenamefont {Xu}\ and\ \citenamefont
  {Cao}(2016)}]{xuNoncanonicalDistributionNonequilibrium2016}%
  \BibitemOpen
  \bibfield  {author} {\bibinfo {author} {\bibfnamefont {D.}~\bibnamefont
  {Xu}}\ and\ \bibinfo {author} {\bibfnamefont {J.}~\bibnamefont {Cao}},\
  }\bibfield  {title} {\bibinfo {title} {Non-canonical distribution and
  non-equilibrium transport beyond weak system-bath coupling regime: {{A}}
  polaron transformation approach},\ }\href
  {https://doi.org/10.1007/s11467-016-0540-2} {\bibfield  {journal} {\bibinfo
  {journal} {Frontiers of Physics}\ }\textbf {\bibinfo {volume} {11}},\
  \bibinfo {pages} {110308} (\bibinfo {year} {2016})}\BibitemShut {NoStop}%
\bibitem [{\citenamefont {Xu}\ \emph {et~al.}(2016)\citenamefont {Xu},
  \citenamefont {Wang}, \citenamefont {Zhao},\ and\ \citenamefont
  {Cao}}]{xuPolaronEffectsPerformance2016}%
  \BibitemOpen
  \bibfield  {author} {\bibinfo {author} {\bibfnamefont {D.}~\bibnamefont
  {Xu}}, \bibinfo {author} {\bibfnamefont {C.}~\bibnamefont {Wang}}, \bibinfo
  {author} {\bibfnamefont {Y.}~\bibnamefont {Zhao}},\ and\ \bibinfo {author}
  {\bibfnamefont {J.}~\bibnamefont {Cao}},\ }\bibfield  {title} {\bibinfo
  {title} {Polaron effects on the performance of light-harvesting systems: A
  quantum heat engine perspective},\ }\href
  {https://doi.org/10.1088/1367-2630/18/2/023003} {\bibfield  {journal}
  {\bibinfo  {journal} {New Journal of Physics}\ }\textbf {\bibinfo {volume}
  {18}},\ \bibinfo {pages} {023003} (\bibinfo {year} {2016})}\BibitemShut
  {NoStop}%
\bibitem [{\citenamefont
  {Jang}(2022)}]{jangPartiallyPolarontransformedQuantum2022}%
  \BibitemOpen
  \bibfield  {author} {\bibinfo {author} {\bibfnamefont {S.~J.}\ \bibnamefont
  {Jang}},\ }\bibfield  {title} {\bibinfo {title} {Partially
  polaron-transformed quantum master equation for exciton and charge transport
  dynamics},\ }\href {https://doi.org/10.1063/5.0106546} {\bibfield  {journal}
  {\bibinfo  {journal} {J. Chem. Phys.}\ }\textbf {\bibinfo {volume} {157}},\
  \bibinfo {pages} {104107} (\bibinfo {year} {2022})}\BibitemShut {NoStop}%
\bibitem [{\citenamefont {Lee}\ \emph {et~al.}(2012)\citenamefont {Lee},
  \citenamefont {Moix},\ and\ \citenamefont
  {Cao}}]{leeAccuracySecondOrder2012}%
  \BibitemOpen
  \bibfield  {author} {\bibinfo {author} {\bibfnamefont {C.~K.}\ \bibnamefont
  {Lee}}, \bibinfo {author} {\bibfnamefont {J.}~\bibnamefont {Moix}},\ and\
  \bibinfo {author} {\bibfnamefont {J.}~\bibnamefont {Cao}},\ }\bibfield
  {title} {\bibinfo {title} {Accuracy of second order perturbation theory in
  the polaron and variational polaron frames},\ }\href
  {https://doi.org/10.1063/1.4722336} {\bibfield  {journal} {\bibinfo
  {journal} {J. Chem. Phys.}\ }\textbf {\bibinfo {volume} {136}},\ \bibinfo
  {pages} {204120} (\bibinfo {year} {2012})}\BibitemShut {NoStop}%
\bibitem [{\citenamefont {Tanimura}\ and\ \citenamefont
  {Wolynes}(1991)}]{tanimuraQuantumClassicalFokkerPlanck1991}%
  \BibitemOpen
  \bibfield  {author} {\bibinfo {author} {\bibfnamefont {Y.}~\bibnamefont
  {Tanimura}}\ and\ \bibinfo {author} {\bibfnamefont {P.~G.}\ \bibnamefont
  {Wolynes}},\ }\bibfield  {title} {\bibinfo {title} {Quantum and classical
  {{Fokker-Planck}} equations for a {{Gaussian-Markovian}} noise bath},\ }\href
  {https://doi.org/10.1103/PhysRevA.43.4131} {\bibfield  {journal} {\bibinfo
  {journal} {Physical Review A}\ }\textbf {\bibinfo {volume} {43}},\ \bibinfo
  {pages} {4131} (\bibinfo {year} {1991})}\BibitemShut {NoStop}%
\bibitem [{\citenamefont {Duan}\ \emph {et~al.}(2017)\citenamefont {Duan},
  \citenamefont {Wang}, \citenamefont {Tang},\ and\ \citenamefont
  {Wu}}]{duanStudyExtendedHierarchy2017}%
  \BibitemOpen
  \bibfield  {author} {\bibinfo {author} {\bibfnamefont {C.}~\bibnamefont
  {Duan}}, \bibinfo {author} {\bibfnamefont {Q.}~\bibnamefont {Wang}}, \bibinfo
  {author} {\bibfnamefont {Z.}~\bibnamefont {Tang}},\ and\ \bibinfo {author}
  {\bibfnamefont {J.}~\bibnamefont {Wu}},\ }\bibfield  {title} {\bibinfo
  {title} {The study of an extended hierarchy equation of motion in the
  spin-boson model: {{The}} cutoff function of the sub-{{Ohmic}} spectral
  density},\ }\href {https://doi.org/10.1063/1.4997669} {\bibfield  {journal}
  {\bibinfo  {journal} {The Journal of Chemical Physics}\ }\textbf {\bibinfo
  {volume} {147}},\ \bibinfo {pages} {164112} (\bibinfo {year}
  {2017})}\BibitemShut {NoStop}%
\bibitem [{\citenamefont {Popescu}\ \emph {et~al.}(2015)\citenamefont
  {Popescu}, \citenamefont {Rahman},\ and\ \citenamefont
  {Kleinekath{\"o}fer}}]{popescuUsingChebychevExpansion2015}%
  \BibitemOpen
  \bibfield  {author} {\bibinfo {author} {\bibfnamefont {B.}~\bibnamefont
  {Popescu}}, \bibinfo {author} {\bibfnamefont {H.}~\bibnamefont {Rahman}},\
  and\ \bibinfo {author} {\bibfnamefont {U.}~\bibnamefont
  {Kleinekath{\"o}fer}},\ }\bibfield  {title} {\bibinfo {title} {Using the
  {{Chebychev}} expansion in quantum transport calculations},\ }\href
  {https://doi.org/10.1063/1.4917198} {\bibfield  {journal} {\bibinfo
  {journal} {J. Chem. Phys.}\ }\textbf {\bibinfo {volume} {142}},\ \bibinfo
  {pages} {154103} (\bibinfo {year} {2015})}\BibitemShut {NoStop}%
\bibitem [{\citenamefont {Tian}\ and\ \citenamefont
  {Chen}(2013)}]{tianApplicationHierarchicalEquations2013}%
  \BibitemOpen
  \bibfield  {author} {\bibinfo {author} {\bibfnamefont {H.}~\bibnamefont
  {Tian}}\ and\ \bibinfo {author} {\bibfnamefont {G.}~\bibnamefont {Chen}},\
  }\bibfield  {title} {\bibinfo {title} {Application of hierarchical equations
  of motion ({{HEOM}}) to time dependent quantum transport at zero and finite
  temperatures},\ }\href {https://doi.org/10.1140/epjb/e2013-40333-7}
  {\bibfield  {journal} {\bibinfo  {journal} {The European Physical Journal B}\
  }\textbf {\bibinfo {volume} {86}},\ \bibinfo {pages} {411} (\bibinfo {year}
  {2013})}\BibitemShut {NoStop}%
\bibitem [{\citenamefont {Liu}\ \emph {et~al.}(2014)\citenamefont {Liu},
  \citenamefont {Zhu}, \citenamefont {Bai},\ and\ \citenamefont
  {Shi}}]{liuReducedQuantumDynamics2014}%
  \BibitemOpen
  \bibfield  {author} {\bibinfo {author} {\bibfnamefont {H.}~\bibnamefont
  {Liu}}, \bibinfo {author} {\bibfnamefont {L.}~\bibnamefont {Zhu}}, \bibinfo
  {author} {\bibfnamefont {S.}~\bibnamefont {Bai}},\ and\ \bibinfo {author}
  {\bibfnamefont {Q.}~\bibnamefont {Shi}},\ }\bibfield  {title} {\bibinfo
  {title} {Reduced quantum dynamics with arbitrary bath spectral densities:
  {{Hierarchical}} equations of motion based on several different bath
  decomposition schemes},\ }\href {https://doi.org/10.1063/1.4870035}
  {\bibfield  {journal} {\bibinfo  {journal} {J. Chem. Phys.}\ }\textbf
  {\bibinfo {volume} {140}},\ \bibinfo {pages} {134106} (\bibinfo {year}
  {2014})}\BibitemShut {NoStop}%
\bibitem [{\citenamefont {Dunn}\ \emph {et~al.}(2019)\citenamefont {Dunn},
  \citenamefont {Tempelaar},\ and\ \citenamefont
  {Reichman}}]{dunnRemovingInstabilitiesHierarchical2019}%
  \BibitemOpen
  \bibfield  {author} {\bibinfo {author} {\bibfnamefont {I.~S.}\ \bibnamefont
  {Dunn}}, \bibinfo {author} {\bibfnamefont {R.}~\bibnamefont {Tempelaar}},\
  and\ \bibinfo {author} {\bibfnamefont {D.~R.}\ \bibnamefont {Reichman}},\
  }\bibfield  {title} {\bibinfo {title} {Removing instabilities in the
  hierarchical equations of motion: {{Exact}} and approximate projection
  approaches},\ }\href {https://doi.org/10.1063/1.5092616} {\bibfield
  {journal} {\bibinfo  {journal} {J. Chem. Phys.}\ }\textbf {\bibinfo {volume}
  {150}},\ \bibinfo {pages} {184109} (\bibinfo {year} {2019})}\BibitemShut
  {NoStop}%
\bibitem [{\citenamefont {Ishizaki}\ and\ \citenamefont
  {Fleming}(2009{\natexlab{a}})}]{ishizakiTheoreticalExaminationQuantum2009}%
  \BibitemOpen
  \bibfield  {author} {\bibinfo {author} {\bibfnamefont {A.}~\bibnamefont
  {Ishizaki}}\ and\ \bibinfo {author} {\bibfnamefont {G.~R.}\ \bibnamefont
  {Fleming}},\ }\bibfield  {title} {\bibinfo {title} {Theoretical examination
  of quantum coherence in a photosynthetic system at physiological
  temperature},\ }\href {https://doi.org/10.1073/pnas.0908989106} {\bibfield
  {journal} {\bibinfo  {journal} {Proceedings of the National Academy of
  Sciences}\ }\textbf {\bibinfo {volume} {106}},\ \bibinfo {pages} {17255}
  (\bibinfo {year} {2009}{\natexlab{a}})}\BibitemShut {NoStop}%
\bibitem [{\citenamefont {Adolphs}\ and\ \citenamefont
  {Renger}(2006)}]{adolphsHowProteinsTrigger2006}%
  \BibitemOpen
  \bibfield  {author} {\bibinfo {author} {\bibfnamefont {J.}~\bibnamefont
  {Adolphs}}\ and\ \bibinfo {author} {\bibfnamefont {T.}~\bibnamefont
  {Renger}},\ }\bibfield  {title} {\bibinfo {title} {How {{Proteins Trigger
  Excitation Energy Transfer}} in the {{FMO Complex}} of {{Green Sulfur
  Bacteria}}},\ }\href {https://doi.org/10.1529/biophysj.105.079483} {\bibfield
   {journal} {\bibinfo  {journal} {Biophysical Journal}\ }\textbf {\bibinfo
  {volume} {91}},\ \bibinfo {pages} {2778} (\bibinfo {year}
  {2006})}\BibitemShut {NoStop}%
\bibitem [{\citenamefont {Walters}\ and\ \citenamefont
  {Makri}(2016)}]{waltersIterativeQuantumclassicalPath2016}%
  \BibitemOpen
  \bibfield  {author} {\bibinfo {author} {\bibfnamefont {P.~L.}\ \bibnamefont
  {Walters}}\ and\ \bibinfo {author} {\bibfnamefont {N.}~\bibnamefont
  {Makri}},\ }\bibfield  {title} {\bibinfo {title} {Iterative quantum-classical
  path integral with dynamically consistent state hopping},\ }\href
  {https://doi.org/10.1063/1.4939950} {\bibfield  {journal} {\bibinfo
  {journal} {The Journal of Chemical Physics}\ }\textbf {\bibinfo {volume}
  {144}},\ \bibinfo {pages} {44108} (\bibinfo {year} {2016})}\BibitemShut
  {NoStop}%
\bibitem [{\citenamefont {Ishizaki}\ and\ \citenamefont
  {Fleming}(2009{\natexlab{b}})}]{ishizakiUnifiedTreatmentQuantum2009}%
  \BibitemOpen
  \bibfield  {author} {\bibinfo {author} {\bibfnamefont {A.}~\bibnamefont
  {Ishizaki}}\ and\ \bibinfo {author} {\bibfnamefont {G.~R.}\ \bibnamefont
  {Fleming}},\ }\bibfield  {title} {\bibinfo {title} {Unified treatment of
  quantum coherent and incoherent hopping dynamics in electronic energy
  transfer: {{Reduced}} hierarchy equation approach},\ }\href
  {https://doi.org/10.1063/1.3155372} {\bibfield  {journal} {\bibinfo
  {journal} {The Journal of Chemical Physics}\ }\textbf {\bibinfo {volume}
  {130}},\ \bibinfo {pages} {234111} (\bibinfo {year}
  {2009}{\natexlab{b}})}\BibitemShut {NoStop}%
\bibitem [{\citenamefont {Hjorth~Larsen}\ \emph {et~al.}(2017)\citenamefont
  {Hjorth~Larsen}, \citenamefont {J{\o}rgen~Mortensen}, \citenamefont
  {Blomqvist}, \citenamefont {Castelli}, \citenamefont {Christensen},
  \citenamefont {Du{\l}ak}, \citenamefont {Friis}, \citenamefont {Groves},
  \citenamefont {Hammer}, \citenamefont {Hargus}, \citenamefont {Hermes},
  \citenamefont {Jennings}, \citenamefont {Bjerre~Jensen}, \citenamefont
  {Kermode}, \citenamefont {Kitchin}, \citenamefont {Leonhard~Kolsbjerg},
  \citenamefont {Kubal}, \citenamefont {Kaasbjerg}, \citenamefont {Lysgaard},
  \citenamefont {Bergmann~Maronsson}, \citenamefont {Maxson}, \citenamefont
  {Olsen}, \citenamefont {Pastewka}, \citenamefont {Peterson}, \citenamefont
  {Rostgaard}, \citenamefont {Schi{\o}tz}, \citenamefont {Sch{\"u}tt},
  \citenamefont {Strange}, \citenamefont {Thygesen}, \citenamefont {Vegge},
  \citenamefont {Vilhelmsen}, \citenamefont {Walter}, \citenamefont {Zeng},\
  and\ \citenamefont {Jacobsen}}]{hjorthlarsenAtomicSimulationEnvironment2017}%
  \BibitemOpen
  \bibfield  {author} {\bibinfo {author} {\bibfnamefont {A.}~\bibnamefont
  {Hjorth~Larsen}}, \bibinfo {author} {\bibfnamefont {J.}~\bibnamefont
  {J{\o}rgen~Mortensen}}, \bibinfo {author} {\bibfnamefont {J.}~\bibnamefont
  {Blomqvist}}, \bibinfo {author} {\bibfnamefont {I.~E.}\ \bibnamefont
  {Castelli}}, \bibinfo {author} {\bibfnamefont {R.}~\bibnamefont
  {Christensen}}, \bibinfo {author} {\bibfnamefont {M.}~\bibnamefont
  {Du{\l}ak}}, \bibinfo {author} {\bibfnamefont {J.}~\bibnamefont {Friis}},
  \bibinfo {author} {\bibfnamefont {M.~N.}\ \bibnamefont {Groves}}, \bibinfo
  {author} {\bibfnamefont {B.}~\bibnamefont {Hammer}}, \bibinfo {author}
  {\bibfnamefont {C.}~\bibnamefont {Hargus}}, \bibinfo {author} {\bibfnamefont
  {E.~D.}\ \bibnamefont {Hermes}}, \bibinfo {author} {\bibfnamefont {P.~C.}\
  \bibnamefont {Jennings}}, \bibinfo {author} {\bibfnamefont {P.}~\bibnamefont
  {Bjerre~Jensen}}, \bibinfo {author} {\bibfnamefont {J.}~\bibnamefont
  {Kermode}}, \bibinfo {author} {\bibfnamefont {J.~R.}\ \bibnamefont
  {Kitchin}}, \bibinfo {author} {\bibfnamefont {E.}~\bibnamefont
  {Leonhard~Kolsbjerg}}, \bibinfo {author} {\bibfnamefont {J.}~\bibnamefont
  {Kubal}}, \bibinfo {author} {\bibfnamefont {K.}~\bibnamefont {Kaasbjerg}},
  \bibinfo {author} {\bibfnamefont {S.}~\bibnamefont {Lysgaard}}, \bibinfo
  {author} {\bibfnamefont {J.}~\bibnamefont {Bergmann~Maronsson}}, \bibinfo
  {author} {\bibfnamefont {T.}~\bibnamefont {Maxson}}, \bibinfo {author}
  {\bibfnamefont {T.}~\bibnamefont {Olsen}}, \bibinfo {author} {\bibfnamefont
  {L.}~\bibnamefont {Pastewka}}, \bibinfo {author} {\bibfnamefont
  {A.}~\bibnamefont {Peterson}}, \bibinfo {author} {\bibfnamefont
  {C.}~\bibnamefont {Rostgaard}}, \bibinfo {author} {\bibfnamefont
  {J.}~\bibnamefont {Schi{\o}tz}}, \bibinfo {author} {\bibfnamefont
  {O.}~\bibnamefont {Sch{\"u}tt}}, \bibinfo {author} {\bibfnamefont
  {M.}~\bibnamefont {Strange}}, \bibinfo {author} {\bibfnamefont {K.~S.}\
  \bibnamefont {Thygesen}}, \bibinfo {author} {\bibfnamefont {T.}~\bibnamefont
  {Vegge}}, \bibinfo {author} {\bibfnamefont {L.}~\bibnamefont {Vilhelmsen}},
  \bibinfo {author} {\bibfnamefont {M.}~\bibnamefont {Walter}}, \bibinfo
  {author} {\bibfnamefont {Z.}~\bibnamefont {Zeng}},\ and\ \bibinfo {author}
  {\bibfnamefont {K.~W.}\ \bibnamefont {Jacobsen}},\ }\bibfield  {title}
  {\bibinfo {title} {The atomic simulation environment\textemdash a {{Python}}
  library for working with atoms},\ }\href
  {https://doi.org/10.1088/1361-648X/aa680e} {\bibfield  {journal} {\bibinfo
  {journal} {Journal of Physics: Condensed Matter}\ }\textbf {\bibinfo {volume}
  {29}},\ \bibinfo {pages} {273002} (\bibinfo {year} {2017})}\BibitemShut
  {NoStop}%
\bibitem [{\citenamefont {White}\ and\ \citenamefont
  {Feiguin}(2004)}]{whiteRealTimeEvolutionUsing2004}%
  \BibitemOpen
  \bibfield  {author} {\bibinfo {author} {\bibfnamefont {S.~R.}\ \bibnamefont
  {White}}\ and\ \bibinfo {author} {\bibfnamefont {A.~E.}\ \bibnamefont
  {Feiguin}},\ }\bibfield  {title} {\bibinfo {title} {Real-{{Time Evolution
  Using}} the {{Density Matrix Renormalization Group}}},\ }\href
  {https://doi.org/10.1103/physrevlett.93.076401} {\bibfield  {journal}
  {\bibinfo  {journal} {Physical Review Letters}\ }\textbf {\bibinfo {volume}
  {93}},\ \bibinfo {pages} {076401} (\bibinfo {year} {2004})}\BibitemShut
  {NoStop}%
\bibitem [{\citenamefont
  {Schollw{\"o}ck}(2005)}]{schollwockDensitymatrixRenormalizationGroup2005}%
  \BibitemOpen
  \bibfield  {author} {\bibinfo {author} {\bibfnamefont {U.}~\bibnamefont
  {Schollw{\"o}ck}},\ }\bibfield  {title} {\bibinfo {title} {The density-matrix
  renormalization group},\ }\href {https://doi.org/10.1103/revmodphys.77.259}
  {\bibfield  {journal} {\bibinfo  {journal} {Reviews of Modern Physics}\
  }\textbf {\bibinfo {volume} {77}},\ \bibinfo {pages} {259} (\bibinfo {year}
  {2005})}\BibitemShut {NoStop}%
\bibitem [{\citenamefont
  {Schollw{\"o}ck}(2011{\natexlab{a}})}]{schollwockDensitymatrixRenormalizationGroup2011}%
  \BibitemOpen
  \bibfield  {author} {\bibinfo {author} {\bibfnamefont {U.}~\bibnamefont
  {Schollw{\"o}ck}},\ }\bibfield  {title} {\bibinfo {title} {The density-matrix
  renormalization group in the age of matrix product states},\ }\href
  {https://doi.org/10.1016/j.aop.2010.09.012} {\bibfield  {journal} {\bibinfo
  {journal} {Annals of Physics}\ }\textbf {\bibinfo {volume} {326}},\ \bibinfo
  {pages} {96} (\bibinfo {year} {2011}{\natexlab{a}})}\BibitemShut {NoStop}%
\bibitem [{\citenamefont
  {Schollw{\"o}ck}(2011{\natexlab{b}})}]{schollwockDensitymatrixRenormalizationGroup2011a}%
  \BibitemOpen
  \bibfield  {author} {\bibinfo {author} {\bibfnamefont {U.}~\bibnamefont
  {Schollw{\"o}ck}},\ }\bibfield  {title} {\bibinfo {title} {The density-matrix
  renormalization group: A short introduction},\ }\href
  {https://doi.org/10.1098/rsta.2010.0382} {\bibfield  {journal} {\bibinfo
  {journal} {Philosophical Transactions of the Royal Society A: Mathematical,
  Physical and Engineering Sciences}\ }\textbf {\bibinfo {volume} {369}},\
  \bibinfo {pages} {2643} (\bibinfo {year} {2011}{\natexlab{b}})}\BibitemShut
  {NoStop}%
\bibitem [{\citenamefont {Paeckel}\ \emph {et~al.}(2019)\citenamefont
  {Paeckel}, \citenamefont {K{\"o}hler}, \citenamefont {Swoboda}, \citenamefont
  {Manmana}, \citenamefont {Schollw{\"o}ck},\ and\ \citenamefont
  {Hubig}}]{paeckelTimeevolutionMethodsMatrixproduct2019}%
  \BibitemOpen
  \bibfield  {author} {\bibinfo {author} {\bibfnamefont {S.}~\bibnamefont
  {Paeckel}}, \bibinfo {author} {\bibfnamefont {T.}~\bibnamefont {K{\"o}hler}},
  \bibinfo {author} {\bibfnamefont {A.}~\bibnamefont {Swoboda}}, \bibinfo
  {author} {\bibfnamefont {S.~R.}\ \bibnamefont {Manmana}}, \bibinfo {author}
  {\bibfnamefont {U.}~\bibnamefont {Schollw{\"o}ck}},\ and\ \bibinfo {author}
  {\bibfnamefont {C.}~\bibnamefont {Hubig}},\ }\bibfield  {title} {\bibinfo
  {title} {Time-evolution methods for matrix-product states},\ }\href
  {https://doi.org/10.1016/j.aop.2019.167998} {\bibfield  {journal} {\bibinfo
  {journal} {Annals of Physics}\ }\textbf {\bibinfo {volume} {411}},\ \bibinfo
  {pages} {167998} (\bibinfo {year} {2019})}\BibitemShut {NoStop}%
\bibitem [{\citenamefont {Bose}\ and\ \citenamefont
  {Walters}(2022{\natexlab{b}})}]{boseEffectTemperatureGradient2022}%
  \BibitemOpen
  \bibfield  {author} {\bibinfo {author} {\bibfnamefont {A.}~\bibnamefont
  {Bose}}\ and\ \bibinfo {author} {\bibfnamefont {P.~L.}\ \bibnamefont
  {Walters}},\ }\bibfield  {title} {\bibinfo {title} {Effect of temperature
  gradient on quantum transport},\ }\href {https://doi.org/10.1039/D2CP03030F}
  {\bibfield  {journal} {\bibinfo  {journal} {Physical Chemistry Chemical
  Physics}\ }\textbf {\bibinfo {volume} {24}},\ \bibinfo {pages} {22431}
  (\bibinfo {year} {2022}{\natexlab{b}})}\BibitemShut {NoStop}%
\bibitem [{\citenamefont {Bose}\ and\ \citenamefont
  {Walters}(2022{\natexlab{c}})}]{boseTensorNetworkPath2022}%
  \BibitemOpen
  \bibfield  {author} {\bibinfo {author} {\bibfnamefont {A.}~\bibnamefont
  {Bose}}\ and\ \bibinfo {author} {\bibfnamefont {P.~L.}\ \bibnamefont
  {Walters}},\ }\bibfield  {title} {\bibinfo {title} {Tensor {{Network Path
  Integral Study}} of {{Dynamics}} in {{B850 LH2 Ring}} with {{Atomistically
  Derived Vibrations}}},\ }\href {https://doi.org/10.1021/acs.jctc.2c00163}
  {\bibfield  {journal} {\bibinfo  {journal} {Journal of Chemical Theory and
  Computation}\ }\textbf {\bibinfo {volume} {18}},\ \bibinfo {pages} {4095}
  (\bibinfo {year} {2022}{\natexlab{c}})}\BibitemShut {NoStop}%
\end{thebibliography}%
\end{document}